\documentclass[12pt]{article}
\usepackage[utf8x]{inputenc}
\usepackage[english]{babel}
\usepackage{amsmath}
\usepackage{amsfonts}
\usepackage{amssymb}
\usepackage{graphicx}
\usepackage[left=2cm,right=2cm,top=2cm,bottom=2cm]{geometry}
\usepackage[colorlinks=true,citecolor=blue,filecolor=magenta]{hyperref}
\usepackage[sort&compress,square,numbers,comma]{natbib}

\title{An Alternative Method for Solving Two Problems of the Standard Model}
\author{Merkotan K.K.$^1$, Zelentsova T.M.$^1$, Chudak N.O.$^1$, Ptashynskyi D.A.$^1$,\\
Urbanevich V.V.$^1$, Potiienko O.S.$^1$, Voitenko V.V.$^1$,\\
Berezovskyi O.D.$^1$, Sharph I.V.$^1$, Rusov V.D.$^1$}
\date{07.05.2017}

\begin{document}

\maketitle

\begin{center}
\begin{minipage}{15cm}
$^1$ \textit{Odessa National Polytechnic University, \\
     \mbox{~~}Shevchenko Ave. 1, Odessa, 65044, Ukraine}

%$^2$ \textit{Department of High Energy Nuclear Physics, Institute of Modern Physics, \\
%     \mbox{~~}Nanchang Road 509, 730000 Lanzhou, China}

%$^3$ \textit{D\'{e}partement de physique nucl\'{e}aire et corpusculaire, Universit\'{e} de Gen\'{e}ve, \\
 %    \mbox{~~}CH-1211 Geneva 4, Switzerland}

%$^4$ \textit{Department of Experimental Particle Physics, Jo\v{z}ef Stefan Institute, \\
%     \mbox{~~}Jamova 39, SI-1000 Ljubljana, Slovenia}
\end{minipage}
\end{center}

\begin{abstract}
Two problems of the Standard Model, associated with the introduction of non-gauge interactions and with the introduction of an electromagnetic field as a linear combination of fields on which various gauge groups are implemented, are analyzed. It is noticed that the existing model contains $U\left( 1 \right)-$ phase uncertainty of the matrix elements of the raising and lowering generators of the $SU\left( 2 \right)$ group. This uncertainty creates the condition for the additional local $U\left( 1 \right)-$ symmetry of the Standard Model Lagrangian with respect to the choice of various equivalent generator representations of the $SU\left( 2 \right)$ group, which is provided by the electromagnetic field. In this case, due to the different action of the raising and lowering generators on the fields of each generation of leptons and quarks, these fields interact with the electromagnetic field in different ways. It is also shown that considering the multi-particle gauge field a description of the Higgs mechanism can be obtained, free from the shortcomings of the well-known single-particle description, the main of which is the introduction of the non-gauge \mbox{<<phi-four>>} interaction, that is not reduced to the fundamental one. In the proposed model, the spontaneous symmetry breaking is achieved due to the same fundamental interaction, the mediating particle mass of which it provides. 


\

\end{abstract}

\section{Introduction}
In our opinion, the theory of the electroweak interaction \cite{WeinbergPhysRevLett.19.1264, SALAM1964168, GLASHOW1961579} and the Higgs mechanism associated with it \cite{HIGGS1964132, PhysRevLett.13.508} have a number of theoretical problems that are accompanied by a certain disagreement with recent experimental results \cite{Chatrchyan:2012xdj,Aad2016,Heinemeyer:2013tqa}. In particular, in the mentioned papers, the results on the experimental observations of the Higgs boson decay channels are given. The currently known experimental data on the decay channels are collected in the review \cite{Olive:2016xmw}. From the experimental results presented in these papers it is clear that the total weak isospin of the particles in the final decay state can only be an integer. At the same time, the Higgs field of the Standard Model \cite{WeinbergPhysRevLett.19.1264,Olive:2016xmw} is transformed by the fundamental representation of the $SU\left( 2 \right)$ group and is two-component. That is, according to the Standard Model, the Higgs boson has a weak isospin equal to ${1}/{2}\;$ \cite{Ryder:1996,PeskinSchroeder:16449}. An analysis of the $\beta -$ decay processes as well as the various channels of the leptonic decays \cite{Olive:2016xmw} leads to the conclusion that the weak interaction conserves the weak isospin. Since the particles mediating the other interactions have a zero weak isospin, it should also be conserved in the processes that occur due to the other interactions. 
 
From our point of view, the essential theoretical problem of the Standard Model is the introduction of {\flqq new\frqq} non-gauge interactions. In particular, the nonzero vacuum expectation value of the Higgs field is achieved through the non-gauge interaction ${{\phi }^{4}}$. The only demonstration of the such interaction, which is discussed, is the spontaneous symmetry breaking \cite{GoldstoneSalamWeinbergPhysRev}. Therefore, it is unclear how the existence of this interaction can be confirmed experimentally. The same can be said about the Yukawa interaction of fermion fields with the Higgs field \cite{WeinbergPhysRevLett.19.1264,Ryder:1996,PeskinSchroeder:16449}, that provides a mass to these fields. These interactions are not a consequence of the localization of any symmetry and therefore are not reduced to any of the known interactions. 

An another point is the unnatural sign of the term, which is quadratic by the Higgs field, in the Lagrangian. Although it does not result to the nonphysical consequences, the question still arises why there is one single field with such properties. The only argument in favor of the such Lagrangian notation is that the goal is achieved in this way, i.e. the appearance of the mass in the gauge particles \cite{Grojean:2007}.

At the same time, considering the multi-particle fields in our previous works \cite{Chudak:2016, Volkotrub:2015laa} we noticed, that the dynamic equation of the two-particle gauge field was similar to the dynamic equation of the \mbox{\flqq phi-three\frqq} theory, but in the place of the squared mass there was an operator, which could have negative eigenvalues under certain conditions. In this case, the operators of the two-particle gauge field describe the creation and annihilation of particles, which are bound states of gauge bosons. Therefore, a self-action of the two-particle gauge field is the demonstration of the interaction between the quanta of the non-Abelian gauge field and does not require the introduction of any \mbox{\flqq new\frqq} interaction. Hence, In the model of the multi-particle fields there are both components that are necessary for the spontaneous symmetry breaking. In this case, the Higgs boson, as in some other models, is not the elemental particle, but the bound state. In particular, P. Higgs himself pointed to this possibility \cite{PhysRevLett.13.508}, but he expected that the scalar field, which made the symmetry be broken, would contain the fermion fields, and not the gauge bosons. In the paper \cite{Hoh:2016} the Higgs boson is considered as a bound state of gauge bosons, which are in a confinement state. In the papers \cite{Chudak:2016, Volkotrub:2015laa} the two-particle gauge field describes the confinement of quarks and gluons. But the field, for which the symmetry is spontaneously broken, describes the creation and annihilation of the bound states of the gauge bosons without the confinement. That is, there are the bound states with a finite binding energy. However, such fields, although they have a nonzero vacuum expectation value, can not lead to the appearance of the mass in gauge bosons, because a scalar representation of the internal symmetry group is realized on them. As a result, they can not interact with the single-particle gauge field and therefore can not create its mass. Therefore, one of the problems of the presented paper is to consider the two-particle gauge field on which the vector representation of the internal symmetry group is realized. Since we are interested in the Higgs mechanism, as this group we will consider the $SU\left( 2 \right)$ group.

Also, in our opinion, the consideration of the electromagnetic field as a linear combination \cite{WeinbergPhysRevLett.19.1264,SALAM1964168,GLASHOW1961579,Ryder:1996,PeskinSchroeder:16449} leads to a number of problems
\begin{equation}
{{A}_{{{a}_{1}}}}\left( x \right)=\sin \left( {{\theta }_{W}} \right){{A}_{{{a}_{1}},{{g}_{1}}=3}}\left( x \right)+\cos \left( {{\theta }_{W}} \right){{B}_{{{a}_{1}}}}\left( x \right).
\label{eq:A_kak_lin_komb_teta}
\end{equation}
Here ${{A}_{{{a}_{1}}}}\left( x \right)-$ field functions of the electromagnetic field, $a_{1}-$ a Lorentz index, ${{A}_{{{a}_{1}},{{g}_{1}}}}\left( x \right)-$ field functions of the gauge field, that restores the local $SU\left( 2 \right)-$ symmetry, $ {g}_{1}-$ an internal index, ${{B}_{{{a}_{1}}}}\left( x \right)-$ a gauge field that restores the local $U\left( 1 \right)-$ symmetry in the Weinberg-Salam-Glashow model, ${{\theta }_{W}}-$ a Weinberg angle. The problems, that arise from the introduction of the electromagnetic field in the form of the linear combination of fields (\ref{eq:A_kak_lin_komb_teta}) with different transformation laws, are associated to the fact, that in this way the electromagnetic field stops being related to a definite group of the Lagrangian transformations. It thus loses the basic physical function of the gauge field, which is to provide the Lagrangian invariance with respect to a certain local symmetry group, since such a function in the Standard Model Lagrangian is performed by the ${{A}_{{{a}_{1}},{{g}_{1}}}}\left( x \right)$ and ${{B}_{{{a}_{1}}}}\left( x \right)$ fields. Let's consider the mentioned problems in detail.

The first problem that accompanies the expansion of (\ref{eq:A_kak_lin_komb_teta}) is, in our opinion, that the dynamic equations for the electromagnetic field will depend on the gauge choice of the $SU\left( 2 \right)-$ fields. Indeed, a part of the Lagrangian of the electroweak theory containing the field functions of the electromagnetic field can be written as:
\begin{equation}
L_{A}^{\operatorname{int}}=L_{A}^{0}+\sum\limits_{k=1}^{6}{L_{A,k}^{\operatorname{int}}},  
 \label{eq:L_A}
 \end{equation} 
where
\begin{equation}
\begin{split}
L_{A}^{0}=-\frac{1}{4}{{g}^{{{a}_{1}}{{a}_{11}}}}{{g}^{{{b}_{1}}{{b}_{11}}}}\left( \frac{\partial {{A}_{{{b}_{1}}}}\left( x \right)}{\partial {{x}^{{{a}_{1}}}}}-\frac{\partial {{A}_{{{a}_{1}}}}\left( x \right)}{\partial {{x}^{{{b}_{1}}}}} \right)\left( \frac{\partial {{A}_{{{b}_{11}}}}\left( x \right)}{\partial {{x}^{{{a}_{11}}}}}-\frac{\partial {{A}_{{{a}_{11}}}}\left( x \right)}{\partial {{x}^{{{b}_{11}}}}} \right),  
\end{split}
\end{equation}
\label{eq:L_A0}
\begin{equation}
\begin{split}
L_{A,1}^{\operatorname{int}}=-\frac{i}{2}e{{g}^{{{a}_{1}}{{a}_{11}}}}{{g}^{{{b}_{1}}{{b}_{11}}}}{{A}_{{{a}_{11}}}}\left( x \right)\left( \frac{\partial W_{{{b}_{1}}}^{+}\left( x \right)}{\partial {{x}^{{{a}_{1}}}}}W_{{{b}_{11}}}^{-}\left( x \right)-\frac{\partial W_{{{b}_{1}}}^{-}\left( x \right)}{\partial {{x}^{{{a}_{1}}}}}W_{{{b}_{11}}}^{+}\left( x \right) \right),  
\end{split}
\label{eq:LAint1}
\end{equation}
\begin{equation}
\begin{split}
L_{A,2}^{\operatorname{int}}=-\frac{i}{2}e{{g}^{{{b}_{1}}{{b}_{11}}}}{{g}^{{{a}_{1}}{{a}_{11}}}}{{A}_{{{a}_{1}}}}\left( x \right)\left( W_{{{b}_{11}}}^{+}\left( x \right)\frac{\partial W_{{{a}_{11}}}^{-}\left( x \right)}{\partial {{x}^{{{b}_{1}}}}}-W_{{{b}_{11}}}^{-}\left( x \right)\frac{\partial W_{{{a}_{11}}}^{+}\left( x \right)}{\partial {{x}^{{{b}_{1}}}}} \right),  
\end{split}
\label{eq:LAint2}
\end{equation}
\begin{equation}
\begin{split}
L_{A,3}^{\operatorname{int}}=\frac{1}{2}{{e}^{2}}{{g}^{{{b}_{1}}{{b}_{11}}}}{{g}^{{{a}_{1}}{{a}_{11}}}}{{A}_{{{a}_{1}}}}\left( x \right){{A}_{{{b}_{11}}}}\left( x \right)W_{{{b}_{1}}}^{+}\left( x \right)W_{{{a}_{11}}}^{-}\left( x \right),  
\end{split}
\label{eq:LAint3}
\end{equation}
\begin{equation}
\begin{split}
  & L_{A,4}^{\operatorname{int}}=\frac{1}{2}{{e}^{2}}\operatorname{ctg}\left( {{\theta }_{W}} \right){{g}^{{{b}_{1}}{{b}_{11}}}}{{g}^{{{a}_{1}}{{a}_{11}}}}\left( Z_{{{a}_{1}}}^{0}\left( x \right)W_{{{a}_{11}}}^{-}\left( x \right)W_{{{b}_{1}}}^{+}\left( x \right){{A}_{{{b}_{11}}}}\left( x \right) \right.+ \\ 
 & \left. +Z_{{{b}_{11}}}^{0}\left( x \right)W_{{{b}_{1}}}^{+}\left( x \right)W_{{{a}_{11}}}^{-}\left( x \right){{A}_{{{a}_{1}}}}\left( x \right)-2Z_{{{b}_{1}}}^{0}\left( x \right)W_{{{a}_{1}}}^{+}\left( x \right)W_{{{a}_{11}}}^{-}\left( x \right){{A}_{{{b}_{11}}}}\left( x \right) \right), \\   
\end{split}
\label{eq:LAint4}
\end{equation}
\begin{equation}
\begin{split}
 L_{A,5}^{\operatorname{int}}=-\frac{1}{2}{{e}^{2}}{{g}^{{{a}_{1}}{{a}_{11}}}}{{g}^{{{b}_{1}}{{b}_{11}}}}{{A}_{{{b}_{11}}}}\left( x \right){{A}_{{{b}_{1}}}}\left( x \right)W_{{{a}_{1}}}^{+}\left( x \right)W_{{{a}_{11}}}^{-}\left( x \right),   
\end{split}
\label{eq:LAint5}
\end{equation}
\begin{equation}
\begin{split}
L_{A,6}^{\operatorname{int}}=\frac{i}{4}e{{g}^{{{a}_{1}}{{a}_{11}}}}{{g}^{{{b}_{1}}{{b}_{11}}}}\left( \frac{\partial {{A}_{{{b}_{1}}}}\left( x \right)}{\partial {{x}^{{{a}_{1}}}}}-\frac{\partial {{A}_{{{a}_{1}}}}\left( x \right)}{\partial {{x}^{{{b}_{1}}}}} \right)\left( W_{{{a}_{11}}}^{+}\left( x \right)W_{{{b}_{11}}}^{-}\left( x \right)-W_{{{a}_{11}}}^{-}\left( x \right)W_{{{b}_{11}}}^{+}\left( x \right) \right).   
\end{split}
\label{eq:LAint6}
\end{equation}
Here $ e -$ an electromagnetic coupling constant, ${{a}_{1}},{{a}_{11}},{{b}_{1}},{{b}_{11}}-$ Lorentz indices, $ {{g}^{{{a}_{1}}{{a}_{11}}}},{{g}^{{{b}_{1}}{{b}_{11}}}}- $ components of the Minkowski tensor,  $ W_{{{a}_{11}}}^{+}\left( x \right)$ and $W_{{{b}_{11}}}^{-}\left( x \right)-$ functions describing the field of $W-$ bosons:
\begin{equation}
\begin{split}
& W_{{{a}_{1}}}^{+}\left( x \right)={{A}_{{{a}_{1}},{{g}_{1}}=1}}\left( x \right)-i{{A}_{{{a}_{1}},{{g}_{1}}=2}}\left( x \right), \\ 
 & W_{{{a}_{1}}}^{-}\left( x \right)={{A}_{{{a}_{1}},{{g}_{1}}=1}}\left( x \right)+i{{A}_{{{a}_{1}},{{g}_{1}}=2}}\left( x \right). \\      
\end{split}
\label{eq:Wpm}
\end{equation}
As it is seen from (\ref{eq:LAint1}-\ref{eq:LAint6}), the electromagnetic field is included to the interaction Lagrangian with factors that are not invariant under the local $SU\left( 2 \right)-$ transformations. The remaining terms that restore the local $SU\left( 2 \right)\otimes U\left( 1 \right)-$ invariance of the total Lagrangian do not depend on the ${{A}_{{{a}_{1}}}}\left( x \right)$ field and therefore will not affect the dynamic equations of this field. In addition, the term (\ref{eq:LAint6}), that can be considered as a term in the Lagrangian of the interaction of the electromagnetic field with the ${{W}^{\pm }}-$ boson field, can not be interpreted as the result of the derivative {\flqq extension\frqq}. This is a logical consequence of the introduction of the electromagnetic field (\ref{eq:A_kak_lin_komb_teta}), because in this way it loses the meaning of the field that establishes the local $U\left( 1 \right)-$ invariance, since in the Weinberg-Salam-Glashow model such a meaning is given to the ${{B}_{{{a}_{1}}}}\left( x \right)$ field. However, we were not able to find any experimental facts about observations of the quanta of this field. At the same time, in terms of the quantum-mechanical superposition principle, a photon with a probability of ${{\cos }^{2}}\left( {{\theta }_{W}} \right)\approx 0.8$ during the measurement can be converted to a state corresponding to a quantum of the ${{B}_{{{a}_{1}}}}\left( x \right)$ field. 

Further in this paper, we will try to {\flqq return\frqq} to the electromagnetic field the status of the field {\flqq lost\frqq} in the Standard Model, which restores the local $U\left( 1 \right)-$ invariance.      

Another problem is related to the transformation law of the electromagnetic field under the local $SU\left( 2 \right)\otimes U\left( 1 \right)-$ transformation. We apply this transformation to the fields that enter into (\ref{eq:A_kak_lin_komb_teta}). In this case, we denote the set of three $SU\left( 2 \right)-$ transformation parameters as $\vec{\alpha }\left( x \right)=\left( {{\alpha }_{1}}\left( x \right),{{\alpha }_{2}}\left( x \right),{{\alpha }_{3}}\left( x \right) \right)$, and the $U\left( 1 \right)-$ transformation parameter as $\beta \left( x \right)$. Let us write the new gauge relation analogous to (\ref{eq:A_kak_lin_komb_teta}), denoting the corresponding field configurations as well as in (\ref{eq:A_kak_lin_komb_teta}), but {\flqq with a dash\frqq}:
\begin{equation}
{{{A}'}_{{{a}_{1}}}}\left( x \right)=\sin \left( {{\theta }_{W}} \right){{{A}'}_{{{a}_{1}},3}}\left( x \right)+\cos \left( {{\theta }_{W}} \right){{{B}'}_{{{a}_{1}}}}\left( x \right).
\label{eq:A_kak_lin_komb_teta_htrih}
\end{equation}  
Using the transformation laws of the non-Abelian and Abelian fields, we obtain:
\begin{equation}
\begin{split}
    & {{{{A}'}}_{{{a}_{1}}}}\left( x \right)=\sin \left( {{\theta }_{W}} \right){{D}_{3,3}}\left( \vec{\theta }\left( x \right) \right)\left( \frac{\partial {{\theta }_{3}}\left( x \right)}{\partial {{x}^{{{a}_{1}}}}}+{{A}_{{{a}_{1}},3}}\left( x \right) \right)+ \\ 
 & +\sin \left( {{\theta }_{W}} \right){{D}_{3,2}}\left( \vec{\theta }\left( x \right) \right)\left( \frac{\partial {{\theta }_{2}}\left( x \right)}{\partial {{x}^{{{a}_{1}}}}}+{{A}_{{{a}_{1}},2}}\left( x \right) \right) \\ 
 & +\sin \left( {{\theta }_{W}} \right){{D}_{3,1}}\left( \vec{\theta }\left( x \right) \right)\left( \frac{\partial {{\theta }_{1}}\left( x \right)}{\partial {{x}^{{{a}_{1}}}}}+{{A}_{{{a}_{1}},1}}\left( x \right) \right) \\ 
 & +\cos \left( {{\theta }_{W}} \right){{B}_{{{a}_{1}}}}\left( x \right)+\cos \left( {{\theta }_{W}} \right)\frac{\partial \theta \left( x \right)}{\partial {{x}^{{{a}_{1}}}}}. \\   
\end{split}
\label{eq:Ahtrix}
\end{equation}
Here ${{D}_{{{g}_{1}},{{g}_{2}}}}\left( \vec{\theta }\left( x \right) \right),{{g}_{1}},{{g}_{2}}=1,2,3-$ matrix elements of the adjoint $SU\left( 2 \right)$ group representation. Taking (\ref{eq:A_kak_lin_komb_teta}) into account, we obtain for the electromagnetic field a complex transformation law
\begin{equation}
\begin{split}
    & {{{{A}'}}_{{{a}_{1}}}}\left( x \right)={{A}_{{{a}_{1}}}}\left( x \right)+\sin \left( {{\theta }_{W}} \right){{D}_{3,2}}\left( \vec{\theta }\left( x \right) \right)\left( \frac{\partial {{\theta }_{2}}\left( x \right)}{\partial {{x}^{{{a}_{1}}}}}+{{A}_{{{a}_{1}},2}}\left( x \right) \right) \\ 
 & +\sin \left( {{\theta }_{W}} \right){{D}_{3,1}}\left( \vec{\theta }\left( x \right) \right)\left( \frac{\partial {{\theta }_{1}}\left( x \right)}{\partial {{x}^{{{a}_{1}}}}}+{{A}_{{{a}_{1}},1}}\left( x \right) \right) \\ 
 & +\left( 1-{{D}_{3,3}}\left( \vec{\theta }\left( x \right) \right) \right)\cos \left( {{\theta }_{W}} \right){{B}_{{{a}_{1}}}}\left( x \right) \\ 
 & +\cos \left( {{\theta }_{W}} \right)\frac{\partial \theta \left( x \right)}{\partial {{x}^{{{a}_{1}}}}}+\sin \left( {{\theta }_{W}} \right){{D}_{3,3}}\left( \vec{\theta }\left( x \right) \right)\frac{\partial {{\theta }_{3}}\left( x \right)}{\partial {{x}^{{{a}_{1}}}}}, \\    
\end{split}
\label{eq:skladnij_zakon}
\end{equation}
which differs significantly from the electromagnetic field transformation law known from the electrodynamics, that is a natural consequence of the presence in (\ref{eq:A_kak_lin_komb_teta}) of the non-Abelian term. Therefore, there is a question, how the observed values, which are the strengths of the electric and magnetic fields, can be constructed, because for this it is necessary to find expressions that would not depend on either the gauge $SU\left( 2 \right)-$ field or the gauge $U\left( 1 \right)-$ field.

Thus, to introduce an electromagnetic field using the relation (\ref{eq:A_kak_lin_komb_teta}) without the {\flqq destruction\frqq} of the well-known and repeatedly verified description of the electromagnetic field, it seems impossible. Therefore, in this paper we propose a slightly different approach to the description of the electroweak interaction, which does not {\flqq destruct\frqq} the usual description of the electromagnetic field. Let us consider this approach. In doing so, we review it first within the {\flqq old\frqq} Standard Model in which neutrinos were considered as massless particles, which are represented only by left-handed spinors. Next, we will discuss how this approach can be changed due to the established fact of neutrino oscillations \cite{SuperCamiokaNDEPhysRevLett.81.1562,SNOcollaborationPhysRevLett.87.071301,SNO_1_PhysRevLett.89.011301}.  

\section{An alternative way to describe the electromagnetic interaction}
The main problem of introducing the electromagnetic interaction into the Standard Model is that the various components of the generations of leptons and quarks have different electric charges \cite{Olive:2016xmw} and, consequently, interact differently with the electromagnetic field. For example, for the electron-electron neutrino generation, it is necessary to {\flqq force\frqq} the electromagnetic field to interact with the electron field and at the same time not to interact with the neutrino field. Generating the considered problems the realation \eqref{eq:A_kak_lin_komb_teta} is a consequence of the method chosen in the Standard Model for solving this problem. An alternative to \eqref{eq:A_kak_lin_komb_teta} way of introducing the electromagnetic interaction is based on the fact that usually the consideration of non-Abelian fields is implemented in one certain representation for the generators of the gauge group. This ignores the possibility of the review of the equivalent generator representations. 

We consider a non-Abelian gauge field in the matrix form (in contrast to numbers, matrices will be denoted by a {\flqq hat\frqq}):
\begin{equation}\label{Amatrix}
{{\hat{A}}_{{{a}_{1}}}}\left( x \right)={{A}_{{{a}_{1}},{{g}_{1}}}}\left( x \right){{\hat{t}}_{{{g}_{1}}}}.
\end{equation}
Here ${{\hat{t}}_{{{g}_{1}}}}-$ generators of arbitrary representation of the gauge group. If one chooses as ${{\hat{t}}_{{{g}_{1}}}}$ different sets of generators, then only the expression for the interaction Lagrangian of the gauge field with fermion fields will change, but not the Lagrangian of the non-Abelian gauge field itself. The selection of the generators ${{\hat{t}}_{{{g}_{1}}}}$ is caused by the representation of the gauge group on the fermion fields. But this selection is fixed up to an equivalent representation. From a physical point of view, it is natural to require that dynamic models, in which fields \eqref{Amatrix} are {\flqq extended\frqq} to generators of equivalent representations, have led to physically equivalent results. Formally, these models can not be reduced to each other by any transformation of the local gauge group to which the field \eqref{Amatrix} is associated. Indeed, the laws of such transformations are such that the Lagrangian does not change under these transformations \textit{\textbf{with a certain selection of generators}}. Therefore, by any transformation of the local gauge group, it is impossible to change the expression for the generators of this group, that is included in the Lagrangian. So it becomes necessary to change the Lagrangian in a way to provide the symmetry with respect to the transition from generators of one equivalent representation to another. The selection of this or that equivalent representation can be local, i.e. at different points of space-time one can use generators of various equivalent representations. This means that we are talking about the additional local symmetry of the Lagrangian. It will be shown further, that in the case of the local $SU\left( 2 \right)$ group associated with the weak interaction, such additional symmetry is the local $U\left( 1 \right)-$ symmetry, that allows to provide it by introducing the electromagnetic interaction. Let us consider in more detail the implementation of this plan. First, we consider the introduction of the electromagnetic interaction for leptons, and then for quarks. 

First we make the following denotations. The bispinor field is denited by ${{\psi }_{{{s}_{1}}}}\left( x \right)$, where ${{s}_{1}}=1,2,3,4-$ a bispinor index. The set of all four components of this field will be reviewed as a column, which we denote by $\hat{\psi }\left( x \right)$. We consider this bispinor field in the chiral representation:
\begin{equation}
\begin{split}
  \hat{\psi }\left( x \right)=\left( \begin{matrix}
   {{{\hat{\psi }}}_{R}}\left( x \right)  \\
   {{{\hat{\psi }}}_{L}}\left( x \right)  \\
\end{matrix} \right)  
\end{split}
\label{eq:CiR}
\end{equation}
Here ${{\hat{\psi }}_{R}}\left( x \right)$ and ${{\hat{\psi }}_{ L}}\left( x \right)$ - right-handed and left-handed two-component spinors, which under the Lorentz transformations that do not contain inversions of the spatial axes, transform according to the right-handed and left-handed spinor representations respectively. The bispinor (\ref{eq:CiR}) is denoted in the form
\begin{equation}
\hat{\psi }\left( x \right)={{\hat{\psi }}^{R}}\left( x \right)+{{\hat{\psi }}^{L}}\left( x \right),
\label{eq:CR1}
\end{equation}
where
\begin{equation}
{{\hat{\psi }}^{R}}\left( x \right)=\left( \begin{matrix}
   {{{\hat{\psi }}}_{R}}\left( x \right)  \\
   {\hat{0}}  \\
\end{matrix} \right),{{\hat{\psi }}^{L}}\left( x \right)=\left( \begin{matrix}
   {\hat{0}}  \\
   {{{\hat{\psi }}}_{L}}\left( x \right)  \\
\end{matrix} \right),
\label{eq:CRLplusR}
\end{equation}
and $\hat{0}$ - a two-component column with zero elements. We note that here and below we use the upper indices $R$ and $L$ to denote four-component bispinor columns, in which the left-handed components for $R$ and the right-handed ones for $L$ are equal to zero. The lower indices $R$ and $L$ will be used to denote the two-component values that are transformed by the right-handed and left-handed spinor representations of the proper Lorentz group. Next, we consider two types of left bispinor fields $\hat{\psi }_{{{I}_{3}}}^{L}\left( x \right),{{I}_{3}}=\pm {1}/{2}\;$, where ${{I}_{3}}-$ denotes the third component of a weak isospin. In this case, we assume that the field $\hat{\psi }_{{{I}_{3}}={1}/{2}\;}^{L}\left( x \right)$ corresponds to the electron neutrino, and $ \hat{\psi }_{{{I}_{3}}=-{1}/{2}\;}^{L}\left( x \right)-$ to the left-handed side of the electron bispinor from the point of the expansion (\ref{eq:CRLplusR}). Therefore, in addition to the mentioned denotations, we will also use
\begin{equation}
  \hat{\psi }_{{{I}_{3}}={1}/{2}\;}^{L}\left( x \right)={{\hat{\nu }}_{e}}\left( x \right),\hat{\psi }_{{{I}_{3}}={-1}/{2}\;}^{L}\left( x \right)={{\hat{e}}^{L}}\left( x \right). 
   \label{eq:poznachenna_nu_e}
   \end{equation}   
We denote as well $ {{\hat{e}}^{R}}\left( x \right)-$ the right-handed side of the electron bispinor, and the entire bispinor as
 \begin{equation}
\hat{e}\left( x \right)={{\hat{e}}^{L}}\left( x \right)+{{\hat{e}}^{R}}\left( x \right)
 \label{eq:electron}
 \end{equation}
 
Then the Lagrangian $ {{L}^{\left( 0 \right)}} $ of the free bispinor fields in the Weinberg-Salam-Glashow model can be written, using the mentioned denotations, in the form:
\begin{equation}
\begin{split}
   & {{L}^{\left( 0 \right)}}=\frac{i}{2}\left( {{{\hat{\bar{e}}}}^{R}}{{{\hat{\gamma }}}^{{{a}_{1}}}}\frac{\partial {{{\hat{e}}}^{R}}}{\partial {{x}^{{{a}_{1}}}}}-\frac{\partial {{{\hat{\bar{e}}}}^{R}}}{\partial {{x}^{{{a}_{1}}}}}{{{\hat{\gamma }}}^{{{a}_{1}}}}{{{\hat{e}}}^{R}} \right)+ \\ 
 & +\frac{i}{2}\left( \hat{\bar{\psi }}_{{{\left( {{I}_{3}} \right)}_{1}}}^{L}{{{\hat{\gamma }}}^{{{a}_{1}}}}\frac{\partial \hat{\psi }_{{{\left( {{I}_{3}} \right)}_{1}}}^{L}}{\partial {{x}^{{{a}_{1}}}}}-\frac{\partial \hat{\bar{\psi }}_{{{\left( {{I}_{3}} \right)}_{1}}}^{L}}{\partial {{x}^{{{a}_{1}}}}}{{{\hat{\gamma }}}^{{{a}_{1}}}}\hat{\psi }_{{{\left( {{I}_{3}} \right)}_{1}}}^{L} \right). \\     
\end{split}
\label{eq:L0VSG}
\end{equation}
Here $\hat{\bar{\psi }}$ as usual denotes the bispinor which is Dirac adjoint to the bispinor $\hat{\psi }$. The Lagrangian (\ref{eq:L0VSG}) is symmetric with respect to the global $SU\left( 2 \right)-$ transformation on the subspace of left-handed bispinors:
\begin{equation}
\begin{split}
        & {{{\hat{{\psi }'}}}_{{{\left( {{I}_{3}} \right)}_{1}}}}={{\left( \exp \left( \frac{i}{2}g\left( {{\theta }_{1}}{{{\hat{\sigma }}}_{1}}+{{\theta }_{2}}{{{\hat{\sigma }}}_{2}}+{{\theta }_{3}}{{{\hat{\sigma }}}_{3}} \right) \right) \right)}_{{{\left( {{I}_{3}} \right)}_{1}},{{\left( {{I}_{3}} \right)}_{2}}}}{{{\hat{\psi }}}_{{{\left( {{I}_{3}} \right)}_{2}}}}, \\ 
 & {{{\hat{\bar{{\psi }'}}}}_{{{\left( {{I}_{3}} \right)}_{2}}}}={{{\hat{\bar{\psi }}}}_{{{\left( {{I}_{3}} \right)}_{1}}}}{{\left( \exp \left( -\frac{i}{2}g\left( {{\theta }_{1}}{{{\hat{\sigma }}}_{1}}+{{\theta }_{2}}{{{\hat{\sigma }}}_{2}}+{{\theta }_{3}}{{{\hat{\sigma }}}_{3}} \right) \right) \right)}_{{{\left( {{I}_{3}} \right)}_{1}},{{\left( {{I}_{3}} \right)}_{2}}}}. \\ 
\end{split}
\label{eq:SU2Fondamental}
\end{equation}
Here $ g- $ the coupling constant, $ {{{{\hat{\sigma }}}_{1}}}/{2}\;,{{{{\hat{\sigma }}}_{2}}}/{2}\;,{{{{\hat{\sigma }}}_{3}}}/{2}\;$ - generators of the $SU\left( 2 \right)$ group representation by $SU\left( 2 \right)-$ matrices themselves, ${{\theta }_{1}},{{\theta }_{2}},{{\theta }_{3}}-$ global transformation parameters. The generators $ {{\hat{\sigma }}_{1}},{{\hat{\sigma }}_{2}},{{\hat{\sigma }}_{3}}- $ can be represented by the Pauli matrices, but for our further purposes it will be more convenient not to fix the explicit form of the generators as long as possible, and to use only their properties, which follow from the group multiplication law. In particular, that are commutation relations.

Symmetries with respect to the local transformations (\ref{eq:SU2Fondamental}) with parameters ${{\theta }_{{{g}_{1}}}}\left( x \right),{{g}_{1}}=1,2,3$ arbitrarily dependent on coordinates are obtained by the derivative extensions:
\begin{equation}
\begin{split}
     & {{L}_{{\hat{A}}}}=\frac{i}{2}\left( {{{\hat{\bar{e}}}}^{R}}\left( x \right){{{\hat{\gamma }}}^{{{a}_{1}}}}\frac{\partial {{{\hat{e}}}^{R}}\left( x \right)}{\partial {{x}^{{{a}_{1}}}}}-\frac{\partial {{{\hat{\bar{e}}}}^{R}}\left( x \right)}{\partial {{x}^{{{a}_{1}}}}}{{{\hat{\gamma }}}^{{{a}_{1}}}}{{{\hat{e}}}^{R}}\left( x \right) \right)+ \\ 
 & +\frac{i}{2}\left( \hat{\bar{\psi }}_{{{\left( {{I}_{3}} \right)}_{1}}}^{L}\left( x \right){{{\hat{\gamma }}}^{{{a}_{1}}}}\left( \frac{\partial \hat{\psi }_{{{\left( {{I}_{3}} \right)}_{1}}}^{L}\left( x \right)}{\partial {{x}^{{{a}_{1}}}}}-\frac{i}{2}g{{A}_{{{a}_{1}},{{g}_{1}}}}\left( x \right){{\left( {{{\hat{\sigma }}}_{{{g}_{1}}}} \right)}_{{{\left( {{I}_{3}} \right)}_{1}},{{\left( {{I}_{3}} \right)}_{2}}}}\hat{\psi }_{{{\left( {{I}_{3}} \right)}_{2}}}^{L}\left( x \right) \right) \right.- \\ 
 & \left. -\left( \frac{\partial \hat{\bar{\psi }}_{{{\left( {{I}_{3}} \right)}_{1}}}^{L}\left( x \right)}{\partial {{x}^{{{a}_{1}}}}}+\frac{i}{2}g{{A}_{{{a}_{1}},{{g}_{1}}}}\left( x \right)\hat{\bar{\psi }}_{{{\left( {{I}_{3}} \right)}_{2}}}^{L}\left( x \right){{\left( {{{\hat{\sigma }}}_{{{g}_{1}}}} \right)}_{{{\left( {{I}_{3}} \right)}_{2}},{{\left( {{I}_{3}} \right)}_{1}}}} \right){{{\hat{\gamma }}}^{{{a}_{1}}}}\hat{\psi }_{{{\left( {{I}_{3}} \right)}_{1}}}^{L}\left( x \right) \right), \\  
\end{split}
\label{eq:LAlocal}
\end{equation}  
or
\begin{equation}
\begin{split}
    & L=\frac{i}{2}\left( {{{\hat{\bar{e}}}}^{R}}\left( x \right){{{\hat{\gamma }}}^{{{a}_{1}}}}\frac{\partial {{{\hat{e}}}^{R}}\left( x \right)}{\partial {{x}^{{{a}_{1}}}}}-\frac{\partial {{{\hat{\bar{e}}}}^{R}}\left( x \right)}{\partial {{x}^{{{a}_{1}}}}}{{{\hat{\gamma }}}^{{{a}_{1}}}}{{{\hat{e}}}^{R}}\left( x \right) \right)+ \\ 
 & +\frac{i}{2}\left( {{{\hat{\bar{e}}}}^{L}}\left( x \right){{{\hat{\gamma }}}^{{{a}_{1}}}}\frac{\partial {{{\hat{e}}}^{L}}\left( x \right)}{\partial {{x}^{{{a}_{1}}}}}-\frac{\partial {{{\hat{\bar{e}}}}^{L}}\left( x \right)}{\partial {{x}^{{{a}_{1}}}}}{{{\hat{\gamma }}}^{{{a}_{1}}}}{{{\hat{e}}}^{L}}\left( x \right) \right)+ \\ 
 & +\frac{i}{2}\left( {{{\hat{\bar{\nu }}}}_{e}}\left( x \right){{{\hat{\gamma }}}^{{{a}_{1}}}}\frac{\partial {{{\hat{\nu }}}_{e}}\left( x \right)}{\partial {{x}^{{{a}_{1}}}}}-\frac{\partial {{{\hat{\bar{\nu }}}}_{e}}\left( x \right)}{\partial {{x}^{{{a}_{1}}}}}{{{\hat{\gamma }}}^{{{a}_{1}}}}{{{\hat{\nu }}}_{e}}\left( x \right) \right)+L_{{\hat{A}}}^{\operatorname{int}}, \\     
\end{split}
\label{eq:LAlocal_e_nue}
\end{equation}
where
\begin{equation}
L_{{\hat{A}}}^{\operatorname{int}}=\left( {g}/{2}\; \right){{A}_{{{a}_{1}},{{g}_{1}}}}\left( x \right){{\left( {{{\hat{\sigma }}}_{{{g}_{1}}}} \right)}_{{{\left( {{I}_{3}} \right)}_{1}},{{\left( {{I}_{3}} \right)}_{2}}}}\hat{\bar{\psi }}_{{{\left( {{I}_{3}} \right)}_{1}}}^{L}{{\hat{\gamma }}^{{{a}_{1}}}}\hat{\psi }_{{{\left( {{I}_{3}} \right)}_{2}}}^{L}.
\label{eq:LhatA_int}
\end{equation}
As above, here a non-Abelian gauge field ${{A}_{{{a}_{1}},g}}\left( x \right)$, that restores the local $SU\left( 2 \right)-$ symmetry under the condition that the transformation (\ref{eq:SU2Fondamental}) with the parameters depending on the coordinates, is added by the gauge field transformation: 
\begin{equation}
{{A}_{{{a}_{1}},{{g}_{1}}}}\left( x \right)={{D}_{{{g}_{1}},{{g}_{2}}}}\left( \vec{\theta }\left( x \right) \right){{{A}'}_{{{a}_{1}},{{g}_{2}}}}\left( x \right)-\frac{\partial {{\theta }_{{{g}_{1}}}}\left( x \right)}{\partial {{x}^{{{a}_{1}}}}}.
\label{eq:Kalibr_pole_peretvor}
\end{equation}
Here $ {{D}_{{{g}_{1}},{{g}_{2}}}}\left( \vec{\theta }\left( x \right) \right) $ is the matrix of the adjoint representation of the $SU\left( 2 \right)-$ group already considered above.

We now consider a subgroup of local transformations in the form:
\begin{equation}
\begin{split}
   & {{{\hat{{\psi }'}}}_{{{\left( {{I}_{3}} \right)}_{1}}}}={{\left( \exp \left( \frac{i}{2}g{{\theta }_{3}}\left( x \right){{{\hat{\sigma }}}_{3}} \right) \right)}_{{{\left( {{I}_{3}} \right)}_{1}},{{\left( {{I}_{3}} \right)}_{2}}}}{{{\hat{\psi }}}_{{{\left( {{I}_{3}} \right)}_{2}}}}, \\ 
 & {{{\hat{\bar{{\psi }'}}}}_{{{\left( {{I}_{3}} \right)}_{2}}}}={{{\hat{\bar{\psi }}}}_{{{\left( {{I}_{3}} \right)}_{1}}}}{{\left( \exp \left( -\frac{i}{2}g{{\theta }_{3}}\left( x \right){{{\hat{\sigma }}}_{3}} \right) \right)}_{{{\left( {{I}_{3}} \right)}_{1}},{{\left( {{I}_{3}} \right)}_{2}}}}, \\ 
 & {{A}_{{{a}_{1}},3}}\left( x \right)={{{{A}'}}_{{{a}_{1}},3}}\left( x \right)-\frac{\partial {{\theta }_{3}}\left( x \right)}{\partial {{x}^{{{a}_{1}}}}}, \\  
 & {{A}_{{{a}_{1}},{{g}_{1}}}}\left( x \right)={{D}_{{{g}_{1}},{{g}_{2}}}}\left( \vec{\theta }\left( x \right)=\left( 0,0,{{\theta }_{3}}\left( x \right) \right) \right){{{{A}'}}_{{{a}_{1}},{{g}_{2}}}}\left( x \right),{{g}_{1}},{{g}_{2}}=1,2. \\     
\end{split}
\label{eq:Obertanna_vocrug_3_visi}
\end{equation}

Since the fields $ {{\hat{\nu }}_{e}}\left( x \right) $ and $ {{\hat{e}}^{L}}\left( x \right) $ are the eigenvectors of the generator $\left( {1}/{2}\; \right){{\hat{\sigma }}_{3}}$ corresponding to the eigenvalues $\left( {\pm 1}/{2}\; \right)$ respectively, the representation of the subgroup (\ref{eq:Obertanna_vocrug_3_visi}) on a linear space, in which the fields $ \hat{\psi }_{{{I}_{3}}}^{L} $ take values, is reducible. And its contraction to the invariant subspaces $ \hat{\psi }_{{{I}_{3}}={1}/{2}\;}^{L}={{\hat{\nu }}_{e}}\left( x \right) $ and $ \hat{\psi }_{{{I}_{3}}=-{1}/{2}\;}^{L}={{\hat{e}}^{L}}\left( x \right) $ agree with the $U\left( 1 \right)-$ group representation. Thus, for the neutrino field the local $U\left( 1 \right)-$ symmetry is the special limit of the $SU\left( 2 \right)-$ symmetry. Moreover, the extra terms, that appear from the derivatives in the Lagrangian due to the dependence on the coordinates of the $U\left( 1 \right)-$ transformation parameter, are balanced not by the electromagnetic field, but by the component ${{A}_{{{a}_{1}},{{g}_{1}}=3}}\left( x \right)$ that transforms on the subgroup (\ref{eq:Obertanna_vocrug_3_visi}) as well as the electromagnetic field. But the field ${{A}_{{{a}_{1}},{{g}_{1}}=3}}\left( x \right)$ can be easily distinguished from the electromagnetic one, since the agreement of the transformation laws takes place only on the subgroup (\ref{eq:Obertanna_vocrug_3_visi}). So within the {\flqq old\frqq} Standard Model that only considers neutrinos with left-handed components, we can say that the neutrino field {\flqq does not require\frqq} the electromagnetic field to restore the local $U\left( 1 \right)-$ symmetry, because this symmetry has already been restored by the gauge field associated with the local $SU\left( 2 \right)-$ symmetry.  
 
 There is a similar situation with ${{W}^{\pm }}-$ bosons. Since, on the one hand, the gauge field $ {{A}_{{{a}_{1}},{{g}_{1}}}}\left( x \right) $ should be real, otherwise it will not be able to implement its balancing function with respect to the local $SU\left( 2 \right)-$ transformation, and on the other hand, as it is known from the experiment, the decays of two of its three mediating particles indicate that they should be be charged, we should consider the complex combinations (\ref{eq:Wpm}). But it is seen from the relations (\ref{eq:Wpm}), that the $U\left( 1 \right)-$ transformation of the ${{W}^{\pm }}$ fields is equivalent to the rotation over the third axis \cite{Rotation_over_3_axe_Salam1959} in the internal space of the $ {{A}_{{{a}_{1}},{{g}_{1}}}}\left( x \right)$ fields. But such a rotation belongs to the adjoint representation of the $SU\left( 2 \right)$ group. If we consider such a local rotation with a parameter dependent on the coordinates, then the transformation of fields is described by formulas (\ref{eq:Obertanna_vocrug_3_visi}). It is seen from them that the ${{A}_{{{a}_{1}},{{g}_{1}}=1}}\left( x \right)$ and ${{A}_{{{a}_{1}},{{g}_{1}}=2}}\left( x \right)$ fields are transformed exactly by the local adjoint representation of the $SU\left( 2 \right)$ group, i.e. without a non-uniform contribution. As seen from (\ref{eq:Obertanna_vocrug_3_visi}), this contribution is included only in the transformation of the ${{A}_{{{a}_{1}},{{g}_{1}}=3}}\left( x \right)$ field.
 Since the total Lagrangian of the Weinberg-Salam-Glashow model is invariant under an arbitrary local $SU\left( 2 \right)-$ transformation, it will be invariant under (\ref{eq:Obertanna_vocrug_3_visi}) as well. That is, for the fields (\ref{eq:Wpm}) again the local $U\left( 1 \right)-$ symmetry is a consequence of the local $SU\left( 2 \right)-$ symmetry. Therefore, for the field (\ref{eq:Wpm}), it is not required to interact with the electromagnetic field in order to restore this symmetry. Thus we obtain the conclusion: \textbf{\textit{despite the fact that ${{W}^{\pm }}-$ bosons have an electric charge, they do not interact with the electromagnetic field. The physical demonstration of the charge of these particles is only that the channels of their decays are formed by particles with a total electric charge $\pm 1$}}.      
      
Of course this conclusion is in a contradiction with existing conceptions of the Standard Model \cite{Altarelli:1698688}. However there are number of works, for example \cite{Abbiendi1999293,Chatrchyan2013,Aad:2013izg,Chatrchyan:2013fya, ChatrchyanPhysRevD.90.032008}, in which the possible interaction of ${{W}^{\pm }}-$ bosons with photons is being studied experimentally. The events of the proton-proton and electron-positron scattering process that include these particles are being reconstructed in these works using results of the lepton decays of the ${{W}^{\pm }}-$bosons. Then these processes is simulated by a Monte-Carlo generations using low-order Feynman diagrams of the Standard Model. The fit of the simulation result to the experiment makes it possible to determite the quantities of the nonabelian verticies. But this experiments can't be considered as a prove of ${{W}^{\pm }}-$ boson-photon interaction because the existence of the vertices for such interaction is built in during experimental data processing. In addition, the purely virtual processes with the ${{W}^{\pm }}-$boson-photon interaction vertices are also considered. It means that the particles which participate in such process could not be directly observed in the initial or final state. For example, such processes are considered in \cite{Szleper:2014xxa}. But according to the Standard Model the two $W-$bosons in the virtual phase of the process may be converted into two photons which could be directly detected. However there are no works in which such process is studied experimentally. The different mechanisms of photon formation in $p-p$ collisions are studied experimentally in particular in work \cite{diphotonChatrchyan2014}, but among them there are no mechanisms of two photons formation through the $W-$bosons. Furthermore the ${{W}^{\pm}}-$-bosons decay channels which are known from the experiment and are given in \cite{Olive:2016xmw} do not contain the photons with leptons together. Photons are formed only together with the hadrons so one can assume that the photons are emitted by quarks during the hadronization process but not by the $W-$bosons. In addition, according to the data \cite{Olive:2016xmw} the decays of $\mu$ and $\tau-$leptons and possibly $t-$quark which occur with leptons formation also contain the photons. Therefore let us accept the model in which the ${{W}^{\pm }}-$bosons do not interact with the electromagnetic field. It is important that the theory with the local $SU\left( 2 \right)$ symmetry and fields (\ref{eq:Wpm}) in addition conains also $U\left( 1 \right)-$ symmetry. This $U\left( 1 \right)-$ symmetry is connected with the $SU\left( 2 \right)$ generators and appears in the different ways for the leptons and their corresponding neutrinos. Let's consider this model in details.
 
During the transition from the real fields ${{A}_{{{a}_{1}},{{g}_{1}}=1}}\left( x \right)$ i ${{A}_{{{a}_{1}},{{g}_{1}}=2}}\left( x \right)$ to the complex fields (\ref{eq:Wpm}) the Lagrangian (\ref{eq:LhatA_int}) becomes the next form:
 
\begin{equation}
	   \begin{split}
	  & L_{{\hat{A}}}^{\operatorname{int}}=\frac{g}{2}W_{{{a}_{1}}}^{+}\left( x \right)\left( \hat{\bar{\psi }}_{{{\left( {{I}_{3}} \right)}_{1}}}^{L}\left( x \right){{\left( {{{\hat{\sigma }}}_{+}} \right)}_{{{\left( {{I}_{3}} \right)}_{1}},{{\left( {{I}_{3}} \right)}_{2}}}}{{{\hat{\gamma }}}^{{{a}_{1}}}}\hat{\psi }_{{{\left( {{I}_{3}} \right)}_{2}}}^{L}\left( x \right) \right) \\ 
 & +\frac{g}{2}W_{{{a}_{1}}}^{-}\left( x \right)\left( \hat{\bar{\psi }}_{{{\left( {{I}_{3}} \right)}_{1}}}^{L}\left( x \right){{\left( {{{\hat{\sigma }}}_{-}} \right)}_{{{\left( {{I}_{3}} \right)}_{1}},{{\left( {{I}_{3}} \right)}_{2}}}}{{{\hat{\gamma }}}^{{{a}_{1}}}}\hat{\psi }_{{{\left( {{I}_{3}} \right)}_{2}}}^{L}\left( x \right) \right) \\ 
 & +\frac{g}{2}{{A}_{{{a}_{1}},3}}\left( x \right)\left( \hat{\bar{\psi }}_{{{\left( {{I}_{3}} \right)}_{1}}}^{L}\left( x \right){{\left( {{{\hat{\sigma }}}_{3}} \right)}_{{{\left( {{I}_{3}} \right)}_{1}},{{\left( {{I}_{3}} \right)}_{2}}}}{{{\hat{\gamma }}}^{{{a}_{1}}}}\hat{\psi }_{{{\left( {{I}_{3}} \right)}_{2}}}^{L}\left( x \right) \right). \\      
	   \end{split}
	   \label{eq:LAint_sigma_plus_minus}
	   \end{equation}	  
Here the standard designations are introduced
\begin{equation}
{{\hat{\sigma }}_{+}}=\frac{1}{2}\left( {{{\hat{\sigma }}}_{1}}+i{{{\hat{\sigma }}}_{2}} \right),{{\hat{\sigma }}_{-}}=\frac{1}{2}\left( {{{\hat{\sigma }}}_{1}}-i{{{\hat{\sigma }}}_{2}} \right).
\label{eq:sigma_plus_minus}
\end{equation}

The operators ${{\hat{\sigma}}_{+}}$ and ${{\hat{\sigma }}_{-}}$ introduced in such way coincide with the raising and lowering operators which are usualy used to build the $SU\left( 2 \right)$ group representations.

 We use as usual \cite{GelfandMinlosShapiro,InuiTanabeOnodera,ElliottDawber} the eigenvectors of the ${{\hat{\sigma }}_{3}}$ generator as the basis set of the representation space. Let's denote by ${{\psi }_{m}}$ the scaled eigenvector which correspond to the eigenvalue $m$ and by $j$ the maximum value (the representation weight) of the eigenvalue $m$. In this case the action of the raising and lowering operators is defined by the following formulas.
 
\begin{equation}
\begin{split}
  & {{{\hat{\sigma }}}_{+}}\left( {{\psi }_{m}} \right)={{\alpha }_{m}}{{\psi }_{m+1}},m<j, \\ 
 & {{{\hat{\sigma }}}_{+}}\left( {{\psi }_{m}} \right)=0,m=j, \\ 
 & {{{\hat{\sigma }}}_{-}}\left( {{\psi }_{m}} \right)={{\beta }_{m}}{{\psi }_{m-1}},m>\left( -j \right), \\ 
 & {{{\hat{\sigma }}}_{-}}\left( {{\psi }_{m}} \right)=0,m=-j. \\   
\end{split}
\label{eq:pidvihuchij_i_ponujuuchij}
\end{equation}

The commutation relations between the generators of the $SU\left( 2 \right)$ representation and  condition of the Hermitian conjugation allows one to find only modulus of the coefficients ${{\alpha }_{m}}$ and ${{\beta }_{m}}$ while their arguments remain arbitrary \cite{GelfandMinlosShapiro,InuiTanabeOnodera,ElliottDawber}. These agruments may always be eliminated due to the basis transformation in the representation space i.e. by transition to the equivalent representation. Therefore these arguments are insignificant to the ordinary problem statement in the group theory which lies in the search for the nonequivalent irreducible group representations. In formula \ref{eq:LAint_sigma_plus_minus} we of course mean one concrete representation of $SU\left( 2 \right)$ group generators. Although the representation of $SU\left( 2 \right)$ group generators may be selected in any arbitrary way, so we could use another one from the multitude of equivalent representations. But it is natural for result to be independent of the selection of the concrete representation. It means that the system has the additional $U\left( 1 \right)-$ symmetry which could be explained more obviously in the following way.

It is known that the arbitrary $SU\left( 2 \right)$ matrix $\hat{u}$ may be represented in the following form

\begin{equation}
\hat{u}=\exp \left( i\hat{h} \right),
\label{eq:uexpih}
\end{equation}
 where $\hat{h}-$ - is self-adjoint traceless matrix. This matrix may be parametrized  in the next way
\begin{equation}
 \begin{split}
 \hat{h}=\left( \begin{matrix}
   a/2 & \left( r/2 \right)\exp \left( -i\left( \phi -{{\phi }_{0}} \right) \right)  \\
   \left( r/2 \right)\exp \left( i\left( \phi -{{\phi }_{0}} \right) \right) & -\left( a/2 \right)  \\
\end{matrix} \right),
  \end{split}
  \label{eq:h}
  \end{equation}  
 
where $a,t$ and $\Delta \phi =\phi -{{\phi }_{0}}$ are three real parameters of the $SU\left( 2 \right)$ group. The uncertainty of the arguments ${{\alpha }_{m}}$ and ${{\beta }_{m}}$ appears due to the arbitrariness of the point of reference for these arguments ${{\phi }_{0}}$ in (\ref{eq:h}).
  
Indeed, the matrix $ \hat{h} $ may be represented as:    
\begin{equation}
\hat{h}=r\cos \left( \phi  \right)\left( {{{\hat{\sigma }}}_{x}}/2 \right)+r\sin \left( \phi  \right)\left( {{{\hat{\sigma }}}_{y}}/2 \right)+a\left( {{{\hat{\sigma }}}_{z}}/2 \right),
\label{eq:hsigma}
\end{equation}
where 
\begin{equation}
\begin{split}
    & {{{\hat{\sigma }}}_{x}}=\left( \begin{matrix}
   0 & \exp \left( -i{{\phi }_{0}} \right)  \\
   \exp \left( i{{\phi }_{0}} \right) & 0  \\
\end{matrix} \right), \\ 
 & {{{\hat{\sigma }}}_{y}}=\left( \begin{matrix}
   0 & \exp \left( -i{{\phi }_{0}} \right)\left( -i \right)  \\
   \exp \left( i{{\phi }_{0}} \right)i & 0  \\
\end{matrix} \right), \\ 
 & {{{\hat{\sigma }}}_{z}}=\left( \begin{matrix}
   1 & 0  \\
   0 & -1  \\
\end{matrix} \right)- \\   
\end{split}
\label{eq:eqvivalent_sigma}
\end{equation}
- an equivalent representation of the Pauli matrices. The corresponding raising and lowering generators (\ref{eq:sigma_plus_minus}) become the form:
\begin{equation}
{{\hat{\sigma }}_{+}}=\left( \begin{matrix}
   0 & \exp \left( -i{{\phi }_{0}} \right)  \\
   0 & 0  \\
\end{matrix} \right),{{\hat{\sigma }}_{-}}=\left( \begin{matrix}
   0 & 0  \\
   \exp \left( i{{\phi }_{0}} \right) & 0  \\
\end{matrix} \right).
\label{eq:sigma_plus_minus_eqivalent}
\end{equation}
Using the simplest basis in the representation space: 
\begin{equation}
{{\psi }_{1/2}}=\left( \begin{matrix}
   1  \\
   0  \\
\end{matrix} \right),{{\psi }_{-1/2}}=\left( \begin{matrix}
   0  \\
   1  \\
\end{matrix} \right), 
 \label{eq:naipristichij_basis}
 \end{equation} 
we obtain:
\begin{equation}
\begin{split}
    & {{\alpha }_{-1/2}}=\left\langle  {{\psi }_{1/2}} | {{{\hat{\sigma }}}_{+}}\left( {{\psi }_{-1/2}} \right) \right\rangle =\left( \begin{matrix}
   1 & 0  \\
\end{matrix} \right)\left( \begin{matrix}
   0 & \exp \left( -i{{\phi }_{0}} \right)  \\
   0 & 0  \\
\end{matrix} \right)\left( \begin{matrix}
   0  \\
   1  \\
\end{matrix} \right)=\exp \left( -i{{\phi }_{0}} \right), \\ 
 & {{\beta }_{1/2}}=\left\langle  {{\psi }_{-1/2}} | {{{\hat{\sigma }}}_{-}}\left( {{\psi }_{1/2}} \right) \right\rangle =\left( \begin{matrix}
   0 & 1  \\
\end{matrix} \right)\left( \begin{matrix}
   0 & 0  \\
   \exp \left( i{{\phi }_{0}} \right) & 0  \\
\end{matrix} \right)\left( \begin{matrix}
   1  \\
   0  \\
\end{matrix} \right)=\exp \left( i{{\phi }_{0}} \right). \\    
\end{split}
\label{eq:equivalent_alfa_i_beta}
\end{equation}

These values differ from the usualy used values ${{\beta }_{1/2}}={{\alpha }_{-1/2}}=1$ \cite{GelfandMinlosShapiro,ElliottDawber,Zelobenko:1164864} only in the arbitrary phase multiplier which connected with the point of reference of the argument $\phi_{0}$. As you can see from the expression \eqref{eq:equivalent_alfa_i_beta}, this excess phase multiplier may be eliminated by selecting the orthonormal basis instead of \eqref{eq:naipristichij_basis}

\begin{equation}\label{new_basis}
{{{\psi }'}_{1/2}}=\left( \begin{matrix}
1  \\
0  \\
\end{matrix} \right),{{{\psi }'}_{-1/2}}=\left( \begin{matrix}
0  \\
\exp \left( i{{\phi }_{0}} \right)  \\
\end{matrix} \right).
\end{equation}  

In this basis we obtain the equivalent representation of the $SU\left( 2 \right)$ group in which
$ {{\beta }_{1/2}}={{\alpha }_{-1/2}}=1. $

It is important for us that the transition from the basis \eqref{eq:naipristichij_basis} to the \eqref{new_basis} is non-symmetrical relative to the components of the columns from the linear space on which the group $SU\left( 2 \right)$ acts. It is important because in the Standard Model all fields, which are the components of such columns, have different charge i.e. interact with the electromagnetic field in the different ways. In particular, let's turn to the Lagrangian\eqref{eq:LAlocal_e_nue} and its part \eqref{eq:LhatA_int}. Also take into account the considered uncertainty of the coefficients $ {{\alpha }_{-1/2}} $ and ${{\beta }_{1/2}} $ and the fact that such uncertanity may be local (i.e. value $ \phi_{0} $ in \eqref{eq:equivalent_alfa_i_beta} may be arbitrary function of coordinates). Then the action of the raising and lowering generators may be represented in the following way:

\begin{equation}
\begin{split}
& {{{\hat{\sigma }}}_{+}}\left( \begin{matrix}
{{{\hat{\nu }}}_{e}}\left( x \right)  \\
{{\hat{e}}^{L}}\left( x \right)  \\
\end{matrix} \right)={{{\hat{\nu }}}_{e}}\left( x \right){{{\hat{\sigma }}}_{+}}\left( \begin{matrix}
1  \\
0  \\
\end{matrix} \right)+{{{\hat{e}}}^{L}}\left( x \right){{{\hat{\sigma }}}_{+}}\left( \begin{matrix}
0  \\
1  \\
\end{matrix} \right)=\exp \left( -i{{\phi }_{0}}\left( x \right) \right)\left( \begin{matrix}
{{{\hat{e}}}^{L}}\left( x \right)  \\
0  \\
\end{matrix} \right), \\ 
& {{{\hat{\sigma }}}_{-}}\left( \begin{matrix}
{{{\hat{\nu }}}_{e}}\left( x \right)  \\
{{{\hat{e}}}^{L}}\left( x \right)  \\
\end{matrix} \right)={{{\hat{\nu }}}_{e}}\left( x \right){{{\hat{\sigma }}}_{-}}\left( \begin{matrix}
1  \\
0  \\
\end{matrix} \right)+{{{\hat{e}}}^{L}}\left( x \right){{{\hat{\sigma }}}_{-}}\left( \begin{matrix}
0  \\
1  \\
\end{matrix} \right)=\exp \left( i{{\phi }_{0}}\left( x \right) \right)\left( \begin{matrix}
0  \\
{{{\hat{\nu }}}_{e}}\left( x \right)  \\
\end{matrix} \right). \\ 
\end{split}
\label{eq:sigma_plus_minus_na_stolbec}
\end{equation}
 Taking to account these results the Lagrangian \eqref{eq:LAlocal_e_nue} may be rewrited in the next form:
 \begin{equation}
 \begin{split}
 & L=\frac{i}{2}\left( {{{\hat{\bar{e}}}}^{R}}\left( x \right){{{\hat{\gamma }}}^{{{a}_{1}}}}\frac{\partial {{{\hat{e}}}^{R}}\left( x \right)}{\partial {{x}^{{{a}_{1}}}}}-\frac{\partial {{{\hat{\bar{e}}}}^{R}}\left( x \right)}{\partial {{x}^{{{a}_{1}}}}}{{{\hat{\gamma }}}^{{{a}_{1}}}}{{{\hat{e}}}^{R}}\left( x \right) \right)+ \\ 
 & +\frac{i}{2}\left( {{{\hat{\bar{e}}}}^{L}}\left( x \right){{{\hat{\gamma }}}^{{{a}_{1}}}}\frac{\partial {{{\hat{e}}}^{L}}\left( x \right)}{\partial {{x}^{{{a}_{1}}}}}-\frac{\partial {{{\hat{\bar{e}}}}^{L}}\left( x \right)}{\partial {{x}^{{{a}_{1}}}}}{{{\hat{\gamma }}}^{{{a}_{1}}}}{{{\hat{e}}}^{L}}\left( x \right) \right) \\ 
 & +\frac{i}{2}\left( {{{\hat{\bar{\nu }}}}_{e}}\left( x \right){{{\hat{\gamma }}}^{{{a}_{1}}}}\frac{\partial {{{\hat{\nu }}}_{e}}\left( x \right)}{\partial {{x}^{{{a}_{1}}}}}-\frac{\partial {{{\hat{\bar{\nu }}}}_{e}}\left( x \right)}{\partial {{x}^{{{a}_{1}}}}}{{{\hat{\gamma }}}^{{{a}_{1}}}}{{{\hat{\nu }}}_{e}}\left( x \right) \right)+ \\ 
 & +\frac{g}{2}\exp \left( -i{{\phi }_{0}}\left( x \right) \right)W_{{{a}_{1}}}^{+}\left( x \right)\left( {{{\hat{\bar{\nu }}}}_{e}}\left( x \right){{{\hat{\gamma }}}^{{{a}_{1}}}}{{{\hat{e}}}^{L}}\left( x \right) \right) \\ 
 & +\frac{g}{2}\exp \left( i{{\phi }_{0}}\left( x \right) \right)W_{{{a}_{1}}}^{-}\left( x \right)\left( {{{\hat{\bar{e}}}}^{L}}\left( x \right){{{\hat{\gamma }}}^{{{a}_{1}}}}{{{\hat{\nu }}}_{e}}\left( x \right) \right) \\ 
 & +\frac{g}{2}{{A}_{{{a}_{1}},3}}\left( x \right)\left( {{{\hat{\bar{\nu }}}}_{e}}\left( x \right){{{\hat{\gamma }}}^{{{a}_{1}}}}{{{\hat{\nu }}}_{e}}\left( x \right) \right)-\frac{g}{2}{{A}_{{{a}_{1}},3}}\left( x \right)\left( {{{\hat{\bar{e}}}}^{L}}\left( x \right){{{\hat{\gamma }}}^{{{a}_{1}}}}{{{\hat{e}}}^{L}}\left( x \right) \right). \\ 
 \end{split}
 \label{eq:Lagrangian_s_fi0}
 \end{equation}

It is needed to compensate the multipliers $ \exp \left( \mp i{{\phi }_{0}}\left( x \right) \right) $ in \eqref{eq:Lagrangian_s_fi0} to provide the Lagrangian independence of the selection of equivalent $SU\left( 2 \right)$ group representation. For this purpose let's introduce the corresponding local $U\left( 1 \right)-$ field transformation:
\begin{equation}\label{peretvorenna_el }
{{\hat{e}}^{L}}\left( x \right)=\exp \left( i{{\phi }_{0}}\left( x \right) \right){{{\hat{e}}}'^{L}}\left( x \right),{{\hat{\bar{e}}}^{L}}\left( x \right)=\exp \left( -i{{\phi }_{0}}\left( x \right) \right){{{\hat{\bar{e}}}}'^{L}}\left( x \right)
\end{equation} 

Nevertheless the part of the Lagrangian \eqref{eq:Lagrangian_s_fi0} which contain derivatives of the field $ {{\hat{e}}^{L}}\left( x \right)$ requires both the covariant derivatives and the gauge field.
Hence the interction between the electromagnetic field and the left components of field which correspond to the electrons may be introduced in this way. But the electromagnetic field should interact in the same way with both the left-handed and the right-handed components of the field which corresponds to the electrons. Note that the Lagrangian \eqref{eq:Lagrangian_s_fi0} will remain invariant if consider the same transformation of the right-handed components inroducing the covariant derivatives and the same gauge field  as for the left components with the same transformation law. Thereby let's join the left-handed and the right-handed fields which correspond to the electron into the one field $ \hat{e}\left( x \right)$ using the formula \eqref{eq:electron}. Then if introduce the covariant derivatives for this field and append the Lagrangians for the free gauge fields  we will obtain the Lagrangian:

\begin{equation}
\begin{split}
    & {{L}_{\left( E\otimes U\left( 1 \right) \right)\circ \left( SU\left( 2 \right)\otimes E \right)}}=\frac{i}{2}\left( \hat{\bar{e}}\left( x \right){{{\hat{\gamma }}}^{{{a}_{1}}}}\left( \frac{\partial \hat{e}\left( x \right)}{\partial {{x}^{{{a}_{1}}}}}-i{{g}^{em}}{{A}_{{{a}_{1}}}}\left( x \right)\hat{e}\left( x \right) \right) \right.+ \\ 
  & -\left( \frac{\partial \hat{\bar{e}}\left( x \right)}{\partial {{x}^{{{a}_{1}}}}}+i{{g}^{em}}{{A}_{{{a}_{1}}}}\left( x \right)\hat{\bar{e}}\left( x \right) \right){{{\hat{\gamma }}}^{{{a}_{1}}}}\hat{e}\left( x \right)+ \\ 
  & +\frac{i}{2}\left( {{{\hat{\bar{\nu }}}}_{e}}\left( x \right){{{\hat{\gamma }}}^{{{a}_{1}}}}\frac{\partial {{{\hat{\nu }}}_{e}}\left( x \right)}{\partial {{x}^{{{a}_{1}}}}}-\frac{\partial {{{\hat{\bar{\nu }}}}_{e}}\left( x \right)}{\partial {{x}^{{{a}_{1}}}}}{{{\hat{\gamma }}}^{{{a}_{1}}}}{{{\hat{\nu }}}_{e}}\left( x \right) \right)+ \\ 
  & +\frac{g}{4}{{\alpha }_{-1/2}}\left( x \right)W_{{{a}_{1}}}^{+}\left( x \right)\left( {{{\hat{\bar{\nu }}}}_{e}}\left( x \right){{{\hat{\gamma }}}^{{{a}_{1}}}}\left( \hat{I}-{{{\hat{\gamma }}}^{5}} \right)\hat{e}\left( x \right) \right) \\ 
  & +\frac{g}{4}{{\beta }_{1/2}}\left( x \right)W_{{{a}_{1}}}^{-}\left( x \right)\left( \hat{\bar{e}}\left( x \right)\left( \hat{I}+{{{\hat{\gamma }}}^{5}} \right){{{\hat{\gamma }}}^{{{a}_{1}}}}{{{\hat{\nu }}}_{e}}\left( x \right) \right) \\ 
  & +\frac{g}{2}{{A}_{{{a}_{1}},3}}\left( x \right)\left( \left( {{{\hat{\bar{\nu }}}}_{e}}\left( x \right){{{\hat{\gamma }}}^{{{a}_{1}}}}{{{\hat{\nu }}}_{e}}\left( x \right) \right)-\frac{1}{4}\left( \hat{\bar{e}}\left( x \right)\left( \hat{I}+{{{\hat{\gamma }}}^{5}} \right){{{\hat{\gamma }}}^{{{a}_{1}}}}\left( \hat{I}-{{{\hat{\gamma }}}^{5}} \right)\hat{e}\left( x \right) \right) \right)- \\ 
  & -\frac{1}{4}{{g}^{{{a}_{1}}{{a}_{11}}}}{{g}^{{{b}_{1}}{{b}_{11}}}}\left( {{F}_{{{a}_{1}}{{b}_{1}},+}}\left( x \right){{F}_{{{a}_{11}}{{b}_{11}},-}}\left( x \right)+{{F}_{{{a}_{1}}{{b}_{1}},3}}\left( x \right){{F}_{{{a}_{11}}{{b}_{11}},3}}\left( x \right) \right)- \\ 
  & -\frac{1}{4}{{g}^{{{a}_{1}}{{a}_{11}}}}{{g}^{{{b}_{1}}{{b}_{11}}}}{{F}_{{{a}_{1}}{{b}_{1}}}}\left( x \right){{F}_{{{a}_{11}}{{b}_{11}}}}\left( x \right). \\
\end{split}
\label{eq:LagrangianSU2naU1}
\end{equation}
Here the next designations are introduced: $\hat{I}-$ unity matrix $4\times 4$ of bispinor indices,
${{\hat{\gamma }}^{5}}=i{{\hat{\gamma }}^{0}}{{\hat{\gamma }}^{1}}{{\hat{\gamma }}^{2}}{{\hat{\gamma }}^{3}}$ (matrix $\left( {1}/{2}\; \right)\left( \hat{I}-{{\gamma }^{5}} \right)$ takes the column 
\eqref{eq:electron} into the column which has the same left-handed components and zero right-handed components, 
$ {{g}^{em}}- $ electromagnetic interaction constant,

\begin{equation}\label{poznachenna_em}
 {{\alpha }_{-1/2}}\left( x \right)=\exp \left( -i{{g}^{em}}{{\phi }_{em}}\left( x \right) \right),{{\beta }_{1/2}}\left( x \right)=\exp \left( i{{g}^{em}}{{\phi }_{em}}\left( x \right) \right) ,  {{\phi }_{em}}\left( x \right)\equiv {{{\phi }_{0}}\left( x \right)}/{{{g}^{em}}}\; . 
\end{equation}
 Here we have considered \eqref{eq:equivalent_alfa_i_beta}. Furthermore:
\begin{equation}
\begin{split}
  & {{F}_{ab,+}}\left( x \right)=\frac{\partial W_{b}^{+}\left( x \right)}{\partial {{x}^{a}}}-\frac{\partial W_{a}^{+}\left( x \right)}{\partial {{x}^{b}}}+igW_{a}^{+}\left( x \right){{A}_{b,3}}\left( x \right)-igW_{b}^{+}\left( x \right){{A}_{a,3}}\left( x \right) \\ 
& {{F}_{ab,-}}\left( x \right)=\frac{\partial W_{b}^{-}\left( x \right)}{\partial {{x}^{a}}}-\frac{\partial W_{a}^{-}\left( x \right)}{\partial {{x}^{b}}}-igW_{a}^{-}\left( x \right){{A}_{b,3}}\left( x \right)+igW_{b}^{-}\left( x \right){{A}_{a,3}}\left( x \right) \\ 
& {{F}_{ab,3}}\left( x \right)=\frac{\partial {{A}_{b,3}}}{\partial {{x}^{a}}}-\frac{\partial {{A}_{a,3}}}{\partial {{x}^{b}}}-\frac{i}{2}g\left( W_{a}^{+}\left( x \right)W_{b}^{-}\left( x \right)-W_{a}^{-}\left( x \right)W_{b}^{+}\left( x \right) \right), \\ 
\end{split}
\label{eq:Fab_neabelev}
\end{equation}
- the components of non-abelian field strength tensor and
\begin{equation}\label{Fab_electromag}
{{F}_{ab}}\left( x \right)=\frac{\partial {{A}_{b}}\left( x \right)}{\partial {{x}^{a}}}-\frac{\partial {{A}_{a}}\left( x \right)}{\partial {{x}^{b}}},
\end{equation}
- the components of the electromagnetic field strength tensor. The designation of the Lagrangian \eqref{eq:LagrangianSU2naU1} is connected with its invariance under the consistent acts of the local transformation (let's designate such operation by {\flqq$\circ $\frqq}). The first such transformateion has form:
\begin{equation}
 \begin{split} & \left( \begin{matrix} 
{{{\hat{\nu }}}_{e}}\left( x \right)  \\
{{{\hat{e}}}^{L}}\left( x \right)  \\
\end{matrix} \right)=\exp \left( -\frac{i}{2}g{{\theta }_{{{g}_{1}}}}\left( x \right){{{\hat{\sigma }}}_{{{g}_{1}}}} \right)\left( \begin{matrix}
{{{{\hat{\nu }}'}}_{e}}\left( x \right)  \\
{{{{\hat{e}}}}'^{L}}\left( x \right)  \\
\end{matrix} \right),{{{\hat{e}}}^{R}}\left( x \right)={{{{\hat{e}}}}'^{R}}\left( x \right), \\ 
& \left( \begin{matrix}
{{{\hat{\bar{\nu }}}}_{e}}\left( x \right) & {{{\hat{\bar{e}}}}^{L}}\left( x \right)  \\
\end{matrix} \right)=\left( \begin{matrix}
{{{{\hat{\bar{\nu }}}'}}_{e}}\left( x \right) & {{{{\hat{\bar{e}}}}}'^{L}}\left( x \right)  \\
\end{matrix} \right)\exp \left( -\frac{i}{2}g{{\theta }_{{{g}_{1}}}}\left( x \right){{{\hat{\sigma }}}_{{{g}_{1}}}} \right),{{{\hat{\bar{e}}}}_{R}}\left( x \right)={{{{\hat{\bar{e}}}}}'^{R}}\left( x \right), \\ 
& {{A}_{{{a}_{1}},{{g}_{1}}}}\left( x \right)={{D}_{{{g}_{1}},{{g}_{2}}}}\left( \vec{\theta }\left( x \right) \right){{{{A}'}}_{{{a}_{1}},{{g}_{2}}}}\left( x \right)-\frac{\partial {{\theta }_{{{g}_{1}}}}\left( x \right)}{\partial {{x}^{{{a}_{1}}}}},{{A}_{{{a}_{1}}}}\left( x \right)={{{{A}'}}_{{{a}_{1}}}}\left( x \right) ,\\ 
& {W'}_{{{a}_{1}}}^{+}\left( x \right)={{A'}_{{{a}_{1}},{{g}_{1}}=1}}\left( x \right)-i{{A'}_{{{a}_{1}},{{g}_{1}}=2}}\left( x \right),  {W'}_{{{a}_{1}}}^{-}\left( x \right)={{A'}_{{{a}_{1}},{{g}_{1}}=1}}\left( x \right)+i{{A'}_{{{a}_{1}},{{g}_{1}}=2}}\left( x \right), \\ 
& {{\alpha }_{-{1}/{2}\;}}\left( x \right)={{{{\alpha }'}}_{-{1}/{2}\;}}\left( x \right),{{\beta }_{{1}/{2}\;}}\left( x \right)={{{{\beta }'}}_{{1}/{2}\;}}\left( x \right) .\\ 
\end{split}
\label{eq:SU2_E}
\end{equation}

It is naturally to denote this transformation by $SU\left( 2 \right)\otimes E$. Here $E-$ denotes the identity transformation. And the designation $SU\left( 2 \right)\otimes E$ reflects the fact that the part of the quantities transforms by the local $SU\left( 2 \right)-$transformation and another part - by the identity transformation \eqref{eq:SU2_E}. The second transformation we will denote by the $E\otimes U\left( 1 \right)$ and it has the following form:

\begin{equation}
\begin{split}
  & {{{{\hat{\nu }}'}}_{e}}\left( x \right)={{{{\hat{\nu }}''}}_{e}}\left( x \right),{{{{\hat{\bar{\nu }}}'}}_{e}}\left( x \right)={{{{\hat{\bar{\nu }}}''}}_{e}}\left( x \right), \\ 
& {{{{\hat{e}}}}'^{L}}\left( x \right)=\exp \left( i{{g}^{em}}{{\phi }_{em}}\left( x \right) \right){{{{\hat{e}}}}''^{L}}\left( x \right),{{{{\hat{\bar{e}}}'}}^{L}}\left( x \right)={{{{\hat{\bar{e}}}}}''^{L}}\left( x \right)\exp \left( -i{{g}^{em}}{{\phi }_{em}}\left( x \right) \right), \\ 
& {{{{\hat{e}}}}'^{R}}\left( x \right)=\exp \left( i{{g}^{em}}{{\phi }_{em}}\left( x \right) \right){{{{\hat{e}}}}''^{R}}\left( x \right),{{{{\hat{\bar{e}}}}}'^{R}}\left( x \right)={{{{\hat{\bar{e}}}}}''^{R}}\left( x \right)\exp \left( -i{{g}^{em}}{{\phi }_{em}}\left( x \right) \right) \\ 
& {{{{\alpha }'}}_{-{1}/{2}\;}}\left( x \right)=\exp \left( -i{{g}^{em}}{{\phi }_{em}}\left( x \right) \right){{{{\alpha }''}}_{-{1}/{2}\;}}\left( x \right),{{{{\beta }'}}_{{1}/{2}\;}}\left( x \right)=\exp \left( i{{g}^{em}}{{\phi }_{em}}\left( x \right) \right){{{{\beta }''}}_{{1}/{2}\;}}\left( x \right) \\ 
& {{{{A'}}}_{{{a}_{1}}}}\left( x \right)={{{{A''}}}_{{{a}_{1}}}}\left( x \right)+\frac{\partial {{\phi }_{em}}\left( x \right)}{\partial {{x}^{{{a}_{1}}}}},{{{{A'}}}_{{{a}_{1}},{{g}_{1}}}}\left( x \right)={{{{A''}}}_{{{a}_{1}},{{g}_{1}}}}\left( x \right), \\ 
& {W'}_{{{a}_{1}}}^{+}\left( x \right)={W''}_{{{a}_{1}}}^{+}\left( x \right),{W'}_{{{a}_{1}}}^{-}\left( x \right)={W''}_{{{a}_{1}}}^{-}\left( x \right). \\ 
\end{split}
\label{eq:E_U1}
\end{equation}

Note that the transformations  $SU\left( 2 \right)\otimes E$ and $E\otimes U\left( 1 \right)$ do not commute. Nevertheless at first should be applied the $SU\left( 2 \right)\otimes E-$ transformation. 
The order of the transformations is important because under the  $E\otimes U\left( 1 \right)$ transformation the transformed quantities with the different weak isospin ${{I}_{3}}$ projections are transfrormed by the different laws. While the $SU\left( 2 \right)\otimes E-$ transformation {\flqq mix up \frqq} these components. For example in the considered case $E\otimes U\left( 1 \right)$-transformaion acts nontrivially on the component with the smaller value of ${{I}_{3}}$. Therefore it is necessary first to determine this component using $SU\left( 2 \right)\otimes E-$ transformation and after that apply the transformation $E\otimes U\left( 1 \right)$. The Lagrangian \eqref{eq:LagrangianSU2naU1} will describe both the electromagnetic and weak interaction of the leptons if append the contributions analogous to the described below but for another lepton generations. Moreover as one can see at \eqref{eq:LagrangianSU2naU1} the Lagrangian doesn't contain the summands which contain the product of the electromagnetic field functions with  $ W_{{{a}_{1}}}^{\pm }\left( x \right),{{A}_{{{a}_{1}},3}}\left( x \right)$ field functions. As a result the equations for the electromagnetic field won't depend on the gauge selection for the non-abelian field unlike the Standard Model as we considered before and we can left the standard description for the electromagnetic field. It is commonly approved that the main experimental evidence of the Glashow-Weinberg-Salam model is the discovery of the reactions with the neutral currents \cite{HASERT1973121,HASERT1973138}. But the experimental observation of such processes only proves that the one of the weak interaction mediators is neutral but this observation doesn't connected with the theoretical ideas about the {\flqq mixing \frqq} of the fields which describe the weak and electromagnetic interactions according to the formula \eqref{eq:A_kak_lin_komb_teta}. From the theoretical point of view the existance of the neutral currents is connected with the fact that one of the three generators of $SU\left( 2 \right)$ group may be adjusted to the diagonal form (this fact follows from the commutation relations between these generators). By virtue of the fact that the the representation is usualy used, in which the $\left( {1}/{2}\; \right){{\hat{\sigma }}_{3}}$ is diagonal, so the reactions with the neutral currents are connected with the exchange of the ${{A}_{{{a}_{1}},{{g}_{1}}=3}}\left( x \right)$ field particle. In the considered model one can associate the field ${{A}_{{{a}_{1}},{{g}_{1}}=3}}\left( x \right)$ with the ${{Z}^{0}}-$ bosons - the neutral mediators of the weak interaction.

Introducing the electromagnetic interaction in the considered way we didn't take into account that in  the {\flqq old \frqq} Standard Model neutrino is massless and so on should be represented only by the left component. It is hard to explain th existence of the neutrino oscillation \cite{SuperCamiokaNDEPhysRevLett.81.1562,SNOcollaborationPhysRevLett.87.071301,SNO_1_PhysRevLett.89.011301} with help of some interaction between the different neutrino species because in this case it means the tangible change of the Standard Model and building the new theory based on the some new symmetry group. Therefore the simplest way now to introduce the neutrino osclillations into the model is to introduce the mass summand of the neutrino field into the Lagrangian of the model \cite{Olive:2016xmw}. This mass summand is described by the matrix which is nondiagonal on the different neutrino species. It gives the possibility to obtain in the theory the processes of neutrino transformations between species. However as a result the Dirac equation for neutrino field may not be satisfied now using zero right-handed components as it could be done when the mass summand equals zero. But the existance of the neutrinos right-handed components leads to the inappreciable changes to the ideas given above. Let the right-handed neutrino field transforms in a trivial way under the both  \eqref{eq:SU2_E} and \eqref{eq:E_U1} transformation considered above.

\begin{equation}
{{\hat{\nu }}^{R}}_{e}\left( x \right)={{{\hat{\nu }}}}'^{R}_{e}\left( x \right)={{{\hat{\nu }}}''^{R}}_{e}\left( x \right),{{\hat{\bar{\nu }}}}^{R}_{e}\left( x \right)={{{\hat{\bar{\nu }}}}}'^{R}_{e}\left( x \right)={{{\hat{\bar{\nu }}}}}''^{R}_{e}\left( x \right).
\label{eq:Pravoe_neitrino}
\end{equation}
The designation $ \hat{\nu_{e}} $ was introduced in \eqref{eq:poznachenna_nu_e}. Let's replace it similar to \eqref{eq:electron}:
 \begin{equation}
\hat{\nu_{e}}\left( x \right)={{\hat{\nu_{e}}}^{L}}\left( x \right)+{{\hat{\nu_{e}}}^{R}}\left( x \right)
\label{eq:nu_e_now}
\end{equation}

Considering this designation in the Lagrangian \eqref{eq:LagrangianSU2naU1} it is necessary to append only the projections to the left-handed subspace for the neutrino fields in such summands which contain the interaction with the gauge field components for the local $SU\left( 2 \right)-$ symmetry.

Let's append an analogous Lagrangians for the other lepton generations and consider the interaction with the Higgs field which provides the mass to leptons. As the result we obtain the complete Lagrangian of the model in which the function of the electromagnetic interaction to recover the local $U\left( 1 \right)-$ symmetry is {\flqq restored \frqq} and the known description of this interaction is {\flqq not destroyed\frqq}. The relation \eqref{eq:A_kak_lin_komb_teta} leads to {\flqq destruction \frqq} of the standard desctiption of the electromagnetic interaction. This relation apperas in the Standard Model as the corollary of the problem solution. The problem is that how to {\flqq constrain \frqq} the electromagnetic field to interact with the electron field and do not interact with the neutrino field at the same time. In our case it comes up due to the fact that the {\flqq redundant \frqq} phase multipliers in \eqref{eq:Lagrangian_s_fi0} may be eliminated through the local $U\left( 1 \right)-$ transformation of the one of two weak isospin components. Therefore the only one of two components will interact with the electromagnetic field so this component should be considered as electron field and another component - as neutrino field. So as we can see such approach is not opposed to both the existance of the non-zero masses of neutrino, and the presence of its right-handed components.

Let's pay attention to the Lagrangian \eqref{eq:LagrangianSU2naU1} with the considered modifications which connected with the possible presence of the right-handed components of neutrino. This Lagrangian beside the $\left( E\otimes U\left( 1 \right) \right)\circ \left( SU\left( 2 \right)\otimes E \right)$ symmetry has also the global $U\left( 1 \right)-$ symmetry relative to the transformation:
\begin{equation}
\begin{split}
  & \hat{e}\left( x \right)=\exp \left( -i\chi  \right){\hat{e}}'\left( x \right),\hat{\bar{e}}\left( x \right)=\exp \left( i\chi  \right){\hat{\bar{e}}}'\left( x \right), \\ 
& W_{{{a}_{1}}}^{-}\left( x \right)=\exp \left( -i\chi  \right){W}_{{{a}_{1}}}'^{-}\left( x \right),W_{{{a}_{1}}}^{+}\left( x \right)=\exp \left( i\chi  \right){W}_{{{a}_{1}}}'^{+}\left( x \right). \\ 
\end{split}
\label{eq:Global_U_1}
\end{equation}  
Here $\chi-$ parameter of the global $U\left( 1 \right)-$ transformation. The symmetry under the transformation \eqref{eq:Global_U_1} provides the electric charge conservation in the processes which connected with the weak interaction with the accounting of its mediators charge. Let's pay attention to the fact that the local $U\left( 1 \right)-$ transformation is connected with the introduction of the electromagnetic interaction. So this local $U\left( 1 \right)-$ transformation and the global transformation \eqref{eq:Global_U_1} are not connected in a conventional manner in the considered model. First of them can't be obtained from the second replacing the global parameter by the arbitrary function of the coordinates.

Also the feature of this model is that the local $U\left( 1 \right)-$ symmetry can't be introduced independently from the local $SU\left( 2 \right)-$ symmetry. It is because the local $U\left( 1 \right)-$ transformation of the electron field is connected with the transformation of the generators of the local $SU\left( 2 \right)-$ transformation. Furthermore the local $ \left( E\otimes U\left( 1 \right) \right) $- transformation may be applied only after the local $ \left( SU\left( 2 \right)\otimes E \right)- $ transformation. From such point of view it is meaningful to speak about consistent electroweak interaction.

To apply the considered model to the weak interaction of quarks let's note that the phase multipliers inside the coefficients ${{\alpha }_{-{1}/{2}\;}}$ and ${{\beta }_{{1}/{2}\;}}$ may be eliminated not only by the local $U\left( 1 \right)-$ transformation of the one of weak isospin doublet components but also by the mutual transformation of the both components. Let's consider the arbitrary isospin doublet
\begin{equation}\label{Dovilnij_izospinovij_dublet}
 \left( \begin{matrix}
\hat{\psi }_{{{I}_{3}}={1}/{2}\;}^{L}\left( x \right)  \\
\hat{\psi }_{{{I}_{3}}={-1}/{2}\;}^{L}\left( x \right)  \\
\end{matrix} \right).
\end{equation}  
For example, field $ \hat{\psi }_{{{I}_{3}}={1}/{2}\;}^{L}\left( x \right) $ may correspond to the left components of the $u-$quark field and $ \hat{\psi }_{{{I}_{3}}=-{1}/{2}\;}^{L}\left( x \right) $ - to the $ d- $quark field. Or instead of these fields an analogous fields may be considered for another generations of the quarks. Considering \eqref{eq:pidvihuchij_i_ponujuuchij}, the interaction Lagrangian \eqref{eq:LAint_sigma_plus_minus} may be represented in the following form:
\begin{equation}
\begin{split}
& L_{{\hat{A}}}^{\operatorname{int}}=\frac{g}{2}W_{{{a}_{1}}}^{+}\left( x \right)\exp \left( -i{{g}^{em}}{{\phi }_{em}}\left( x \right) \right)\left( x \right)\left( \hat{\bar{\psi }}_{{1}/{2}\;}^{L}\left( x \right){{{\hat{\gamma }}}^{{{a}_{1}}}}\hat{\psi }_{{-1}/{2}\;}^{L}\left( x \right) \right)+ \\ 
& +\frac{g}{2}W_{{{a}_{1}}}^{-}\left( x \right)\exp \left( i{{g}^{em}}{{\phi }_{em}}\left( x \right) \right)\left( \hat{\bar{\psi }}_{{-1}/{2}\;}^{L}\left( x \right){{{\hat{\gamma }}}^{{{a}_{1}}}}\hat{\psi }_{{1}/{2}\;}^{L}\left( x \right) \right) \\ 
& +\frac{g}{2}{{A}_{{{a}_{1}},3}}\left( x \right)\left( \hat{\bar{\psi }}_{{1}/{2}\;}^{L}\left( x \right){{{\hat{\gamma }}}^{{{a}_{1}}}}\hat{\psi }_{{1}/{2}\;}^{L}\left( x \right)-\hat{\bar{\psi }}_{{-1}/{2}\;}^{L}\left( x \right){{{\hat{\gamma }}}^{{{a}_{1}}}}\hat{\psi }_{{-1}/{2}\;}^{L}\left( x \right) \right). \\ 
\end{split}
\label{eq:Lagrangian_v_obhem_sluchae}
\end{equation}
Here the designations are the same as in the \eqref{poznachenna_em}. Let's consider the local $U\left( 1 \right)$-transformation:
\begin{equation}
\begin{split}
& \hat{\psi }_{{-1}/{2}\;}^{L}\left( x \right)=\exp \left( -i{{q}_{{-1}/{2}\;}}{{g}^{em}}{{\phi }_{em}}\left( x \right) \right){\hat{\psi }}_{{-1}/{2}\;}'^{L}\left( x \right), \\ 
& \hat{\psi }_{{1}/{2}\;}^{L}\left( x \right)=\exp \left( -i{{q}_{{1}/{2}\;}}{{g}^{em}}{{\phi }_{em}}\left( x \right) \right){\hat{\psi }}_{{1}/{2}\;}'^{L}\left( x \right), \\ 
& \hat{\bar{\psi }}_{{-1}/{2}\;}^{L}\left( x \right)=\exp \left( i{{q}_{{-1}/{2}\;}}{{g}^{em}}{{\phi }_{em}}\left( x \right) \right){\hat{\bar{\psi }}'^{L}}_{{-1}/{2}\;}\left( x \right), \\ 
& \hat{\bar{\psi }}_{{1}/{2}\;}^{L}\left( x \right)=\exp \left( i{{q}_{{1}/{2}\;}}{{g}^{em}}{{\phi }_{em}}\left( x \right) \right){\hat{\bar{\psi }}'^{L}}_{{1}/{2}\;}\left( x \right). \\ 
\end{split}
\label{eq:U1_quarks}
\end{equation}
The designations $ {{q}_{{\pm 1}/{2}\;}}$ are the charges of the corresponding field quantums measured in the units of the electromagnetic interaction constant with consideration of the sign.
For example, if the fields ${{I}_{3}}={\pm1}/{2}\;$ correspond to the quarks of the same generation, then \cite{Olive:2016xmw}
\begin{equation}\label{zaradi_quarkiv}
{{q}_{{1}/{2}\;}}={2}/{3}\;,{{q}_{{-1}/{2}\;}}=-{1}/{3}\;. 
\end{equation} 
Substituting the \eqref{eq:U1_quarks} inside the \eqref{eq:Lagrangian_v_obhem_sluchae} we obtain:
\begin{equation}
\begin{split}
& L_{{\hat{A}}}^{\operatorname{int}}=\frac{g}{2}W_{{{a}_{1}}}^{+}\left( x \right)\exp \left( i\left( {{q}_{{1}/{2}\;}}-{{q}_{{-1}/{2}\;}}-1 \right){{g}^{em}}{{\phi }_{em}}\left( x \right) \right)\left( {\hat{\bar{\psi }}'^{L}}_{{1}/{2}\;}\left( x \right){{{\hat{\gamma }}}^{{{a}_{1}}}}{\hat{\psi }'^{L}}_{{-1}/{2}\;}\left( x \right) \right)+ \\ 
& +\frac{g}{2}W_{{{a}_{1}}}^{-}\left( x \right)\exp \left( i\left( 1+{{q}_{{-1}/{2}\;}}-{{q}_{{1}/{2}\;}} \right){{g}^{em}}{{\phi }_{em}}\left( x \right) \right)\left( {\hat{\bar{\psi }}'^{L}}_{{-1}/{2}\;}\left( x \right){{{\hat{\gamma }}}^{{{a}_{1}}}}{\hat{\psi }}_{{1}/{2}\;}'^{L}\left( x \right) \right) \\ 
& +\frac{g}{2}{{A}_{{{a}_{1}},3}}\left( x \right)\left( {\hat{\bar{\psi }}'^{L}}_{{1}/{2}\;}\left( x \right){{{\hat{\gamma }}}^{{{a}_{1}}}}{\hat{\psi }}_{{1}/{2}\;}'^{L}\left( x \right)-{\hat{\bar{\psi }}'^{L}}_{{-1}/{2}\;}\left( x \right){{{\hat{\gamma }}}^{{{a}_{1}}}}{\hat{\psi }}_{{-1}/{2}\;}'^{L}\left( x \right) \right). \\ 
\end{split}
\label{eq:Laint_s_q}
\end{equation}

Taking into account the \eqref{zaradi_quarkiv} we see that the transformation \eqref{eq:U1_quarks} compensate {\flqq redundant  \frqq} phase multipliers which come from the coefficients ${{\alpha }_{-{1}/{2}\;}}$ and ${{\beta }_{{1}/{2}\;}}$. To present the local symmetry for the transformation \eqref{eq:U1_quarks} we should extend the derivatives for each field and so on introduce its interaction with the electromagnetic field. Let's amplify the transformation \eqref{eq:U1_quarks} by the analogous transformation for the right-handed components of corresponding field and extend the derivatives for this components to present the intrinsic for this field inverse symmetry.

In the next section we consider the second problem of the Standard Model which connected with the introduction of the non-gauge interactions.

\section{A two-particle gauge field equation}
Let’s consider two instances of the gauge field  ${{A}_{{{a}_{1}},{{g}_{1}}}}\left( {{x}_{1}} \right)$ and ${{A}_{{{a}_{2}},{{g}_{2}}}}\left( {{x}_{2}} \right)$. Here ${a}_{1}$ and ${a}_{2}$ - four vector Lorentz indices, which are equal  0,1,2,3, ${g}_{1}$ and  ${g}_{2}$- internal indices which are equal 1,2,3 for the $SU\left( 2 \right)$ group, $ {x}_{1} $ and $ {x}_{2} $ - four vectors of the Minkowski space. Further we will denote Lorentz indices with the letter $a$ with different sub-indexes, and we will denote the internal indices with the letter-  $ g $ with different sub-indexes. At the same time we will denote a coupling constant with $g$ without sub-indexes. Besides a usual summation by repetitive indexes is used.

We construct the corresponding strength tensors ${{F}_{{{a}_{1}}{{a}_{2}},{{g}_{1}}}}\left( {{x}_{1}} \right)$ and ${{F}_{{{a}_{3}}{{a}_{4}},{{g}_{2}}}}\left( {{x}_{2}} \right)$ for these fields and build Lagrangians on them. If we allow the Levi-Civita symbol properties ${{\varepsilon }_{{{g}_{1}}{{g}_{2}}{{g}_{3}}}}$, through which the adjoint representation $SU\left( 2 \right)$ group generators are expressed we will get the Lagrange-Euler equation for each of these fields:
\begin{equation}
\begin{split}
   & g{{\varepsilon }_{{{g}_{1}}{{g}_{13}}{{g}_{5}}}}\frac{\partial {{A}_{{{a}_{1}},{{g}_{1}}}}\left( {{x}_{b}} \right)}{\partial x_{b}^{{{a}_{2}}}}{{A}_{{{a}_{4}},{{g}_{13}}}}\left( {{x}_{b}} \right){{g}^{{{a}_{1}}{{a}_{4}}}}{{g}^{{{a}_{2}}{{a}_{5}}}} \\ 
 & -g{{\varepsilon }_{{{g}_{5}}{{g}_{2}}{{g}_{3}}}}\frac{\partial {{A}_{{{a}_{4}},{{g}_{3}}}}\left( {{x}_{b}} \right)}{\partial x_{b}^{{{a}_{6}}}}{{A}_{{{a}_{3}},{{g}_{2}}}}\left( {{x}_{b}} \right){{g}^{{{a}_{3}}{{a}_{5}}}}{{g}^{{{a}_{6}}{{a}_{4}}}} \\ 
 & -{{g}^{2}}\left( {{A}_{{{a}_{1}},{{g}_{12}}}}\left( {{x}_{b}} \right){{A}_{{{a}_{3}},{{g}_{12}}}}\left( {{x}_{b}} \right) \right){{A}_{{{a}_{4}},{{g}_{5}}}}\left( {{x}_{b}} \right){{g}^{{{a}_{1}}{{a}_{3}}}}{{g}^{{{a}_{5}}{{a}_{4}}}} \\ 
 & +{{g}^{2}}\left( {{A}_{{{a}_{1}},{{g}_{13}}}}\left( {{x}_{b}} \right){{A}_{{{a}_{4}},{{g}_{13}}}}\left( {{x}_{b}} \right) \right){{A}_{{{a}_{3}},{{g}_{5}}}}\left( {{x}_{b}} \right){{g}^{{{a}_{1}}{{a}_{3}}}}{{g}^{{{a}_{5}}{{a}_{4}}}} \\ 
 & +\frac{{{\partial }^{2}}{{A}_{{{a}_{3}},{{g}_{5}}}}\left( {{x}_{b}} \right)}{\partial x_{b}^{{{a}_{6}}}\partial x_{b}^{{{a}_{4}}}}{{g}^{{{a}_{5}}{{a}_{3}}}}{{g}^{{{a}_{6}}{{a}_{4}}}}-\frac{{{\partial }^{2}}{{A}_{{{a}_{4}},{{g}_{5}}}}\left( {{x}_{b}} \right)}{\partial x_{b}^{{{a}_{6}}}\partial x_{b}^{{{a}_{3}}}}{{g}^{{{a}_{5}}{{a}_{3}}}}{{g}^{{{a}_{6}}{{a}_{4}}}}=0. \\
\end{split}
\label{eq:Lagrang_Eiler}
\end{equation}
Here $g$-a coupling constant, $b$ - an index which equals respectively 1 and 2 for each instances of the gauge field, ${{g}^{{{a}_{1}}{{a}_{2}}}}$ - Minkowski tensor components. The indexes ${{a}_{5}}$ and ${{g}_{5}}$ are not summation indexes among the indexes included in (\ref{eq:Lagrang_Eiler}). In order to get the equation for a scalar two-particle field relatively Lorentz group let’s fold the first equation for which  $b=1$ from ${{A}_{{{a}_{5}},{{g}_{15}}}}\left( {{x}_{2}} \right)$. Herewith ${g}_{15}$ is an arbitrary value of internal index which are not related to the values included in (\ref{eq:Lagrang_Eiler}). Let’s choose the equation depends on $ {x}_{2} $ from equation system which for the second instance of the field. This equation corresponds to the internal index ${g}_{15} $ and we will fold it with  ${{A}_{{{a}_{5}},{{g}_{5}}}}\left( {{x}_{1}} \right)$.  The left parts of the resulting equations will depend on a pair of indexes ${g}_{5},{g}_{15}$. Herewith the functions from ${x}_{1}$ will be introduced under the sign of the derivatives of ${x}_{2}$.The obtained expressions are violated only on the subset ${{x}_{1}}={{x}_{2}}$ of the two Minkowski spaces tensor product. This subset has a zero measure. There are no reasons to consider for this subset, unaccounted members turn to infinity. These members haven’t contribute a non-zero contribution to the observed dimensions. 
After described transformations the first equation from (\ref{eq:Lagrang_Eiler}) becomes:
\begin{equation}
\begin{split}
    & 2g{{\varepsilon }_{{{g}_{1}}{{g}_{5}}{{g}_{3}}}}\frac{\partial \left( {{A}_{{{a}_{2}},{{g}_{1}}}}\left( {{x}_{1}} \right){{A}_{{{a}_{5}},{{g}_{15}}}}\left( {{x}_{2}} \right) \right)}{\partial x_{1}^{{{a}_{1}}}}{{A}_{{{a}_{4}},{{g}_{3}}}}\left( {{x}_{1}} \right){{g}^{{{a}_{1}}{{a}_{4}}}}{{g}^{{{a}_{2}}{{a}_{5}}}}+ \\ 
 & +g{{\varepsilon }_{{{g}_{1}}{{g}_{13}}{{g}_{5}}}}\frac{\partial \left( {{A}_{{{a}_{1}},{{g}_{1}}}}\left( {{x}_{1}} \right){{A}_{{{a}_{5}},{{g}_{15}}}}\left( {{x}_{2}} \right) \right)}{\partial x_{1}^{{{a}_{2}}}}{{A}_{{{a}_{4}},{{g}_{13}}}}\left( {{x}_{1}} \right){{g}^{{{a}_{1}}{{a}_{4}}}}{{g}^{{{a}_{2}}{{a}_{5}}}}- \\ 
 & -g{{\varepsilon }_{{{g}_{5}}{{g}_{2}}{{g}_{3}}}}\frac{\partial A\left( _{{{a}_{4}},{{g}_{3}}}\left( {{x}_{1}} \right){{A}_{{{a}_{5}},{{g}_{15}}}}\left( {{x}_{2}} \right) \right)}{\partial x_{1}^{{{a}_{6}}}}{{A}_{{{a}_{3}},{{g}_{2}}}}\left( {{x}_{1}} \right){{g}^{{{a}_{3}}{{a}_{5}}}}{{g}^{{{a}_{6}}{{a}_{4}}}}- \\ 
 & -{{g}^{2}}\left( {{A}_{{{a}_{1}},{{g}_{12}}}}\left( {{x}_{1}} \right){{A}_{{{a}_{3}},{{g}_{12}}}}\left( {{x}_{1}} \right) \right)\left( {{A}_{{{a}_{4}},{{g}_{5}}}}\left( {{x}_{1}} \right){{A}_{{{a}_{5}},{{g}_{15}}}}\left( {{x}_{2}} \right) \right){{g}^{{{a}_{1}}{{a}_{3}}}}{{g}^{{{a}_{5}}{{a}_{4}}}}+ \\ 
 & +{{g}^{2}}\left( {{A}_{{{a}_{1}},{{g}_{13}}}}\left( {{x}_{1}} \right){{A}_{{{a}_{4}},{{g}_{13}}}}\left( {{x}_{1}} \right) \right)\left( {{A}_{{{a}_{3}},{{g}_{5}}}}\left( {{x}_{1}} \right){{A}_{{{a}_{5}},{{g}_{15}}}}\left( {{x}_{2}} \right) \right){{g}^{{{a}_{1}}{{a}_{3}}}}{{g}^{{{a}_{5}}{{a}_{4}}}}+ \\ 
 & +\frac{{{\partial }^{2}}\left( {{A}_{{{a}_{3}},{{g}_{5}}}}\left( {{x}_{1}} \right){{A}_{{{a}_{5}},{{g}_{15}}}}\left( {{x}_{2}} \right) \right)}{\partial x_{1}^{{{a}_{6}}}\partial x_{1}^{{{a}_{4}}}}{{g}^{{{a}_{5}}{{a}_{3}}}}{{g}^{{{a}_{6}}{{a}_{4}}}}- \\ 
 & -\frac{\partial ^{2}\left({{A}_{{{a}_{4}},{{g}_{5}}}}\left( {{x}_{1}} \right){{A}_{{{a}_{5}},{{g}_{15}}}}\left( {{x}_{2}} \right) \right)}{\partial x_{1}^{{{a}_{6}}}\partial x_{1}^{{{a}_{3}}}}{{g}^{{{a}_{5}}{{a}_{3}}}}{{g}^{{{a}_{6}}{{a}_{4}}}}=0. \\
\end{split}
\label{eq:Zgortka1z2}
\end{equation}
After analogous transformations the second equation from (\ref{eq:Lagrang_Eiler}) will be different from (\ref{eq:Zgortka1z2}) with replace ${{x}_{1}}$ by ${{x}_{2}}$. 

The value $ {{A}_{{{a}_{1}},{{g}_{1}}}}\left( {{x}_{1}} \right){{A}_{{{a}_{2}},{{g}_{2}}}}\left( {{x}_{2}} \right) $   turns  as a two covariant tensor with Lorentz transformations. Let's consider the Lorentz group representation on the linear space of such tensors. We will decompose this linear space into a direct sum of invariant subspaces relatively considered representation. We will select an invariant subspace, on which the scalar irreducible representation is realized:
\begin{equation}
\begin{split}
 {{A}_{{{a}_{1}},{{g}_{1}}}}\left( {{x}_{1}} \right){{A}_{{{a}_{2}},{{g}_{2}}}}\left( {{x}_{2}} \right)={{\phi }_{{{g}_{1}}{{g}_{2}}}}\left( {{x}_{1}},{{x}_{2}} \right){{g}_{{{a}_{1}}{{a}_{2}}}}+\ldots 
\end{split}
\label{eq:Videlenie_skalara }
\end{equation}
Here $ {{\phi }_{{{g}_{1}}{{g}_{2}}}}\left( {{x}_{1}},{{x}_{2}} \right) $ - a tensor projection $ {{A}_{{{a}_{1}},{{g}_{1}}}}\left( {{x}_{1}} \right){{A}_{{{a}_{2}},{{g}_{2}}}}\left( {{x}_{2}} \right) $  on an invariant subspace, on which the scalar Lorentz group representation is realized,  \mbox{\flqq ... \frqq} means projections on the remain invariant subspaces. We want to describe the scalar Higgs field so let’s consider a case when projections equal to zero except the scalar projection. This step also can be considered as the preparation for a quantization of the field in the interaction representation, where rejected terms describe the scalar component interaction with other components. We will consider \mbox{\flqq free \frqq} scalar field. Taken (\ref{eq:Videlenie_skalara  }) instead of  (\ref{eq:Zgortka1z2}) we obtain:
\begin{equation}
\begin{split}
  & 2g{{\varepsilon }_{{{g}_{1}}{{g}_{5}}{{g}_{3}}}}\frac{\partial {{\phi }_{{{g}_{1}}{{g}_{15}}}}\left( {{x}_{1}},{{x}_{2}} \right)}{\partial x_{1}^{{{a}_{1}}}}{{A}_{{{a}_{4}},{{g}_{3}}}}\left( {{x}_{1}} \right){{g}^{{{a}_{1}}{{a}_{4}}}}- \\ 
 & -{{g}^{2}}\left( {{A}_{{{a}_{1}},{{g}_{12}}}}\left( {{x}_{1}} \right){{A}_{{{a}_{3}},{{g}_{12}}}}\left( {{x}_{1}} \right) \right){{\phi }_{{{g}_{5}}{{g}_{15}}}}\left( {{x}_{1}},{{x}_{2}} \right){{g}^{{{a}_{1}}{{a}_{3}}}}+\frac{{{\partial }^{2}}{{\phi }_{{{g}_{5}}{{g}_{15}}}}\left( {{x}_{1}},{{x}_{2}} \right)}{\partial x_{1}^{{{a}_{5}}}\partial x_{1}^{{{a}_{3}}}}{{g}^{{{a}_{5}}{{a}_{3}}}}=0. \\ 
\end{split}
\label{eq:fi_g_5_g_15 }
\end{equation}

The term $ 2g{{\varepsilon }_{{{g}_{1}}{{g}_{5}}{{g}_{3}}}}\left( {\partial {{\phi }_{{{g}_{1}}{{g}_{15}}}}\left( {{x}_{1}},{{x}_{2}} \right)}/{\partial x_{1}^{{{a}_{1}}}}\; \right){{A}_{{{a}_{4}},{{g}_{3}}}}\left( {{x}_{1}} \right){{g}^{{{a}_{1}}{{a}_{4}}}} $ can be considered such as described the interaction of field  $ {{\phi }_{{{g}_{5}}{{g}_{15}}}}\left( {{x}_{1}},{{x}_{2}} \right) $ with other fields. Therefor if we consider  \mbox{\flqq free \frqq} field and future quantization in the interaction representation we will reject this term. Let’s repeat the same action with the second equation of the system (\ref{eq:Lagrang_Eiler}). Taking into account the two-particle field equations have be symmetric relatively replace ${{x}_{1}}$ on ${{x}_{2}}$ from physical considerations. Let’s compose the received equations:
\begin{equation}
\begin{split}
 \frac{{{\partial }^{2}}{{\phi }_{{{g}_{1}}{{g}_{2}}}}\left( {{x}_{1}},{{x}_{2}} \right)}{\partial x_{1}^{{{a}_{1}}}\partial x_{1}^{{{a}_{2}}}}{{g}^{{{a}_{1}}{{a}_{2}}}}+\frac{{{\partial }^{2}}{{\phi }_{{{g}_{1}}{{g}_{2}}}}\left( {{x}_{1}},{{x}_{2}} \right)}{\partial x_{2}^{{{a}_{1}}}\partial x_{2}^{{{a}_{2}}}}{{g}^{{{a}_{1}}{{a}_{2}}}}-{{g}^{2}}\chi \left( {{x}_{1}},{{x}_{2}} \right){{\phi }_{{{g}_{1}}{{g}_{2}}}}\left( {{x}_{1}},{{x}_{2}} \right)=0.
\end{split}
\label{eq:Rivnanna_dla_2prticls_field}
\end{equation}

Here we insert a symbol
\begin{equation}
\begin{split}
 \chi \left( {{x}_{1}},{{x}_{2}} \right)\equiv \left( {{A}_{{{a}_{1}},{{g}_{1}}}}\left( {{x}_{1}} \right){{A}_{{{a}_{2}},{{g}_{1}}}}\left( {{x}_{1}} \right)+{{A}_{{{a}_{1}},{{g}_{1}}}}\left( {{x}_{2}} \right){{A}_{{{a}_{2}},{{g}_{1}}}}\left( {{x}_{2}} \right) \right){{g}^{{{a}_{1}}{{a}_{2}}}}.
\end{split}
\label{eq:Xip}
\end{equation}
The function  $ \chi \left( {{x}_{1}},{{x}_{2}} \right) $ is the same function, which was in  \cite{Chudak:2016,Volkotrub:2015laa} by consideration of equations for the two-particle gauge field scalar part. The equation system \eqref{eq:Rivnanna_dla_2prticls_field} is not determined in full, because the function $ \chi\left( x_{1},x_{2}\right) $ is the unknown function such as the function $ {\phi }_{{{g}_{1}}{{g}_{2}}}\left( {{x}_{1}},{{x}_{2}} \right) $. Further we will consider possibility to impose an additional condition, which determine this system completely.

At first we will confine the consideration global $SU\left(2\right)-$ transformations for the two-particle field ${{\phi }_{{{g}_{1}}{{g}_{2}}}}\left( {{x}_{1}},{{x}_{2}} \right)$. The transition to a local symmetry will be made later by the usual way of derivatives extension and introduction interaction with the gauge field. The field functions complex $ {{\phi }_{{{g}_{1}}{{g}_{2}}}}\left( {{x}_{1}},{{x}_{2}} \right) $ makes a tensor relative to global transformations. The field of values of these field functions constructs a linear space, on which a tensor product of two adjoin representations $SU\left(2\right)$ group is realized. Let’s decompose this linear space into a direct sum of invariant subspaces composed of tensors which aliquot single tensor ${{\delta }_{{{g}_{1}}{{g}_{2}}}}$, symmetric tensors with zero trace $ \phi _{{{g}_{1}}{{g}_{2}}}^{s\left( 0 \right)}\left( {{x}_{1}},{{x}_{2}} \right) $ and antisymmetric  $ \phi _{{{g}_{1}}{{g}_{2}}}^{a}\left( {{x}_{1}},{{x}_{2}} \right): $ 
\begin{equation}
\begin{split}
 {{\phi }_{{{g}_{1}}{{g}_{2}}}}\left( {{x}_{1}},{{x}_{2}} \right)={{\rho }_{0}}\left( {{x}_{1}},{{x}_{2}} \right){{\delta }_{{{g}_{1}}{{g}_{2}}}}+\phi _{{{g}_{1}}{{g}_{2}}}^{s\left( 0 \right)}\left( {{x}_{1}},{{x}_{2}} \right)+\phi _{{{g}_{1}}{{g}_{2}}}^{a}\left( {{x}_{1}},{{x}_{2}} \right).
\end{split}
\label{eq:Rozklad_phi_na_nezvidni_tenzori}
\end{equation}
Here ${{\rho }_{0}}\left( {{x}_{1}},{{x}_{2}} \right)-$ a tensor projection  ${{\phi }_{{{g}_{1}}{{g}_{2}}}}\left( {{x}_{1}},{{x}_{2}} \right)$ on a linear subspace of tensors which aliquot a single tensor.
Substituting (\ref{eq:Rozklad_phi_na_nezvidni_tenzori}) in (\ref{eq:Rivnanna_dla_2prticls_field}) we will get an analogous equation  (\ref{eq:Rivnanna_dla_2prticls_field}) for each accented irreducible parts in  (\ref{eq:Rozklad_phi_na_nezvidni_tenzori}). 
\begin{equation}
\begin{split}
  & {{g}^{{{a}_{1}}{{a}_{2}}}}\frac{{{\partial }^{2}}{{\rho }_{0}}\left( {{x}_{1}},{{x}_{2}} \right)}{\partial x_{1}^{{{a}_{1}}}\partial x_{1}^{{{a}_{2}}}}+{{g}^{{{a}_{1}}{{a}_{2}}}}\frac{{{\partial }^{2}}{{\rho }_{0}}\left( {{x}_{1}},{{x}_{2}} \right)}{\partial x_{2}^{{{a}_{1}}}\partial x_{2}^{{{a}_{2}}}}-{{g}^{2}}\chi_{0} \left( {{x}_{1}},{{x}_{2}} \right){{\rho }_{0}}\left( {{x}_{1}},{{x}_{2}} \right)=0, \\ 
& {{g}^{{{a}_{1}}{{a}_{2}}}}\frac{{{\partial }^{2}}\phi _{{{g}_{1}}{{g}_{2}}}^{s\left( 0 \right)}\left( {{x}_{1}},{{x}_{2}} \right)}{\partial x_{1}^{{{a}_{1}}}\partial x_{1}^{{{a}_{2}}}}+{{g}^{{{a}_{1}}{{a}_{2}}}}\frac{{{\partial }^{2}}\phi _{{{g}_{1}}{{g}_{2}}}^{s\left( 0 \right)}\left( {{x}_{1}},{{x}_{2}} \right)}{\partial x_{2}^{{{a}_{1}}}\partial x_{2}^{{{a}_{2}}}}-{{g}^{2}}\chi_{1} \left( {{x}_{1}},{{x}_{2}} \right)\phi _{{{g}_{1}}{{g}_{2}}}^{s\left( 0 \right)}\left( {{x}_{1}},{{x}_{2}} \right)=0, \\ 
& {{g}^{{{a}_{1}}{{a}_{2}}}}\frac{{{\partial }^{2}}\phi _{{{g}_{1}}{{g}_{2}}}^{a}\left( {{x}_{1}},{{x}_{2}} \right)}{\partial x_{1}^{{{a}_{1}}}\partial x_{1}^{{{a}_{2}}}}+{{g}^{{{a}_{1}}{{a}_{2}}}}\frac{{{\partial }^{2}}\phi _{{{g}_{1}}{{g}_{2}}}^{a}\left( {{x}_{1}},{{x}_{2}} \right)}{\partial x_{2}^{{{a}_{1}}}\partial x_{2}^{{{a}_{2}}}}-{{g}^{2}}\chi_{2} \left( {{x}_{1}},{{x}_{2}} \right)\phi _{{{g}_{1}}{{g}_{2}}}^{a}\left( {{x}_{1}},{{x}_{2}} \right)=0. \\ 
\end{split}
\label{eq:Rivnanna_dla_nezvidnih_chstin}
\end{equation}
Symbols  $ \chi_{j}\left( x_{1},x_{2}\right) ,j=0,1,2 $ reflect that fact, the  function $ \chi\left( x_{1},x_{2}\right)   $ is not given at the equation system \eqref{eq:Rivnanna_dla_2prticls_field}, so it can  {\flqq adapt \frqq} to the appropriate function, the derivatives of which are included in the equation. The global group $SU\left( 2 \right)$ representation on the invariant subspace of the antisymmetric tensors $\phi _{{{g}_{1}}{{g}_{2}}}^{a}\left( {{x}_{1}},{{x}_{2}} \right)$ is equivalent to the vector representation. Indeed the three diagonal components of this tensor of nine components are equal to zero as a result of antisymmetry. The remaining six components can be broken into three pairs of components that differ only in the order of the indexes. In each of these pairs it is enough to specify only one component, because the other one will be differ from him only by a sign. Namely a tensor  $\phi _{{{g}_{1}}{{g}_{2}}}^{a}\left( {{x}_{1}},{{x}_{2}} \right)$ has three independent components. We will choose following three components:
\begin{equation}\label{Poznacenna_dla_Levi_Chivitta}
\phi _{12}^{a}\left( {{x}_{1}},{{x}_{2}} \right)\equiv \frac{1}{\sqrt{2}}{{\phi }_{3}}\left( {{x}_{1}},{{x}_{2}} \right),\phi _{31}^{a}\left( {{x}_{1}},{{x}_{2}} \right)\equiv \frac{1}{\sqrt{2}}{{\phi }_{2}}\left( {{x}_{1}},{{x}_{2}} \right),\phi _{23}^{a}\left( {{x}_{1}},{{x}_{2}} \right)\equiv \frac{1}{\sqrt{2}}{{\phi }_{1}}\left( {{x}_{1}},{{x}_{2}} \right).
\end{equation}   
These symbols are introduced to represent an antisymmetric tensor in such form:
\begin{equation}\label{antisimetricnij_cherez_LeviChivitta}
\phi _{{{g}_{1}}{{g}_{2}}}^{a}\left( {{x}_{1}},{{x}_{2}} \right)=\frac{1}{\sqrt{2}}{{\varepsilon }_{{{g}_{1}}{{g}_{2}}{{g}_{3}}}}{{\phi }_{{{g}_{3}}}}\left( {{x}_{1}},{{x}_{2}} \right),
\end{equation}
Where ${{\varepsilon }_{{{g}_{1}}{{g}_{2}}{{g}_{3}}}}-$ - a Levi-Civita symbol. The multiplier  $\left( {1}/{\sqrt{2}}\; \right)$ provides an execution of equality
\begin{equation}\label{Zruchna_rivnist}
\sum\limits_{{{g}_{1}}=1}^{3}{\sum\limits_{{{g}_{1}}=2}^{3}{{{\left( \phi _{{{g}_{1}}{{g}_{2}}}^{a}\left( {{x}_{1}},{{x}_{2}} \right) \right)}^{2}}}}=\sum\limits_{{{g}_{1}}=1}^{3}{{{\left( {{\phi }_{{{g}_{1}}}}\left( {{x}_{1}},{{x}_{2}} \right) \right)}^{2}}},
\end{equation}
which will be convenient for the construction the Lagrangian of the field  $ \phi_{{g}_{1}}\left({{x}_{1}},{{x}_{2}} \right).$ We will fold the both parts apply to  \eqref{eq:Rivnanna_dla_nezvidnih_chstin} from $\phi _{{{g}_{1}}{{g}_{2}}}^{a}\left( {{x}_{1}},{{x}_{2}} \right)$ we will get equation instead  \eqref{eq:Rivnanna_dla_nezvidnih_chstin}
\begin{equation}
\begin{split}
  & {{g}^{{{a}_{1}}{{a}_{2}}}}\frac{{{\partial }^{2}}{{\rho }_{0}}\left( {{x}_{1}},{{x}_{2}} \right)}{\partial x_{1}^{{{a}_{1}}}\partial x_{1}^{{{a}_{2}}}}+{{g}^{{{a}_{1}}{{a}_{2}}}}\frac{{{\partial }^{2}}{{\rho }_{0}}\left( {{x}_{1}},{{x}_{2}} \right)}{\partial x_{2}^{{{a}_{1}}}\partial x_{2}^{{{a}_{2}}}}-{{g}^{2}}\chi_{0} \left( {{x}_{1}},{{x}_{2}} \right){{\rho }_{0}}\left( {{x}_{1}},{{x}_{2}} \right)=0, \\ 
& {{g}^{{{a}_{1}}{{a}_{2}}}}\frac{{{\partial }^{2}}\phi _{{{g}_{1}}{{g}_{2}}}^{s\left( 0 \right)}\left( {{x}_{1}},{{x}_{2}} \right)}{\partial x_{1}^{{{a}_{1}}}\partial x_{1}^{{{a}_{2}}}}+{{g}^{{{a}_{1}}{{a}_{2}}}}\frac{{{\partial }^{2}}\phi _{{{g}_{1}}{{g}_{2}}}^{s\left( 0 \right)}\left( {{x}_{1}},{{x}_{2}} \right)}{\partial x_{2}^{{{a}_{1}}}\partial x_{2}^{{{a}_{2}}}}-{{g}^{2}}\chi_{1} \left( {{x}_{1}},{{x}_{2}} \right)\phi _{{{g}_{1}}{{g}_{2}}}^{s\left( 0 \right)}\left( {{x}_{1}},{{x}_{2}} \right)=0, \\ 
& {{g}^{{{a}_{1}}{{a}_{2}}}}\frac{{{\partial }^{2}}{{\phi }_{{{g}_{1}}}}\left( {{x}_{1}},{{x}_{2}} \right)}{\partial x_{1}^{{{a}_{1}}}\partial x_{1}^{{{a}_{2}}}}+{{g}^{{{a}_{1}}{{a}_{2}}}}\frac{{{\partial }^{2}}{{\phi }_{{{g}_{1}}}}\left( {{x}_{1}},{{x}_{2}} \right)}{\partial x_{2}^{{{a}_{1}}}\partial x_{2}^{{{a}_{2}}}}-{{g}^{2}}\chi_{2} \left( {{x}_{1}},{{x}_{2}} \right){{\phi }_{{{g}_{1}}}}\left( {{x}_{1}},{{x}_{2}} \right)=0. \\
\end{split}
\label{eq:Rivnanna_dla_nezvidnih_chstin1}
\end{equation}
Let’s consider the second and third equations \eqref{eq:Rivnanna_dla_nezvidnih_chstin1} partial solutions which having the form:
\begin{equation}\label{cherez_ro}
\phi _{{{g}_{1}}{{g}_{2}}}^{s\left( 0 \right)}\left( {{x}_{1}},{{x}_{2}} \right)={{\rho }_{1}}\left( {{x}_{1}},{{x}_{2}} \right)e_{{{g}_{1}}{{g}_{2}}}^{s\left( 0 \right)},{{\phi }_{{{g}_{1}}}}\left( {{x}_{1}},{{x}_{2}} \right)={{\rho }_{2}}\left( {{x}_{1}},{{x}_{2}} \right){{e}_{{{g}_{1}}}}.
\end{equation}
Here $ e_{{{g}_{1}}{{g}_{2}}}^{s\left( 0 \right)} - $ some symmetric tensor with a zero trace and with components independent of coordinates, $ {{e}_{{{g}_{1}}}}- $ a vector with components independent of coordinates, $  {{\rho }_{1}}\left( {{x}_{1}},{{x}_{2}} \right),{{\rho }_{2}}\left( {{x}_{1}},{{x}_{2}} \right) -$ unknown functions, which should be established by further solving the equations. Herewith as will be seen further,  $ e_{{{g}_{1}}{{g}_{2}}}^{s\left( 0 \right)} $ and $ {{e}_{{{g}_{1}}}} $ are normalized by conditions
\begin{equation}\label{umovi_normuvanna}
\sum\limits_{{{g}_{1}}=1}^{3}{\sum\limits_{{{g}_{2}}=1}^{3}{{{\left( e_{{{g}_{1}}{{g}_{2}}}^{s\left( 0 \right)} \right)}^{2}}}}=1,\sum\limits_{{{g}_{1}}=1}^{3}{{{\left( {{e}_{{{g}_{1}}}} \right)}^{2}}}=1.
\end{equation}
Considering \eqref{cherez_ro} and \eqref{eq:Rivnanna_dla_nezvidnih_chstin1} all three functions  $ {{\rho }_{j}}\left( {{x}_{1}},{{x}_{2}} \right), j=0,1,2 $ satisfy the one and the same equation: 
\begin{equation}\label{rivnanna_dla_ro}
{{g}^{{{a}_{1}}{{a}_{2}}}}\frac{{{\partial }^{2}}{{\rho }_{j}}\left( {{x}_{1}},{{x}_{2}} \right)}{\partial x_{1}^{{{a}_{1}}}\partial x_{1}^{{{a}_{2}}}}+{{g}^{{{a}_{1}}{{a}_{2}}}}\frac{{{\partial }^{2}}{{\rho }_{j}}\left( {{x}_{1}},{{x}_{2}} \right)}{\partial x_{2}^{{{a}_{1}}}\partial x_{2}^{{{a}_{2}}}}-{{g}^{2}}\chi_{j} \left( {{x}_{1}},{{x}_{2}} \right){{\rho }_{j}}\left( {{x}_{1}},{{x}_{2}} \right)=0,j=0,1,2.
\end{equation}
The difference between these functions may be related to the statement of different boundary conditions for the equation \eqref{rivnanna_dla_ro}. Let’s consider the solution properties of this equation at different boundary conditions 

\section{The equations analysis for functions $ {{\rho }_{j}}\left( {{x}_{1}},{{x}_{2}} \right), j=0,1,2. $ }
Analogous equations  \eqref{rivnanna_dla_ro} have already been considered in \cite{Chudak:2016,Volkotrub:2015laa}. Here we will analyze them with the same method as in these articles, but pay attention to some properties that remained out of view.

Instead of field functions  $ \rho_{j} \left( {{x}_{1}},{{x}_{2}} \right) $ i $ \chi_{j} \left( {{x}_{1}},{{x}_{2}} \right) $ we will introduce new field functions  $ a_{j}\left( {{x}_{1}},{{x}_{2}} \right)$  and $b_{j}\left( {{x}_{1}},{{x}_{2}} \right) $  as follows
\begin{equation}
\begin{split}
   & \rho_{j} \left( {{x}_{1}},{{x}_{2}} \right)=a_{j}\left( {{x}_{1}},{{x}_{2}} \right)-b_{j}\left( {{x}_{1}},{{x}_{2}} \right) \\ 
 & \chi_{j}\left( {{x}_{1}},{{x}_{2}} \right)=a_{j}\left( {{x}_{1}},{{x}_{2}} \right)+b_{j}\left( {{x}_{1}},{{x}_{2}} \right) .\\ 
\end{split}
\label{eq:aib}
\end{equation}
Then the equation  \eqref{rivnanna_dla_ro} is rewritten in a symmetric form
\begin{equation}
\begin{split}
   & \left( \frac{{{\partial }^{2}}a_{j}\left( {{x}_{1}},{{x}_{2}} \right)}{\partial x_{1}^{{{a}_{1}}}\partial x_{1}^{{{a}_{2}}}}{{g}^{{{a}_{1}}{{a}_{2}}}}+\frac{{{\partial }^{2}}a_{j}\left( {{x}_{1}},{{x}_{2}} \right)}{\partial x_{2}^{{{a}_{1}}}\partial x_{2}^{{{a}_{2}}}}{{g}^{{{a}_{1}}{{a}_{2}}}}-{{g}^{2}}{{a_{j}^{2}}}\left( {{x}_{1}},{{x}_{2}} \right) \right)- \\ 
 & -\left( \frac{{{\partial }^{2}}b_{j}\left( {{x}_{1}},{{x}_{2}} \right)}{\partial x_{1}^{{{a}_{1}}}\partial x_{1}^{{{a}_{2}}}}{{g}^{{{a}_{1}}{{a}_{2}}}}+\frac{{{\partial }^{2}}b_{j}\left( {{x}_{1}},{{x}_{2}} \right)}{\partial x_{2}^{{{a}_{1}}}\partial x_{2}^{{{a}_{2}}}}{{g}^{{{a}_{1}}{{a}_{2}}}}-{{g}^{2}}{{b}_{j}^{2}}\left( {{x}_{1}},{{x}_{2}} \right) \right)=0. \\ 
\end{split}
\label{eq:sym_vid }
\end{equation}
This equation can be imposed by a partial solution
\begin{equation}
\begin{split}
   & \frac{{{\partial }^{2}}a_{j}\left( {{x}_{1}},{{x}_{2}} \right)}{\partial x_{1}^{{{a}_{1}}}\partial x_{1}^{{{a}_{2}}}}{{g}^{{{a}_{1}}{{a}_{2}}}}+\frac{{{\partial }^{2}}a_{j}\left( {{x}_{1}},{{x}_{2}} \right)}{\partial x_{2}^{{{a}_{1}}}\partial x_{2}^{{{a}_{2}}}}{{g}^{{{a}_{1}}{{a}_{2}}}}-{{g}^{2}}{{a}_{j}^{2}}\left( {{x}_{1}},{{x}_{2}} \right)=k_{j}, \\ 
 & \frac{{{\partial }^{2}}b_{j}\left( {{x}_{1}},{{x}_{2}} \right)}{\partial x_{1}^{{{a}_{1}}}\partial x_{1}^{{{a}_{2}}}}{{g}^{{{a}_{1}}{{a}_{2}}}}+\frac{{{\partial }^{2}}b_{j}\left( {{x}_{1}},{{x}_{2}} \right)}{\partial x_{2}^{{{a}_{1}}}\partial x_{2}^{{{a}_{2}}}}{{g}^{{{a}_{1}}{{a}_{2}}}}-{{g}^{2}}{{b}_{j}^{2}}\left( {{x}_{1}},{{x}_{2}} \right)=k_{j}, \\ 
\end{split}
\label{eq:chastkovij_rozvazok }
\end{equation}
Where $k_{j}$- some constants. The condition  \eqref{eq:chastkovij_rozvazok } is an additional condition that determines the equations system for a two-particle gauge field and which was discussed above. The equations are the same for functions   $ a_{j}\left( {{x}_{1}},{{x}_{2}} \right) $ and $ b_{j}\left( {{x}_{1}},{{x}_{2}} \right) $, therefore we will analyze only the equation for   $ a_{j}\left( {{x}_{1}},{{x}_{2}} \right) $.
Further we will introduce the Jacobi coordinates as in  \cite{Chudak:2016,Volkotrub:2015laa}
\begin{equation}
\begin{split}
{{X}^{{{a}_{1}}}}=\frac{1}{2}\left( x_{1}^{{{a}_{1}}}+x_{2}^{{{a}_{1}}} \right),{{y}^{{{a}_{1}}}}=x_{2}^{{{a}_{1}}}-x_{1}^{{{a}_{1}}},\vec{y}=\left( {{y}^{1}},{{y}^{2}},{{y}^{3}} \right).
\end{split}
\label{eq:Jacobi }
\end{equation}
and we will constrict the received equation to a simultaneity subset
\begin{equation}
\begin{split}
x_{1}^{0}=x_{2}^{0},{{y}^{0}}=0. 
\end{split}
\label{eq:Odnochasnist}
\end{equation}
A similar constriction was discussed in detail in the mentioned articles  \cite{Chudak:2016,Volkotrub:2015laa}. After described transformations, we will get the equation
\begin{equation}
\begin{split}
  & {{g}^{{{a}_{1}}{{a}_{2}}}}\frac{{{\partial }^{2}}a_{j}\left( X,\vec{y} \right)}{\partial {{X}^{{{a}_{2}}}}\partial {{X}^{{{a}_{1}}}}}+\left( -4 \right){{\Delta_{{\vec{y}}}} a_{j}}\left( X,\vec{y} \right)-2{{g}^{2}}{{a}_{j}^{2}}\left( X,\vec{y} \right)=2k_{j}, \\ 
 & {{\Delta }_{{\vec{y}}}}\equiv \frac{{{\partial }^{2}}}{{{\left( \partial {{y}^{1}} \right)}^{2}}}+\frac{{{\partial }^{2}}}{{{\left( \partial {{y}^{2}} \right)}^{2}}}+\frac{{{\partial }^{2}}}{{{\left( \partial {{y}^{3}} \right)}^{2}}}. \\ 
\end{split}
\label{eq:Rivnanna_dla_a_Jacobi}
\end{equation}
The equation  (\ref{eq:Rivnanna_dla_a_Jacobi}) obviously has a partial solution  ${{a}_{0,j}}\left( {\vec{y}} \right)$, which depends only on the internal coordinates $\vec{y}$ and satisfies the equation:
\begin{equation}
 \begin{split}
 \left( -2 \right){{\Delta }_{{\vec{y}}}}{{a}_{0,j}}\left( {\vec{y}} \right)-{{g}^{2}}{{\left( {{a}_{0,j}}\left( {\vec{y}} \right) \right)}^{2}}=k_{j}. 
  \end{split}
 \label{eq:Nevelichke_rivnanna}
 \end{equation} 
Let’s introduce a new unknown function ${{a}_{1,j}}\left( X,\vec{y} \right) $   instead the field   $ a_{j}\left( X,\vec{y} \right)$ according the term
\begin{equation}
\begin{split}
a_{j}\left( X,\vec{y} \right)={{a}_{0,j}}\left( {\vec{y}} \right)+{{a}_{1,j}}\left( X,\vec{y} \right).
\end{split}
\label{eq:a0_plus_a1}
\end{equation}
Then we will get the equation for  ${{a}_{1,j}}\left( X,\vec{y} \right) $ 
\begin{equation}
\begin{split}
 -{{g}^{{{a}_{2}}{{a}_{3}}}}\frac{{{\partial }^{2}}{{a}_{1,j}}\left( X,\vec{y} \right)}{\partial {{X}^{{{a}_{2}}}}\partial {{X}^{{{a}_{3}}}}}-{{\left( {{{\hat{H}_{j}}}^{\text{internal}}} \right)}^{2}}\left( {{a}_{1,j}}\left( X,\vec{y} \right) \right)+2{{g}^{2}}{{\left( {{a}_{1,j}}\left( X,\vec{y} \right) \right)}^{2}}=0,
\end{split}
\label{eq:Rivnanna_dla_a1}
\end{equation}
where the denotation is introduced
\begin{equation}
\begin{split}
 {{\left( {{{\hat{H}_{j}}}^{\text{internal}}} \right)}^{2}}\left( {{a}_{1,j}}\left( X,\vec{y} \right) \right)\equiv 4\left( -{{\Delta }_{{\vec{y}}}}{{a}_{1,j}}\left( X,\vec{y} \right)+{{g}^{2}}\left( -{{a}_{0,j}}\left( {\vec{y}} \right) \right){{a}_{1,j}}\left( X,\vec{y} \right) \right).
\end{split}
\label{eq:Hinternal_kvadrat}
\end{equation}

Let’s denote the characteristic length  $l$ of the problem. As you know the one-particle gauge field  $ {{A}_{{{a}_{1}},{{g}_{1}}}}\left( x \right) $ has dimension  $l^{-1}$. Accordingly the two-particles fields should have dimension  $l^{-2}$. The appropriate dimensionless values  ${{\vec{y}}_{1}}$ and ${{a}_{2,j}}\left( {{{\vec{y}}}_{1}} \right)$  instead values $ \vec{y}$ and ${{a}_{0,j}}\left( {\vec{y}} \right)$ are introduced using the terms:
\begin{equation}
\vec{y}=\frac{l}{{{g}^{2}}}{{\vec{y}}_{1}},{{a}_{0,j}}\left( {\vec{y}} \right)={{l}^{-2}}{{a}_{2,j}}\left( {{{\vec{y}}}_{1}} \right).
\label{eq:Bezrazmer}
\end{equation}   
Then the operator  (\ref{eq:Hinternal_kvadrat}) is represented in the form:
\begin{equation}
\begin{split}
   & {{\left( {{{\hat{H}_{j}}}^{\text{internal}}} \right)}^{2}}\left( {{a}_{1,j}}\left( X,\vec{y} \right) \right)\equiv 4{{l}^{-2}}{{g}^{2}}\hat{h}\left( {{a}_{1,j}}\left( X,\vec{y} \right) \right), \\ 
 & \hat{h}\left( {{a}_{1,j}}\left( X,\vec{y}_{1}\right) \right)=-{{\Delta }_{{{{\vec{y}}}_{1}}}}{{a}_{1,j}}\left( X,{{{\vec{y}}}_{1}} \right)+\left( -{{a}_{2,j}}\left( {{{\vec{y}}}_{1}} \right) \right){{a}_{1,j}}\left( X,{{{\vec{y}}}_{1}} \right).\\    
\end{split}
\label{eq:Bezrazmer_h}
\end{equation}
The multiplier  $l^{-2}$ provides the {\flqq right \frqq} dimension for the energy squared. The dimensionless operator $\hat{h}$ formally matches with the particle Hamiltonian which has dimensionless mass 1/2 in the field of a dimensionless potential energy $ V\left( {{{\vec{y}}}_{1}} \right)={{g}^{2}}\left( -{{a}_{2}}\left( {{{\vec{y}}}_{1}} \right) \right)$. This energy can be found as a equation solution (\ref{eq:Nevelichke_rivnanna}) after dimensionlessness (\ref{eq:Bezrazmer}). Let's consider the spherically symmetric solution $ {{a}_{2,j}}\left( \left| {{{\vec{y}}}_{1}} \right| \right)$ of this equation in case  $k_{j}<0, j=0,1,2$. After a standard variables replacement 
\begin{equation}
\left| {{{\vec{y}}}_{1}} \right|={{q}_{j}^{-2}}r,{{a}_{2,j}}\left( \left| {{{\vec{y}}}_{1}} \right| \right)=-\frac{1}{q_{j}}\frac{{{a}_{3,j}}\left( r \right)}{r},
  \label{eq:Zamina}
  \end{equation}  
where $ {{a}_{3,j}}\left( \left| {{{\vec{y}}}_{1}} \right| \right)-$ a new unknown function. We will insert the denotation
\begin{equation}
k_{j}{{l}^{4}}=-{{q}_{j}^{2}}, 
\label{eq:Poznachenna_q}
\end{equation}
which obviously allows the case of negative tems  $k_{j},j=0,1,2$, we obtain :
\begin{equation}
\begin{split}
\frac{{{d}^{2}}{{a}_{3,j}}\left( r \right)}{d{{r}^{2}}}=\frac{\left( {{a}_{3,j}}\left( r \right)-r \right)\left( {{a}_{3,j}}\left( r \right)+r \right)}{2r}.     
\end{split}
\label{eq:Rivnanna_a3}
\end{equation}
If we restrict the consideration of finite equation solutions  (\ref{eq:Nevelichke_rivnanna}) when $\vec{y}=0$  (because there are no physical reasons for singularity at this point), we will add  boundary conditions to the equation  (\ref{eq:Rivnanna_a3})
\begin{align}
 {{\left. {{a}_{3,j}}\left( r \right) \right|}_{r=0}}=0,{{\left. \frac{d{{a}_{3,j}}\left( r \right)}{dr} \right|}_{r=0}}=C_{j}, 
 \label{eq:Granicni_umovi } 
 \end{align} 
Where $C_{j}$ are any constants of which the equation solutions (\ref{eq:Rivnanna_a3}) behavior depends on for different values $ j $. Corresponding analysis of this behavior is given in  \cite{Volkotrub:2015laa}. Let’s consider the analysis of this result. There are two apparent partial solutions in the equation  (\ref{eq:Rivnanna_a3}) 
\begin{align}
  a_{3,j}^{\left( + \right)}\left( r \right)=r,a_{3,j}^{\left( - \right)}\left( r \right)=-r.
  \label{eq:Chastkovi_rozvazki } 
  \end{align}  
The graphs of these solutions separate the half-plane  $\left( r,{{a}_{3,j}} \right),r>0$ into three parts. If solution  ${{a}_{3,j}}\left( r \right)$ gets in the part of half-plane  ${{a}_{3,j}}\left( r \right)>a_{3,j}^{\left( + \right)}\left( r \right)$ (it happens on condition  $C_{j}>1$), this solution ${{a}_{3,j}}\left( r \right)$ will approach to $\left( +\infty  \right)$ on condition  $r\to +\infty $ . The similar solutions described quarks and gluons confinement as shown in \cite{Chudak:2016,Volkotrub:2015laa}. On condition  $C_{j}<1$, the solution asymptotically is approaching to $ a_{3,j}^{\left( - \right)}\left( r \right) $ at $r\to +\infty $  and making  \mbox{\flqq damping oscillation\frqq} about this solution. The solution  $ a_{3,j}^{\left( + \right)}\left( r \right)$ is unstable. The small deflection from this solution to the domain ${{a}_{3,j}}\left( r \right)>a_{3,j}^{\left( + \right)}\left( r \right)$, or ${{a}_{3,j}}\left( r \right)<a_{3,j}^{\left( + \right)}\left( r \right)$ will lead to the asymptotic realization, which is characteristic for each of this domain.

On Pic.1 the examples of numerical equation solutions  (\ref{eq:Rivnanna_a3}) are shown at the constant different values $C$. Herewith graphs for function are given 
\begin{equation}
{{a}_{4,j}}\left( r \right)=\frac{{{a}_{3,j}}\left( r \right)}{r},   
\label{eq:a4}
\end{equation}   
This function matches with the Hamiltonian potential (\ref{eq:Bezrazmer_h}) with accuracy up to the constant multiplier. From pic 1 we can see the operators (\ref{eq:Bezrazmer_h}) and (\ref{eq:Hinternal_kvadrat}) have negative eigenvalues.  These eigenvalues depend on the selected constant $C_{j}$. The eigenfunction of operator which corresponding to the discrete spectrum will approach to zero at $ \left| {{{\vec{y}}}_{1}} \right|\to \infty $. It means these eigenfunction of operator will describe the two gauge bosons bound state. Due to the fact that the potential  ${{a}_{4}}\left( r \right)$ has a finite asymptotic behavior these gauge bosons are not in a confinement state, it differs from \cite{Hoh:2016}.    

\begin{figure}
\center{\includegraphics[scale=0.75]{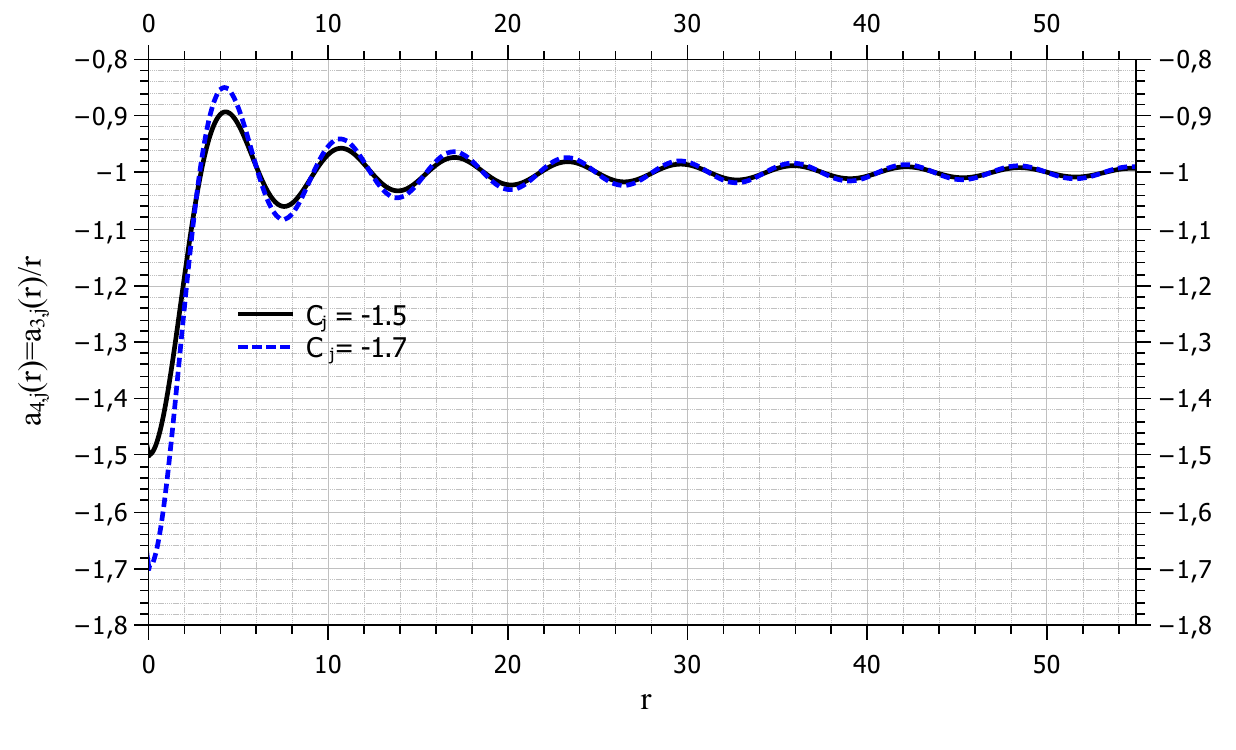}}
%\vskip-3mm\noindent{\footnotesize Results of the numerical equation solution  (\ref{eq:Rivnanna_a3}) at different boundary conditions.}%
%\vskip15pt
\caption{Results of the numerical equation solution  (\ref{eq:Rivnanna_a3}) at different boundary conditions.}
\end{figure}

The negative eigenvalue of operator $ {{\left( {{{\hat{H}_{j}}}^{\text{internal}}} \right)}^{2}}$ with a nonlinear contribution to the equation (\ref{eq:Rivnanna_dla_a1}) for field function  ${{a}_{1,j}}\left( X,\vec{y} \right)$, which are in (\ref{eq:Lagrang_Eiler})-(\ref{eq:Rivnanna_dla_a1}) lead to spontaneous symmetry violation. This nonlinear contribution comes from the self-action of a non-abelian gauge field. Actually the equation (\ref{eq:Rivnanna_dla_a1}) produces the expression for action:
 
\begin{equation}
\begin{split}
  & S_{j}=\int{{{d}^{4}}Xd\vec{y}}\left( \frac{1}{2}{{g}^{{{a}_{5}}{{a}_{6}}}}\frac{\partial {{a}_{1,j}}\left( X,\vec{y} \right)}{\partial {{X}^{{{a}_{5}}}}}\frac{\partial {{a}_{1,j}}\left( X,\vec{y} \right)}{\partial {{X}^{{{a}_{6}}}}} \right.- \\ 
 & \left. -2\sum\limits_{b=1}^{3}{{{\left( \frac{\partial {{a}_{1,j}}\left( X,\vec{y} \right)}{\partial {{y}^{b}}} \right)}^{2}}}+2{{g}^{2}}{{a}_{0,j}}\left( {\vec{y}} \right){{\left( {{a}_{1,j}}\left( X,\vec{y} \right) \right)}^{2}}+\frac{2}{3}{{g}^{2}}{{\left( {{a}_{1,j}}\left( X,\vec{y} \right) \right)}^{3}} \right). \\ 
\end{split}
\label{eq:Dia}
\end{equation}

We integrate by parts the items which contains  $ {{\left( {\partial {{a}_{1,j}}\left( X,\vec{y} \right)}/{\partial {{y}^{b}}}\; \right)}^{2}}$ and $ {{a}_{1,j}}\left( X,\vec{y} \right)$  is a eigenfunction of the operator discrete spectrum  $ {{\left( {{{\hat{H}_{j}}}^{\text{internal}}} \right)}^{2}}$ approaching to zero at  $\left| {\vec{y}} \right|\to +\infty $, we obtain
\begin{equation}
\begin{split}
  & S_{j}=\int{{{d}^{4}}Xd\vec{y}}\left( \frac{1}{2}{{g}^{{{a}_{5}}{{a}_{6}}}}\frac{\partial {{a}_{1,j}}\left( X,\vec{y} \right)}{\partial {{X}^{{{a}_{5}}}}}\frac{\partial {{a}_{1,j}}\left( X,\vec{y} \right)}{\partial {{X}^{{{a}_{6}}}}} \right.- \\ 
 & \left. -\frac{1}{2}{{a}_{1,j}}\left( X,\vec{y} \right){{\left( {{{\hat{H}_{j}}}^{\text{internal}}} \right)}^{2}}{{a}_{1}}\left( X,\vec{y} \right)+\frac{2}{3}{{g}^{2}}{{\left( {{a}_{1,j}}\left( X,\vec{y} \right) \right)}^{3}} \right). \\     
\end{split}
\label{eq:Dia_Hinternal_kv}
\end{equation}

We will denote  $\psi_{a,j} \left( {\vec{y}} \right)$  as a eigenfunction of operator $ {{\left( {{{\hat{H}_{j}}}^{\text{internal}}} \right)}^{2}}$,which is rationed to unit and conforms to the least eigenvalue. This eigenvalue we denote  $\left( -{{\mu _{a,j} }^{2}} \right)$.  Let’s represent the field   $ {{a}_{1,j}}\left( X,\vec{y} \right)$ in the form
\begin{align}
{{a}_{1,j}}\left( X,\vec{y} \right)=\phi_{a,j} \left( X \right)\psi_{a,j} \left( {\vec{y}} \right),
\label{eq:a1cherez_fi} 
\end{align}
Where $\phi_{a,j} \left( X \right)$ - new unknown field functions. We will substitute the term  (\ref{eq:a1cherez_fi}) in (\ref{eq:Dia_Hinternal_kv}), and integrate by internal variables  $\vec{y}$ we obtain:
 \begin{equation}
 \begin{split}
  S_{j}=\int{{{d}^{4}}X}\left( \frac{1}{2}{{g}^{{{a}_{5}}{{a}_{6}}}}\frac{\partial {{\phi }_{a,j}}\left( X \right)}{\partial {{X}^{{{a}_{5}}}}}\frac{\partial {{\phi }_{a,j}}\left( X \right)}{\partial {{X}^{{{a}_{6}}}}} \right.\left. +\frac{\mu _{a,j}^{2}}{2}{{\left( {{\phi }_{a,j}}\left( X \right) \right)}^{2}}+\frac{2}{3}{{g}^{2}}{{Z}_{a,j}}{{\left( \phi_{,ja}\left( X \right) \right)}^{3}} \right),
  \end{split}
 \label{eq:Dia_fi}
 \end{equation}
Where a term is inserted 
\begin{align}
\int{d\vec{y}}{{\left( {{\psi }_{a,j}}\left( {\vec{y}} \right) \right)}^{3}}={{Z}_{a,j}}.
\label{eq:G_bolchoe} 
\end{align}
The function  $\psi_{a,j} \left( {\vec{y}} \right)$ should be subjected to arbitrary  $U\left( 1 \right)-$ transformation. The field  ${{a}_{1,j}}\left( X,\vec{y} \right)$ is built with one-particle real gauge fields and this field is considered as a real field. The field ${{\phi }_{a,j}}\left( X \right)$, is considered as real field too, because Higgs boson decomposes into two photons  \cite{Aad:2015zhl}, and it should be a neutral boson. We expect the field ${{\phi }_{a,j}}\left( X \right)$ will describe processes of creation and annihilation of Higgs bosons after quantization. Therefore the function  $\psi_{a,j} \left( {\vec{y}} \right)$ should be real function too. As a result of oscillation theorem the function $\psi_{a,j} \left( {\vec{y}} \right)$ is an eigenfunction of Hamiltonian (\ref{eq:Bezrazmer_h}), which corresponds to the least eigenvalue. This function takes the value of the same sign for all argument values. This sign can be determined if we introduce the lower boundary of field energy density $ \phi_{a,j} \left( X \right)$. For that we need to agree the sigh for  $\psi_{a,j} \left( {\vec{y}} \right)$ with sigh for  $ \phi_{a,j} \left( X \right)$ so they will be opposite. Let’s introduce  $ \phi_{a,j} \left( X \right)$        
in form  $ {{\phi }_{a,j}}\left( X \right)=\pm \left| {{\phi }_{a,j}}\left( X \right) \right| $, and ${{Z}_{a,j}}=\mp \left| {{Z}_{a,j}} \right|$. The energy density  ${{T}_{00,j}}$ has the form for the field  $ \phi_{a,j} \left( X \right)$:
\begin{equation}
\begin{split}
{{T}_{00,j}}=\frac{1}{2}{{\sum\limits_{{{a}_{5}}=0}^{3}{\left( \frac{\partial {{\phi }_{a,j}}\left( X \right)}{\partial {{X}^{{{a}_{5}}}}} \right)}}^{2}}-\frac{{{\mu_{a,j} }^{2}}}{2}{{\left( {{\phi }_{a,j}}\left( X \right) \right)}^{2}}+\frac{2}{3}{{g}^{2}}\left| {{Z}_{a,j}} \right|{{\left( \left| {{\phi }_{a,j}}\left( X \right) \right| \right)}^{3}},
\end{split}
\label{eq:T00}
\end{equation}
This limitation is achieved by introduction the non-analytic at  ${{T}_{00,j}}$ at $ {{\phi }_{a,j}}\left( X \right)=0.$ But due to a non-zero vacuum value the fields $ \phi_{a,j} \left( X \right)$, we should be quantize small field fluctuations at nearness of this non-zero value. But all values are analytical at the field non-zero value. The appointed non-analytic is insignificant, because the field  {\flqq does not get  \frqq} such values in which this non-analytic will not occur.
 
The value ${{T}_{00,j}}$ has a local minimum at
\begin{align}
{{\phi }_{a,j}}\left( X \right)\equiv {{\phi }_{a0,j}}=\frac{\mu _{a,j}^{2}}{2{{g}^{2}}\left| {{Z}_{a,j}} \right|}.
\label{eq:fi0}
\end{align}
Any deviation $\phi_{a,j} \left( X \right)={{\phi_{a} }_{0,j}}+\delta \phi_{a,j} \left( X \right)$ this field configuration leads to grow energy density ${{T}_{00,j}}$, provided that the function $ \delta \phi_{a,j} \left( X \right)$ will accept sufficiently small values for arbitary values of the argument $X$. More precisely, it is sufficient that the condition $ \delta \phi_{a,j} \left( X \right)>-{{\phi_{a} }_{0,j}} $ is satisfied. Thus, we have field fluctuations around a nonzero vacuum value ${{\phi_{a} }_{0,j}}$. At the same time, as can be seen from (\ref{eq:T00}), expending ${{T}_{00,j}}$ in Tailor series in the vicinity of this value has the form
\begin{align}
  & {{T}_{00,j}}=\frac{1}{2}{{\sum\limits_{{{a}_{5}}=0}^{3}{\left( \frac{\partial \left( {{\phi }_{a,j}}\left( X \right)-{{\phi }_{a0,j}} \right)}{\partial {{X}^{{{a}_{5}}}}} \right)}}^{2}}-\frac{\mu _{a,j}^{6}}{24{{g}^{4}}{{\left| {{Z}_{a,j}} \right|}^{2}}}+ \\
 & +\frac{\mu _{a,j}^{2}}{2}{{\left( {{\phi }_{a,j}}\left( X \right)-{{\phi }_{a0,j}} \right)}^{2}}+\frac{2}{3}{{g}^{2}}\left| {{Z}_{a,j}} \right|{{\left( {{\phi }_{a,j}}\left( X \right)-{{\phi }_{a0,j}} \right)}^{3}}.
\label{eq:Rozclad_T00}
\end{align}

If the value of the vacuum energy of single-particle fields takes a zero energy, then a negative sign of its own value $\left( -\mu _{a,j}^{2} \right)$ of the operator $ {{\left( {{{\hat{H}_{j}}}^{\text{internal}}} \right)}^{2}}$ can be interpreted that the formation of a bound state of two gauge bosons leads to decrease in energy. This is evident from the quadratic field contribution (\ref{eq:T00}). The formation of single-particle fields of a pair of bosons and their binding from vacuum requires no energy expenses, but it passes with the release of energy. We have the gauge bosons condensation process. At the same time, the interaction between the gauge bosons generates the interaction between their bound states. This interaction is described by a cubic field in a plural in (\ref{eq:T00}) and it leads to increase energy, there is based on its sign. The competition between these two processes, and the equilibrium between them, determines the non-zero vacuum expectation (\ref{eq:fi0}) of field $ \phi_{a} \left( X \right)$ and a new minimum of the energy density. Consequently, the value of the mass squared of the bound state of two gauge bosons equals $\mu _{a}^{2}$. This is evident from the expanding (\ref{eq:Rozclad_T00}). This value is determined not only by the {\flqq internal\frqq} energy of the forming particles, but also by the contribution of the {\flqq external\frqq} energy of their interaction with each other, which is not relate to the internal state. Therefore the {\flqq genuine\frqq} operator of the squared energy of the bound state of two gauge bosons can be written in the form:
\begin{equation}
\begin{split}
  & {{\left( \hat{H}_{\text{internal},j}^{\text{true}} \right)}^{2}}={{\left( {{{\hat{H}_{j}}}^{\text{internal}}} \right)}^{2}}+{{\left. \frac{{{d}^{2}}}{d{{\left( {{\phi }_{a,j}}\left( X \right) \right)}^{2}}}\left( \frac{2}{3}{{g}^{2}}\left| {{Z}_{a,j}} \right|{{\left( \left| {{\phi }_{a,j}}\left( X \right) \right| \right)}^{3}} \right) \right|}_{{{\phi }_{a,j}}\left( X \right)={{\phi }_{a0,j}}}}\hat{E}= \\
 & ={{\left( {{{\hat{H}_{j}}}^{\text{internal}}} \right)}^{2}}+2\mu _{a,j}^{2}\hat{E}. \\
\end{split}
\label{eq:Hinternal_true}
\end{equation}
Here the notation $ \hat{E} $  is introduced for a single operator. The corresponding contribution to the \mbox{{\flqq genuine\frqq}} operator is not related to the internal state and therefore in the space of such states must be represented by a single operator.

We now consider the field $b_{j}\left( X,\vec{y} \right)$, which is obtained from the field $b_{j}\left( {{x}_{1}},{{x}_{2}} \right)$ to a subset of the simultaneity (\ref{eq:Odnochasnist}). Having the same equation for $a_{j}\left( X,\vec{y} \right)$, we carry out the same transformations as like as for $a_{j}\left( X,\vec{y} \right)$. In particular, it is presenting in a similar form (\ref{eq:a0_plus_a1})
\begin{equation}
b_{j}\left( X,\vec{y} \right)={{b}_{0,j}}\left( {\vec{y}} \right)+{{b}_{1,j}}\left( X,\vec{y} \right).
  \label{eq:b0_plus_b1}
  \end{equation}
Then for the ${{b}_{0,j}}\left( {\vec{y}} \right)$ we obtain the same equation (\ref{eq:Nevelichke_rivnanna}) as for ${{a}_{0,j}}\left( {\vec{y}} \right)$. We will choose as its solution the same function as in the case of a field $a_{j}\left( X,\vec{y} \right)$ in order to get the simplest situation. We consider again the spherically symmetric solution of the equation (\ref{eq:Nevelichke_rivnanna}) and we will set the same boundary conditions to the (\ref{eq:Granicni_umovi }) with the same value of the constant $C_{j}$ (for each $ j $). Therefore for the field ${{b}_{1,j}}\left( X,\vec{y} \right)$ we get an equation with the same operator $ {{\left( {{{\hat{H}_{j}}}^{\text{internal}}} \right)}^{2}}$ as sure as (\ref{eq:Hinternal_kvadrat}).Presenting ${{b}_{1,j}}\left( X,\vec{y} \right)$ in the form ${{b}_{1,j}}\left( X,\vec{y} \right)={{\phi }_{b,j}}\left( X \right){{\psi }_{b,j}}\left( {\vec{y}} \right)$ where ${{\psi }_{b,j}}\left( {\vec{y}} \right)$ satisfies the equation
\begin{equation}
{{\left( {{{\hat{H}_{j}}}^{\text{internal}}} \right)}^{2}}{{\psi }_{b,j}}\left( {\vec{y}} \right)=\left( -\mu _{b,j}^{2} \right){{\psi }_{b,j}}\left( {\vec{y}} \right),
\label{eq:Hinternal_psi_b}
\end{equation}
use this equation in addition to the operator's own function $ {{\left( {{{\hat{H}_{j}}}^{\text{internal}}} \right)}^{2}}$ also it has a trivial solution ${{\psi }_{b,j}}\left( {\vec{y}} \right)=0.$ We get for the field ${{b}_{1,j}}\left( X,\vec{y} \right)$,choosing exactly this option, partial solution
\begin{equation}
{{b}_{1,j}}\left( X,\vec{y} \right)={{a}_{0,j}}\left( {\vec{y}} \right).
\label{eq:Chastkovij_rozvjazok_b1}
\end{equation}

Taking into account (\ref{eq:aib}), (\ref{eq:a0_plus_a1}),(\ref{eq:a1cherez_fi}), (\ref{eq:Chastkovij_rozvjazok_b1})  contraction the fields ${{\rho }_{j}}\left( {{x}_{1}},{{x}_{2}} \right),j=0,1,2$ on a subset of simultaneity (\ref{eq:Odnochasnist}) take the form of:
\begin{equation}
{\rho }_{j}\left( X,\vec{y} \right)=-\left| {{\phi }_{a,j}}\left( X \right) \right|\left| {{\psi }_{a,j}}\left( {\vec{y}} \right) \right|.
\label{eq:fi_g_1_cherez_psi_a}
\end{equation}
Now we can use the properties of the fields ${\rho }_{j}\left( X,\vec{y} \right) $ set in this section for analysis of interaction with the non-Abelian gauge $SU\left( 2 \right)-$ field various irreplaceable components \eqref{cherez_ro}.

\section{Higgs field associated with an antisymmetric part of a two-particle field \eqref{eq:Rozklad_phi_na_nezvidni_tenzori}.}
Considering the Higgs field
\begin{equation}
{{H}_{{{g}_{1}}}}\left( X \right)=-\left| {{\phi }_{a,2}}\left( X \right) \right|{{e}_{{{g}_{1}}}}.
\label{eq:Higgs_field}
\end{equation}

The action (\ref{eq:Dia_fi}) for the Higgs field can be written in the form:
\begin{equation}
\begin{split}
& S_{2}=\int{{{d}^{4}}X}\left( \frac{1}{2}{{g}^{{{a}_{5}}{{a}_{6}}}}\sum\limits_{{{g}_{1}}=1}^{3}{\left( \frac{\partial {{H}_{{{g}_{1}}}}\left( X \right)}{\partial {{X}^{{{a}_{5}}}}}\frac{\partial {{H}_{{{g}_{1}}}}\left( X \right)}{\partial {{X}^{{{a}_{6}}}}} \right)} \right.- \\
 & \left. +\frac{1}{2}\mu _{a,2}^{2}\sum\limits_{{{g}_{1}}=1}^{3}{{{\left( {{H}_{{{g}_{1}}}}\left( X \right) \right)}^{2}}}-\frac{2}{3}{{g}^{2}}\left| {{Z}_{a,2}} \right|{{\left( \sum\limits_{{{g}_{1}}=1}^{3}{{{\left( {{H}_{{{g}_{1}}}}\left( X \right) \right)}^{2}}} \right)}^{3/2}} \right). \\
\end{split}
\label{eq:Dia_Higgs}
\end{equation}
The presence of the degree ${3}/{2}\;$ does not pose a problem. Even a global transformation of  $SU\left( 2 \right)$ is enough to turn the ${{e}_{{{g}_{1}}}}$ so that it coincides with one of the coordinate meshes. According to (\ref{eq:Higgs_field}),  two of the three components of the Higgs field turned to zero. So far, we have considered expressions for actions as  \eqref{eq:Dia_Higgs} have a global $SU\left( 2 \right)-$ symmetry.
Now it demands the symmetry of the Lagrangian, which corresponds to action (\ref{eq:Dia_Higgs}), which relatives to local $SU\left( 2 \right)-$ transformations we extend the derivatives and add a Lagrangian of a free gauge field.  we will use the fact that because of the validity of the Higgs field, the action  \eqref{eq:Dia_Higgs} can be rewritten in the form
\begin{equation}\label{Dia_Higgs_Hermit}
\begin{split}
& {{S}_{2}}=\int{{{d}^{4}}X}\left( \frac{1}{2}{{g}^{{{a}_{1}}{{a}_{2}}}}\frac{\partial H_{{{g}_{1}}}^{\dagger }\left( X \right)}{\partial {{X}^{{{a}_{1}}}}}\frac{\partial {{H}_{{{g}_{1}}}}\left( X \right)}{\partial {{X}^{{{a}_{2}}}}}+ \right. \\
& \left. +\left( {{{\left( \mu _{a,2}^{\left( 0 \right)} \right)}^{2}}}/{2}\; \right)\left( H_{{{g}_{1}}}^{\dagger }\left( X \right){{H}_{{{g}_{1}}}}\left( X \right) \right)-\frac{2}{3}{{g}^{2}}\left| Z_{a,2}^{\left( 0 \right)} \right|{{\left( H_{{{g}_{1}}}^{\dagger }\left( X \right){{H}_{{{g}_{1}}}}\left( X \right) \right)}^{3/2}} \right). \\
\end{split}
\end{equation}
Here we $ {{H}_{{{g}_{1}}}}\left( X \right), {g}_{1}=1,2,3 $ are considered as column elements and  $ H_{{{g}_{1}}}^{\dagger }\left( X \right) $ denotes the element of the Hermitian conjugate string. The index $ {g} _ {1}, $ is repeating and it usually means summing. Derivatives of the Higgs field are replaced with extended derivatives in \eqref{Dia_Higgs_Hermit}. Derivatives from the Hermitian-conjugate field are replaced on operators that are Hermitian-conjugate to extended derivatives. It got the function of action that we denote $S_{2,A} $:
\begin{equation}
\begin{split}
     & S_{2,A}=\int{{{d}^{4}}X}\left( \frac{1}{2}{{g}^{{{a}_{5}}{{a}_{6}}}}\sum\limits_{{{g}_{1}}=1}^{3}{\left( \left( \frac{\partial {{H}_{{{g}_{1}}}}\left( X \right)}{\partial {{X}^{{{a}_{5}}}}}-g{{A}_{{{a}_{5}},{{g}_{5}}}}\left( X \right){{H}_{{{g}_{2}}}}\left( X \right){{\varepsilon }_{{{g}_{5}}{{g}_{2}}{{g}_{1}}}} \right) \right.} \right.\times  \\
 & \left. \times \left( \frac{\partial {{H}_{{{g}_{1}}}}\left( X \right)}{\partial {{X}^{{{a}_{6}}}}}-g{{A}_{{{a}_{6}},{{g}_{6}}}}\left( X \right){{\varepsilon }_{{{g}_{6}}{{g}_{1}}{{g}_{3}}}}{{H}_{{{g}_{3}}}}\left( X \right) \right) \right)- \\
 & +\frac{1}{2}\mu _{a,2}^{2}\sum\limits_{{{g}_{1}}=1}^{3}{{{\left( {{H}_{{{g}_{1}}}}\left( X \right) \right)}^{2}}}-\frac{2}{3}{{g}^{2}}\left| {{Z}_{a,2}} \right|{{\left( \sum\limits_{{{g}_{1}}=1}^{3}{{{\left( {{H}_{{{g}_{1}}}}\left( X \right) \right)}^{2}}} \right)}^{3/2}}- \\
 & \left. -\frac{1}{4}{{g}^{{{a}_{5}}{{a}_{6}}}}{{g}^{{{a}_{15}}{{a}_{16}}}}\sum\limits_{{{g}_{1}}=1}^{3}{{{F}_{{{a}_{5}}{{a}_{15}},{{g}_{1}}}}\left( X \right){{F}_{{{a}_{6}}{{a}_{16}},{{g}_{1}}}}\left( X \right)} \right). \\
\end{split}
\label{eq:Dia_z_podovjenimi_poxidnimi}
\end{equation}
Here the ${{F}_{{{a}_{1}}{{a}_{2}},{{g}_{1}}}}\left( X \right)$ is a tensor of the non-abelian gauge field.
We select the gauge in which Higgs field looks:
\begin{equation}
 {{H}_{{{g}_{1}}}}\left( X \right)=\left( \begin{matrix}
   0, & 0, & {{H}_{3}}\left( X \right)  \\
\end{matrix} \right)
 \label{eq:Kalibrovka_Higgs}
 \end{equation}

There is the integral expression expands on the degrees of deflection $\delta {{H}_{3}}\left( X \right)$ of the field $ {{H}_{3}}\left( X \right) $ from the value ${{\phi }_{a0}}$ (\ref{eq:fi0}), which delivers a minimum of energy density:
\begin{equation}
 \begin{split}
   & S_{2,A}=\int{{{d}^{4}}X}\left( \frac{1}{2}{{g}^{{{a}_{5}}{{a}_{6}}}}\frac{\partial \delta {{H}_{3}}\left( X \right)}{\partial {{X}^{{{a}_{5}}}}}\frac{\partial \delta {{H}_{3}}\left( X \right)}{\partial {{X}^{{{a}_{6}}}}}-\frac{1}{2}\mu _{a,2}^{2}{{\left( \delta {{H}_{3}}\left( X \right) \right)}^{2}} \right.- \\
 & -\frac{2}{3}{{g}^{2}}\left| {{Z}_{a,2}} \right|{{\left( \delta {{H}_{3}}\left( X \right) \right)}^{3}}- \\
 & -\frac{{{g}^{2}}}{2}\left( 2{{\phi }_{a0,2}}\delta {{H}_{3}}\left( X \right)+{{\left( \delta {{H}_{3}}\left( X \right) \right)}^{2}} \right){{g}^{{{a}_{5}}{{a}_{6}}}}\left( {{A}_{{{a}_{5}},1}}\left( X \right){{A}_{{{a}_{6}},1}}\left( X \right)+{{A}_{{{a}_{5}},2}}\left( X \right){{A}_{{{a}_{6}},2}}\left( X \right) \right) \\
 & -\frac{1}{2}{{\left( g{{\phi }_{a0,2}} \right)}^{2}}{{g}^{{{a}_{5}}{{a}_{6}}}}\left( {{A}_{{{a}_{5}},1}}\left( X \right){{A}_{{{a}_{6}},1}}\left( X \right)+{{A}_{{{a}_{5}},2}}\left( X \right){{A}_{{{a}_{6}},2}}\left( X \right) \right) \\
 & \left. -\frac{1}{4}{{g}^{{{a}_{5}}{{a}_{6}}}}{{g}^{{{a}_{15}}{{a}_{16}}}}\sum\limits_{{{g}_{1}}=1}^{3}{{{F}_{{{a}_{5}}{{a}_{15}},{{g}_{1}}}}\left( X \right){{F}_{{{a}_{6}}{{a}_{16}},{{g}_{1}}}}\left( X \right)} \right). \\
 \end{split}
 \label{eq:Dia_v_okrestnosti_totchki_minimuma}
 \end{equation}
In this expression, we rejected the field-independent term,which derives from the zero contribution in degrees  $\delta {{H}_{3}}\left( X \right)$  in the expansion of a Lagrangian part corresponding to the Higgs field. From (\ref{eq:Dia_v_okrestnosti_totchki_minimuma}) we see that the mass ${{m}_{2}}=g{{\phi }_{a0,2}}$ had got only the components of the gauge field ${{A}_{{{a}_{1}},{{g}_{1}}=1}}\left( x \right)$ and ${{A}_{{{a}_{1}},{{g}_{1}}=2}}\left( x \right),$ but no ${{A}_{{{a}_{1}},{{g}_{1}}=3}}\left( x \right),$ which is inherent in the case when the Higgs field is considered in the vector representation \cite{Georgi_and_Glashow_Priednane_predstavlenna,Slavnov:904821,PeskinSchroeder:16449}

\section{The Higgs field symmetric by internal indices with the zero trace}

We introduce the Higgs tensor field as it was done in \eqref{eq:Higgs_field}:
\begin{equation}\label{Tenzor_Higgs}
{{H}_{{{g}_{1}}{{g}_{2}}}}\left( X \right)=-\left| \phi _{a,2}\left( X \right) \right|e_{{{g}_{1}}{{g}_{2}}}^{s\left( 0 \right)}.
\end{equation}
Taking into account the normalization condition \eqref{umovi_normuvanna} the action \eqref{eq:Dia_fi} can be rewritten through the field \eqref{Tenzor_Higgs}.
\begin{equation}
\begin{split}
  & {S_{1}}=\int{{{d}^{4}}X}\left( \frac{1}{2}{{g}^{{{a}_{1}}{{a}_{2}}}}\frac{\partial {{H}_{{{g}_{1}}{{g}_{2}}}}\left( X \right)}{\partial {{X}^{{{a}_{1}}}}}\frac{\partial {{H}_{{{g}_{1}}{{g}_{2}}}}\left( X \right)}{\partial {{X}^{{{a}_{2}}}}}+ \right. \\
& \left. +\left( {{{\left( \mu _{a,1}^{\left( 0 \right)} \right)}^{2}}}/{2}\; \right)\left( {{H}_{{{g}_{1}}{{g}_{2}}}}\left( X \right){{H}_{{{g}_{1}}{{g}_{2}}}}\left( X \right) \right)-\frac{2}{3}{{g}^{2}}\left| Z_{a,1}^{\left( 0 \right)} \right|{{\left( {{H}_{{{g}_{1}}{{g}_{2}}}}\left( X \right){{H}_{{{g}_{1}}{{g}_{2}}}}\left( X \right) \right)}^{3/2}} \right). \\
\end{split}
\label{eq:Dia_Higgs_tenzor}
\end{equation}

Now we can consider the interaction of the gauge field ${{A}_{{{a}_{1}},{{g}_{1}}}}\left( x \right)$ with the Higgs tensor field  $ {H}{_{{g}_{1}{{g}_{2}}}\left( X \right)} $. It is sufficient to require the Lagrangian's invariance relative to the transformations of the adjoint representation of the local group $SU\left( 2 \right)$, which transforms the Higgs tensor field $ {H}{_{{g}_{1}{{g}_{2}}}\left( X \right)}. $ This Lagrangian corresponds to the action \eqref{eq:Dia_Higgs_tenzor}.

Let us consider the kind of extended derivative of a tensor field. The local transformation of the field has the form:
\begin{equation}\label{Peretvorenna_tenzornogo_pola_Higgsa}
{{{H}'}_{{{g}_{1}}{{g}_{2}}}}\left( X \right)={{D}_{{{g}_{1}}{{g}_{11}}}}\left( \vec{\theta }\left( X \right) \right){{D}_{{{g}_{2}}{{g}_{12}}}}\left( \vec{\theta }\left( X \right) \right){{H}_{{{g}_{11}}{{g}_{12}}}}\left( X \right).
\end{equation}
Here ${{D}_{{{g}_{1}}{{g}_{11}}}}\left( \vec{\theta }\left( x \right) \right),{{D}_{{{g}_{2}}{{g}_{12}}}}\left( \vec{\theta }\left( x \right) \right)$ are elements of the matrices of the adjoint representation of the group $SU\left( 2 \right)$. These matrices can be represented in the form ($ g- $ is the constant of a weak interaction):
\begin{equation}\label{D_cherez_generator}
{{D}_{{{g}_{1}}{{g}_{11}}}}\left( \vec{\theta }\left( x \right) \right)={{\left( \exp \left( {{{g\hat{I}}}_{{{g}_{3}}}}{{\theta }_{{{g}_{3}}}}\left( x \right) \right) \right)}_{{{g}_{1}}{{g}_{11}}}},
\end{equation}
where index ${g}_{3}$ is repeated and it means summation.  ${{\hat{I}}_{{{g}_{3}}}}-$ are generator matrices of the adjoint representation of the group $SU\left( 2 \right)$. Elements of which are expressed through the symbol of Levi-Civita:
\begin{equation}\label{Generator_cherez_Levi_Chivitta}
{{\left( {{{\hat{I}}}_{{{g}_{3}}}} \right)}_{{{g}_{31}}{{g}_{32}}}}={{\varepsilon }_{{{g}_{3}}{{g}_{31}}{{g}_{32}}}}.
\end{equation}
It was calculated the derivative of both parts of equality \eqref{Peretvorenna_tenzornogo_pola_Higgsa}. We got that
the ordinary derivative of the field $ {{H}_{{{g}_{1}}{{g}_{2}}}}\left( X \right) $ behaves in this way, at the local $SU\left( 2 \right)-$ transformation:
\begin{equation}\label{Poxidna_vid_zakonu_tenzora}
\begin{split}
  & \frac{\partial {{{{H}'}}_{{{g}_{1}}{{g}_{2}}}}\left( X \right)}{\partial {{X}^{{{a}_{1}}}}}={{\left( D\left( \vec{\theta }\left( X \right) \right) \right)}_{{{g}_{1}}{{g}_{21}}}}{{\left( D\left( \vec{\theta }\left( X \right) \right) \right)}_{{{g}_{2}}{{g}_{22}}}}\times  \\
& \times \left( \frac{\partial {{H}_{{{g}_{21}}{{g}_{22}}}}\left( X \right)}{\partial {{X}^{{{a}_{1}}}}}+g\frac{\partial {{\theta }_{{{g}_{13}}}}\left( X \right)}{\partial {{X}^{{{a}_{1}}}}}{{\left( {{I}_{{{g}_{13}}}} \right)}_{{{g}_{21}}{{g}_{11}}}}{{H}_{{{g}_{11}}{{g}_{22}}}}\left( X \right)+g\frac{\partial {{\theta }_{{{g}_{14}}}}\left( X \right)}{\partial {{X}^{{{a}_{1}}}}}{{\left( {{I}_{{{g}_{14}}}} \right)}_{{{g}_{22}}{{g}_{12}}}}{{H}_{{{g}_{21}}{{g}_{12}}}}\left( X \right) \right). \\
\end{split}
\end{equation}
Based on the structure  of {\flqq excesses \frqq} additions in  \eqref{Poxidna_vid_zakonu_tenzora}, which contains derivatives of the local transformation and determine the difference in the law of transformation of the field itself $ {H}_{{{g}_{1}}{{g}_{2}}} \left( X\right) $ and its derivative, we define an extended derivative of the tensor  $ {H}_{{{g}_{1}}{{g}_{2}}} \left( X\right) $ as follows:
\begin{equation}\label{Poxidna_vid_tenzora}
\begin{split}
  & {{{\hat{D}}}_{{{a}_{1}}}}\left( A \right){{H}_{{{g}_{1}}{{g}_{2}}}}\left( X \right)=\frac{\partial {{H}_{{{g}_{1}}{{g}_{2}}}}\left( X \right)}{\partial {{X}^{{{a}_{1}}}}}-g{{\left( {{I}_{{{g}_{13}}}} \right)}_{{{g}_{1}}{{g}_{21}}}}{{A}_{{{a}_{1}},{{g}_{13}}}}\left( X \right){{H}_{{{g}_{21}}{{g}_{2}}}}\left( X \right) \\
& -g{{\left( {{I}_{{{g}_{14}}}} \right)}_{{{g}_{2}}{{g}_{22}}}}{{A}_{{{a}_{1}},{{g}_{14}}}}\left( X \right){{H}_{{{g}_{1}}{{g}_{22}}}}\left( X \right) .\\
\end{split}
\end{equation}
Here ${{A}_{{{a}_{1}},{{g}_{1}}}}\left( x \right)$ is the gauge field, the law of transformation of which must compiles {\flqq excesses \frqq} additions in \eqref{Poxidna_vid_tenzora}. After the local $SU\left( 2 \right)-$ transformation  \eqref{Peretvorenna_tenzornogo_pola_Higgsa}, we will have:
\begin{equation}\label{Podovjena_pohidna_strix}
\begin{split}
  & {{{\hat{D}}}_{{{a}_{1}}}}\left( {{A}'} \right){{{{H}'}}_{{{g}_{1}}{{g}_{2}}}}\left( X \right)=\frac{\partial {{{{H}'}}_{{{g}_{1}}{{g}_{2}}}}\left( X \right)}{\partial {{X}^{{{a}_{1}}}}}-g{{\left( {{I}_{{{g}_{13}}}} \right)}_{{{g}_{1}}{{g}_{21}}}}{{{{A}'}}_{{{a}_{1}},{{g}_{13}}}}\left( X \right){{{{H}'}}_{{{g}_{21}}{{g}_{2}}}}\left( X \right) \\
& -g{{\left( {{I}_{{{g}_{14}}}} \right)}_{{{g}_{2}}{{g}_{22}}}}{{{{A}'}}_{{{a}_{1}},{{g}_{14}}}}\left( X \right){{{{H}'}}_{{{g}_{1}}{{g}_{22}}}}\left( X \right) .\\
\end{split}
\end{equation}
Here $ {{{{A}'}}_{{{a}_{1}},{{g}_{1}}}} $ is components of the gauge field after the local $SU\left( 2 \right)-$ transformation,which has the form for the gauge field
\begin{equation}\label{Peretvorenna_A}
{{{A}'}_{{{a}_{1}},{{g}_{12}}}}\left( X \right)=\left( \frac{\partial {{\theta }_{{{g}_{11}}}}\left(X \right)}{\partial {{x}^{{{a}_{1}}}}}+{{A}_{{{a}_{1}},{{g}_{11}}}}\left( X \right) \right)D_{{{g}_{11}}{{g}_{12}}}^{-1}\left( \vec{\theta }\left(X\right) \right).
\end{equation}
Taking into account \eqref{Peretvorenna_tenzornogo_pola_Higgsa} and \eqref{Poxidna_vid_zakonu_tenzora}, we can write
\begin{equation}
\begin{split}
& {{{\hat{D}}}_{{{a}_{1}}}}\left( {{A}'} \right){{{{H}'}}_{{{g}_{1}}{{g}_{2}}}}\left( X \right)={{\left( D\left( \vec{\theta }\left( X \right) \right) \right)}_{{{g}_{1}}{{g}_{21}}}}{{\left( D\left( \vec{\theta }\left( X \right) \right) \right)}_{{{g}_{2}}{{g}_{22}}}}\times  \\
& \times \left( \frac{\partial {{H}_{{{g}_{21}}{{g}_{22}}}}\left( X \right)}{\partial {{X}^{{{a}_{1}}}}}+g\frac{\partial {{\theta }_{{{g}_{13}}}}\left( X \right)}{\partial {{X}^{{{a}_{1}}}}}{{\left( {{I}_{{{g}_{13}}}} \right)}_{{{g}_{21}}{{g}_{11}}}}{{H}_{{{g}_{11}}{{g}_{22}}}}\left( X \right)+g\frac{\partial {{\theta }_{{{g}_{14}}}}\left( X \right)}{\partial {{X}^{{{a}_{1}}}}}{{\left( {{I}_{{{g}_{14}}}} \right)}_{{{g}_{22}}{{g}_{12}}}}{{H}_{{{g}_{21}}{{g}_{12}}}}\left( X \right) \right) \\
& -g{{\left( {{I}_{{{g}_{13}}}} \right)}_{{{g}_{1}}{{g}_{21}}}}{{{{A}'}}_{{{a}_{1}},{{g}_{13}}}}\left( X \right){{\left( D\left( \vec{\theta }\left( X \right) \right) \right)}_{{{g}_{21}}{{g}_{23}}}}{{\left( D\left( \vec{\theta }\left( X \right) \right) \right)}_{{{g}_{2}}{{g}_{22}}}}{{H}_{{{g}_{23}}{{g}_{22}}}}\left( X \right) \\
& -g{{\left( {{I}_{{{g}_{14}}}} \right)}_{{{g}_{2}}{{g}_{22}}}}{{{{A}'}}_{{{a}_{1}},{{g}_{14}}}}\left( X \right){{\left( D\left( \vec{\theta }\left( X \right) \right) \right)}_{{{g}_{1}}{{g}_{21}}}}{{\left( D\left( \vec{\theta }\left( X \right) \right) \right)}_{{{g}_{22}}{{g}_{23}}}}{{H}_{{{g}_{21}}{{g}_{23}}}}\left( X \right). \\
\end{split}
\label{eq:Rozpusana_podovjena_poxidna}
\end{equation}
This expression can be identically rewritten in the form:
\begin{equation}\label{Totognij_vid}
\begin{split}
  & {{{\hat{D}}}_{{{a}_{1}}}}\left( {{A}'} \right){{{{H}'}}_{{{g}_{1}}{{g}_{2}}}}\left( X \right)={{\left( D\left( \vec{\theta }\left( X \right) \right) \right)}_{{{g}_{1}}{{g}_{21}}}}{{\left( D\left( \vec{\theta }\left( X \right) \right) \right)}_{{{g}_{2}}{{g}_{22}}}}\times  \\
& \times \left( \frac{\partial {{H}_{{{g}_{21}}{{g}_{22}}}}\left( X \right)}{\partial {{X}^{{{a}_{1}}}}}+g\frac{\partial {{\theta }_{{{g}_{13}}}}\left( X \right)}{\partial {{X}^{{{a}_{1}}}}}{{\left( {{I}_{{{g}_{13}}}} \right)}_{{{g}_{21}}{{g}_{11}}}}{{H}_{{{g}_{11}}{{g}_{22}}}}\left( X \right)+g\frac{\partial {{\theta }_{{{g}_{14}}}}\left( X \right)}{\partial {{X}^{{{a}_{1}}}}}{{\left( {{I}_{{{g}_{14}}}} \right)}_{{{g}_{22}}{{g}_{12}}}}{{H}_{{{g}_{21}}{{g}_{12}}}}\left( X \right) \right. \\
& -g\left( {{\left( {{D}^{-1}}\left( \vec{\theta }\left( X \right) \right) \right)}_{{{g}_{21}}{{g}_{25}}}}{{\left( {{I}_{{{g}_{13}}}} \right)}_{{{g}_{25}}{{g}_{24}}}}{{\left( D\left( \vec{\theta }\left( X \right) \right) \right)}_{{{g}_{24}}{{g}_{23}}}} \right){{{{A}'}}_{{{a}_{1}},{{g}_{13}}}}\left( X \right){{H}_{{{g}_{23}}{{g}_{22}}}}\left( X \right) \\
& \left. -g{{{{A}'}}_{{{a}_{1}},{{g}_{14}}}}\left( X \right)\left( {{\left( {{D}^{-1}}\left( \vec{\theta }\left( X \right) \right) \right)}_{{{g}_{22}}{{g}_{25}}}}{{\left( {{I}_{{{g}_{14}}}} \right)}_{{{g}_{25}}{{g}_{24}}}}{{\left( D\left( \vec{\theta }\left( X \right) \right) \right)}_{{{g}_{24}}{{g}_{23}}}} \right) \right){{H}_{{{g}_{21}}{{g}_{23}}}}\left( X \right).. \\
\end{split}
\end{equation}
Taking into account the definition of the adjoint representation of the group $SU\left( 2 \right)$. The expression \eqref{Totognij_vid} can be rewritten in this way
\begin{equation}\label{Totognij_vid1}
\begin{split}
  & {{{\hat{D}}}_{{{a}_{1}}}}\left( {{A}'} \right){{{{H}'}}_{{{g}_{1}}{{g}_{2}}}}\left( X \right)={{\left( D\left( \vec{\theta }\left( X \right) \right) \right)}_{{{g}_{1}}{{g}_{21}}}}{{\left( D\left( \vec{\theta }\left( X \right) \right) \right)}_{{{g}_{2}}{{g}_{22}}}}\times  \\
& \times \left( \frac{\partial {{H}_{{{g}_{21}}{{g}_{22}}}}\left( X \right)}{\partial {{X}^{{{a}_{1}}}}}+g\frac{\partial {{\theta }_{{{g}_{13}}}}\left( X \right)}{\partial {{X}^{{{a}_{1}}}}}{{\left( {{I}_{{{g}_{13}}}} \right)}_{{{g}_{21}}{{g}_{11}}}}{{H}_{{{g}_{11}}{{g}_{22}}}}\left( X \right)+g\frac{\partial {{\theta }_{{{g}_{14}}}}\left( X \right)}{\partial {{X}^{{{a}_{1}}}}}{{\left( {{I}_{{{g}_{14}}}} \right)}_{{{g}_{22}}{{g}_{12}}}}{{H}_{{{g}_{21}}{{g}_{12}}}}\left( X \right) \right. \\
& -g{{D}_{{{g}_{13}}{{g}_{14}}}}\left( \vec{\theta }\left( X \right) \right){{\left( {{I}_{{{g}_{14}}}} \right)}_{{{g}_{21}}{{g}_{23}}}}{{{{A}'}}_{{{a}_{1}},{{g}_{13}}}}\left( X \right){{H}_{{{g}_{23}}{{g}_{22}}}}\left( X \right) \\
& \left. -g{{{{A}'}}_{{{a}_{1}},{{g}_{14}}}}\left( X \right){{D}_{{{g}_{14}}{{g}_{13}}}}\left( \vec{\theta }\left( X \right) \right){{\left( {{I}_{{{g}_{13}}}} \right)}_{{{g}_{22}}{{g}_{23}}}} \right){{H}_{{{g}_{21}}{{g}_{23}}}}\left( X \right). .\\
\end{split}
\end{equation}
Taking into account the conversion law of the gauge field \eqref{Peretvorenna_A}, then {\flqq  excesses \frqq} additive components, which contains derivatives of the parameters of the local $SU\left( 2 \right)-$ transformation are compensated. We obtain the same transformation law for the extended derivative \eqref{Poxidna_vid_tenzora} as for the field \eqref{Peretvorenna_tenzornogo_pola_Higgsa}.

We must extend the derivatives in the expression \eqref{eq:Dia_Higgs_tenzor} to construct a locally $SU\left( 2 \right)-$ symmetric expression for an action. In this case, as in the case of \eqref{Dia_Higgs_Hermit}, we need to enter a Hermitian-conjugate field, which coincides with the field ${{H}_{{{g}_{1}}{{g}_{2}}}}\left( X \right)$, because components of the field  ${{H}_{{{g}_{1}}{{g}_{2}}}}\left( X \right)$ is a real and its symmetry relative to the permutation of the indexes ${g}_{1} $ and ${g}_{2} $. The Hermitage conjugation of the extended derivative has the form
\begin{equation}\label{Podovjena_poxidna_Hermit}
\begin{split}
  & {{\left( {{{\hat{D}}}_{{{a}_{1}}}}\left( A \right){{H}_{{{g}_{1}}{{g}_{2}}}}\left( X \right) \right)}^{\dagger }}=\frac{\partial {{H}_{{{g}_{2}}{{g}_{1}}}}\left( X \right)}{\partial {{X}^{{{a}_{1}}}}}-g{{H}_{{{g}_{2}}{{g}_{21}}}}\left( X \right){{\left( {{I}_{{{g}_{13}}}} \right)}_{{{g}_{21}}{{g}_{1}}}}{{A}_{{{a}_{1}},{{g}_{13}}}}\left( X \right) \\
& -g{{H}_{{{g}_{22}}{{g}_{1}}}}\left( X \right){{\left( {{I}_{{{g}_{14}}}} \right)}_{{{g}_{22}}{{g}_{2}}}}{{A}_{{{a}_{1}},{{g}_{14}}}}\left( X \right). \\
\end{split}
\end{equation}
The requirement of the local $SU\left( 2 \right)$ invariance will be fulfilled if we go from the action \eqref{eq:Dia_Higgs_tenzor} to
\begin{equation}\label{Dia1_s_podovjenimi_poxidnimi}
\begin{split}
 & {{S}_{1,A}}=\int{{{d}^{4}}X}\left( \frac{1}{2}{{g}^{{{a}_{1}}{{a}_{2}}}}\left( \frac{\partial {{H}_{{{g}_{2}}{{g}_{1}}}}\left( X \right)}{\partial {{X}^{{{a}_{1}}}}}-g{{H}_{{{g}_{21}}{{g}_{2}}}}\left( X \right){{A}_{{{a}_{1}},{{g}_{11}}}}\left( X \right){{\varepsilon }_{{{g}_{11}}{{g}_{21}}{{g}_{1}}}} \right. \right.- \\
 & \left. -g{{H}_{{{g}_{1}}{{g}_{21}}}}\left( X \right){{A}_{{{a}_{1}},{{g}_{11}}}}\left( X \right){{\varepsilon }_{{{g}_{11}}{{g}_{21}}{{g}_{2}}}} \right)\times  \\
 & \times \left( \frac{\partial {{H}_{{{g}_{1}}{{g}_{2}}}}\left( X \right)}{\partial {{X}^{{{a}_{2}}}}}-g{{A}_{{{a}_{2}},{{g}_{12}}}}\left( X \right){{\varepsilon }_{{{g}_{12}}{{g}_{1}}{{g}_{22}}}}{{H}_{{{g}_{22}}{{g}_{2}}}}\left( X \right)-g{{A}_{{{a}_{2}},{{g}_{12}}}}\left( X \right){{\varepsilon }_{{{g}_{12}}{{g}_{2}}{{g}_{22}}}}{{H}_{{{g}_{1}}{{g}_{22}}}}\left( X \right) \right)+ \\
 & \left. +\left( {{{\left( \mu _{a,1}^{\left( 0 \right)} \right)}^{2}}}/{2}\; \right)\left( {{H}_{{{g}_{1}}{{g}_{2}}}}\left( X \right){{H}_{{{g}_{1}}{{g}_{2}}}}\left( X \right) \right)-\frac{2}{3}{{g}^{2}}\left| Z_{a,1}^{\left( 0 \right)} \right|{{\left( {{H}_{{{g}_{1}}{{g}_{2}}}}\left( X \right){{H}_{{{g}_{1}}{{g}_{2}}}}\left( X \right) \right)}^{3/2}} \right)- \\
 & -\frac{1}{4}{{g}^{{{a}_{1}}{{a}_{11}}}}{{g}^{{{a}_{2}}{{a}_{12}}}}{{F}_{{{a}_{1}}{{a}_{2}},{{g}_{1}}}}\left( X \right){{F}_{{{a}_{11}}{{a}_{12}},{{g}_{1}}}}\left( X \right). \\
\end{split}
\end{equation}
Here, as in \eqref{eq:Dia_z_podovjenimi_poxidnimi}, the value ${{F}_{{{a}_{1}}{{a}_{2}},{{g}_{1}}}}\left( X \right)$ denotes the tensor of the non-abelian gauge field.

After transformations, the expression \eqref{Dia1_s_podovjenimi_poxidnimi} can be written in the form:
\begin{equation}
\begin{split}
& {{S}_{1,A}}=\int{{{d}^{4}}X}\left( \frac{1}{2}{{g}^{{{a}_{1}}{{a}_{2}}}}\left( \frac{\partial {{H}_{{{g}_{2}}{{g}_{1}}}}\left( X \right)}{\partial {{X}^{{{a}_{1}}}}}\frac{\partial {{H}_{{{g}_{1}}{{g}_{2}}}}\left( X \right)}{\partial {{X}^{{{a}_{2}}}}}- \right. \right. \\
& -2{{g}^{2}}\left( \left( {{H}_{{{g}_{21}}{{g}_{2}}}}\left( X \right){{H}_{{{g}_{21}}{{g}_{2}}}}\left( X \right) \right)\left( {{A}_{{{a}_{1}},{{g}_{11}}}}\left( X \right){{A}_{{{a}_{2}},{{g}_{11}}}}\left( X \right) \right)- \right. \\
& -\left( {{A}_{{{a}_{1}},{{g}_{11}}}}\left( X \right){{H}_{{{g}_{11}}{{g}_{2}}}}\left( X \right) \right)\left( {{H}_{{{g}_{21}}{{g}_{2}}}}\left( X \right){{A}_{{{a}_{2}},{{g}_{21}}}}\left( X \right) \right)- \\
& \left. \left. -{{\varepsilon }_{{{g}_{11}}{{g}_{21}}{{g}_{1}}}}{{\varepsilon }_{{{g}_{12}}{{g}_{2}}{{g}_{22}}}}{{H}_{{{g}_{21}}{{g}_{2}}}}\left( X \right){{H}_{{{g}_{1}}{{g}_{22}}}}\left( X \right){{A}_{{{a}_{1}},{{g}_{11}}}}\left( X \right){{A}_{{{a}_{2}},{{g}_{12}}}}\left( X \right) \right) \right) \\
& \left. +\left( {{{\left( \mu _{a,1}^{\left( 0 \right)} \right)}^{2}}}/{2}\; \right)\left( {{H}_{{{g}_{1}}{{g}_{2}}}}\left( X \right){{H}_{{{g}_{1}}{{g}_{2}}}}\left( X \right) \right)-\frac{2}{3}{{g}^{2}}\left| Z_{a,1}^{\left( 0 \right)} \right|{{\left( {{H}_{{{g}_{1}}{{g}_{2}}}}\left( X \right){{H}_{{{g}_{1}}{{g}_{2}}}}\left( X \right) \right)}^{3/2}} \right)- \\
& \left. -\frac{1}{4}{{g}^{{{a}_{1}}{{a}_{11}}}}{{g}^{{{a}_{2}}{{a}_{12}}}}{{F}_{{{a}_{1}}{{a}_{2}},{{g}_{1}}}}\left( X \right){{F}_{{{a}_{11}}{{a}_{12}},{{g}_{1}}}}\left( X \right) \right). \\
\end{split}
\label{eq:Dia1_s_podovjenimi_poxidnimi1}
\end{equation}
This expression can be simplified. We use the symmetric tensor $e_{{{g}_{1}}{{g}_{2}}}^{s\left( 0 \right)}$, which can be reduced to a diagonal form, because it was considered the global $SU\left( 2 \right)-$ symmetry with the corresponding global transformation. At the same time its trace should be stored and therefore equal to zero. Taking into account it, we can write down
\begin{equation}\label{diagonalnij_vid}
e_{{{g}_{1}}{{g}_{2}}}^{s\left( 0 \right)}={{m}_{1}}{{\delta }_{{{g}_{1}}1}}{{\delta }_{{{g}_{2}}1}}+{{m}_{2}}{{\delta }_{{{g}_{1}}2}}{{\delta }_{{{g}_{2}}2}}-\left( {{m}_{1}}+{{m}_{2}} \right){{\delta }_{{{g}_{1}}3}}{{\delta }_{{{g}_{2}}3}}.
\end{equation}
Here ${{m}_{1}},{{m}_{2}}$ are arbitrary coefficients, which due to the condition of normalization \eqref{umovi_normuvanna} must satisfy demand:
\begin{equation}\label{normuvanna_m}
m_{1}^{2}+m_{2}^{2}+{{\left( {{m}_{1}}+{{m}_{2}} \right)}^{2}}=1.
\end{equation}
Taking into account the view of \eqref{diagonalnij_vid} and \eqref{Tenzor_Higgs},the expression for  \eqref{eq:Dia1_s_podovjenimi_poxidnimi1} can be represented as:
\begin{equation}
\begin{split}
& {{S}_{1,A}}=\int{{{d}^{4}}X}\left( \frac{1}{2}{{g}^{{{a}_{1}}{{a}_{2}}}}\left( \frac{\partial {{\phi }_{a,1}}\left( X \right)}{\partial {{X}^{{{a}_{1}}}}}\frac{\partial {{\phi }_{a,1}}\left( X \right)}{\partial {{X}^{{{a}_{2}}}}}- \right. \right. \\
& -2{{g}^{2}}{{\left( {{\phi }_{a,1}}\left( X \right) \right)}^{2}}\left( {{A}_{{{a}_{1}},1}}\left( X \right){{A}_{{{a}_{2}},1}}\left( X \right)\left( 1-{{\left( {{m}_{1}} \right)}^{2}}+2{{m}_{2}}\left( {{m}_{1}}+{{m}_{2}} \right) \right) \right. \\
& +{{A}_{{{a}_{1}},2}}\left( X \right){{A}_{{{a}_{2}},2}}\left( X \right)\left( 1-{{\left( {{m}_{2}} \right)}^{2}}+2{{m}_{1}}\left( {{m}_{1}}+{{m}_{2}} \right) \right) \\
& \left. \left. +{{A}_{{{a}_{1}},3}}\left( X \right){{A}_{{{a}_{2}},3}}\left( X \right)\left( 1-{{\left( {{m}_{1}}+{{m}_{2}} \right)}^{2}}-2{{m}_{2}}{{m}_{1}} \right) \right) \right). \\
\end{split}
\label{eq:Dia_cherez_m1m2}
\end{equation}
Fields $ {{A}_{{{a}_{1}},1}}\left( X \right) $ and $ {{A}_{{{a}_{1}},2}}\left( X \right) $ are connected with ${{W}^{+}}$ and with ${{W}^{-}}-$ bosons, which are antiparticles to each other. When these fields interact with Higgs field, they should receive identical masses. This means that the coefficients for  $ {{A}_{{{a}_{1}},1}}\left( X \right){{A}_{{{a}_{2}},1}}\left( X \right) $ and $ {{A}_{{{a}_{1}},2}}\left( X \right){{A}_{{{a}_{2}},2}}\left( X \right) $ in \eqref{eq:Dia_cherez_m1m2} must be the same. Therefore, we have an equation:
\begin{equation}\label{Rivnanna_dla_m12}
1-{{\left( {{m}_{1}} \right)}^{2}}+2{{m}_{2}}\left( {{m}_{1}}+{{m}_{2}} \right)=1-{{\left( {{m}_{2}} \right)}^{2}}+2{{m}_{1}}\left( {{m}_{1}}+{{m}_{2}} \right).
\end{equation}
This equation has two solutions:
\begin{equation}\label{Dva_rozvazku}
{{m}_{1}}={{m}_{2}},{{m}_{1}}=-{{m}_{2}}.
\end{equation}
At first we consider a case
\begin{equation}\label{m1ravnom2}
{{m}_{1}}={{m}_{2}}=m.
\end{equation}
Then, due to the normalization condition \eqref{normuvanna_m} we have
\begin{equation}\label{mkvadrat_odna_chestaja}
{{m}^{2}}=\frac{1}{6}.
\end{equation}
Taking into account it, instead of \eqref{eq:Dia_cherez_m1m2} we have
\begin{equation}
\begin{split}
& {{S}_{1,A}}=\int{{{d}^{4}}X}\left( \frac{1}{2}{{g}^{{{a}_{1}}{{a}_{2}}}}\left( \frac{\partial {{\phi }_{a,1}}\left( X \right)}{\partial {{X}^{{{a}_{1}}}}}\frac{\partial {{\phi }_{a,1}}\left( X \right)}{\partial {{X}^{{{a}_{2}}}}}- \right. \right. \\
& \left. -{{g}^{2}}{{\left( {{\phi }_{a,1}}\left( X \right) \right)}^{2}}\sum\limits_{{{g}_{1}}=1}^{2}{{{A}_{{{a}_{1}},{{g}_{1}}}}\left( X \right){{A}_{{{a}_{2}},{{g}_{1}}}}\left( X \right)} \right)+ \\
& +\frac{{{\left( \mu _{a,1}^{\left( 0 \right)} \right)}^{2}}}{2}{{\left( {{\phi }_{a,1}}\left( X \right) \right)}^{2}}-\frac{2}{3}{{g}^{2}}\left| Z_{a,1}^{\left( 0 \right)} \right|{{\left| {{\phi }_{a,1}}\left( X \right) \right|}^{3}}- \\
& \left. -\frac{1}{4}{{g}^{{{a}_{1}}{{a}_{11}}}}{{g}^{{{a}_{2}}{{a}_{12}}}}{{F}_{{{a}_{1}}{{a}_{2}},{{g}_{1}}}}\left( X \right){{F}_{{{a}_{11}}{{a}_{12}},{{g}_{1}}}}\left( X \right) \right). \\
\end{split}
\label{eq:Dia_odna_chestaja}
\end{equation}

It is considered the other case
\begin{equation}\label{m1_ravno_minusm2}
{{m}_{1}}=-{{m}_{2}}.
\end{equation}
Taking into account \eqref{normuvanna_m} we get:
\begin{equation}
\begin{split}
& {{S}_{1,A}}=\int{{{d}^{4}}X}\left( \frac{1}{2}{{g}^{{{a}_{1}}{{a}_{2}}}}\left( \frac{\partial {{\phi }_{a,1}}\left( X \right)}{\partial {{X}^{{{a}_{1}}}}}\frac{\partial {{\phi }_{a,1}}\left( X \right)}{\partial {{X}^{{{a}_{2}}}}}- \right. \right. \\
& \left. -{{g}^{2}}{{\left( {{\phi }_{a,1}}\left( X \right) \right)}^{2}}\left( \sum\limits_{{{g}_{1}}=1}^{2}{{{A}_{{{a}_{1}},{{g}_{1}}}}\left( X \right){{A}_{{{a}_{2}},{{g}_{1}}}}\left( X \right)}+4{{A}_{{{a}_{1}},3}}\left( X \right){{A}_{{{a}_{2}},3}}\left( X \right) \right) \right)+ \\
& +\frac{{{\left( \mu _{a,1}^{\left( 0 \right)} \right)}^{2}}}{2}{{\left( {{\phi }_{a,1}}\left( X \right) \right)}^{2}}-\frac{2}{3}{{g}^{2}}\left| Z_{a,1}^{\left( 0 \right)} \right|{{\left| {{\phi }_{a,1}}\left( X \right) \right|}^{3}}- \\
& \left. -\frac{1}{4}{{g}^{{{a}_{1}}{{a}_{11}}}}{{g}^{{{a}_{2}}{{a}_{12}}}}{{F}_{{{a}_{1}}{{a}_{2}},{{g}_{1}}}}\left( X \right){{F}_{{{a}_{11}}{{a}_{12}},{{g}_{1}}}}\left( X \right) \right). \\
\end{split}
\label{eq:S_masoj_Z}
\end{equation}

So, it can be seen from  \eqref{eq:Dia_odna_chestaja} in the case ${{m}_{1}}={{m}_{2}}$ as in the case, which discussed in the previous section, due to interaction with the Higgs field, only fields ${{A}_{{{a}_{1}},{{g}_{1}}=1}}\left( X \right),{{A}_{{{a}_{1}},{{g}_{1}}=2}}\left( X \right)$ receive a mass and ${{W}^{\pm }}-$ bosons , which associated with them. $Z-$ boson doesn't receive a mass, because it associates with the field ${{A}_{{{a}_{1}},{{g}_{1}}=3}}\left( X \right)$.
Instead, the case ${{m}_{1}}={-{m}_{2}}$ is more interesting because, as seen from \eqref{eq:S_masoj_Z}, in this case all three components of the gauge field receive a mass ${{W}^{\pm }}-$ bosons and the $Z-$boson.
This mass can be obtained, as usual, by expanding the field $ {{\phi }_{a,1}}\left( X \right) $ in the Taylor series in the neighborhood of a nonzero vacuum value  $ {{\phi }_{a0,1}}, $ which defined by the formula  \eqref{eq:fi0}. The zero application of this expansion in \eqref{eq:S_masoj_Z} leads to such contributions to the masses $ {{W}^{\pm }} $ and $Z-$ bosons due to interaction with the symmetric zero trace Higgs tensor field ${{H}_{{{g}_{1}}{{g}_{2}}}}\left( X \right)$:
\begin{equation}\label{massi1}
\Delta m_{W}^{\left( 1 \right)}=\sqrt{2}g{{\phi }_{a0,1}}, \Delta m_{Z}^{\left( 1 \right)}=2\sqrt{2}g{{\phi }_{a0,1}}.
\end{equation}
The index $\left( 1 \right)$ means that we are talking about contributions to the mass due to interaction with a symmetric zero-trace part of the expansion \eqref{eq:Rozklad_phi_na_nezvidni_tenzori}. Significantly, the contribution to the mass of the $Z-$ boson is greater than the contribution to the mass of $ {{W}^{\pm }}- $ bosons. As we know from the experiment \cite{Olive:2016xmw} the mass of the $Z-$boson is greater than the mass of $ {{W}^{\pm }}- $ bosons. The interaction of the gauge fields with the antisymmetric part of the decomposition \eqref{eq:Rozklad_phi_na_nezvidni_tenzori} generates only the masses of $ {{W}^{\pm }}- $ bosons, if the specified relation between contributions were not executed, then the model would not correspond to the experiment. It has seen from \eqref{eq:Dia_v_okrestnosti_totchki_minimuma}, that the contributions to the mass due to the interaction of the gauge fields with the antisymmetric part of the decomposition \eqref{eq:Rozklad_phi_na_nezvidni_tenzori} have the form:
\begin{equation}\label{masi_2}
\Delta m_{W}^{\left( 2 \right)}=g{{\phi }_{a0,2}},\Delta m_{Z}^{\left( 2 \right)}=0.
\end{equation}
As the gauge field ${{A}_{{{a}_{1}},{{g}_{1}}}}\left( X \right)$ interacts with both an antisymmetric and a symmetric part of the decomposition \eqref{eq:Rozklad_phi_na_nezvidni_tenzori}, the Lagrangian interaction will contain the sum of the corresponding contributions both in \eqref{eq:S_masoj_Z} and in \eqref{eq:Dia_v_okrestnosti_totchki_minimuma}. Therefore, the masses of $ {{W}^{\pm }}$ and $Z-$bosons will be obtained by adding the contributions \eqref{massi1} and \eqref{masi_2} in the considered model:
\begin{equation}\label{masi_sum}
{{m}_{W}}=g\left( \sqrt{2}{{\phi }_{a0,1}}+{{\phi }_{a0,2}} \right),{{m}_{Z}}=2\sqrt{2}g{{\phi }_{a0,1}}.
\end{equation}
These formulas, as already noted above, allow us to satisfy the experimental correlation ${{m}_{Z}}>{{m}_{W}}.$

A rather specific kind of tensor was needed $e_{{{g}_{1}}{{g}_{2}}}^{s\left( 0 \right)}$ to obtain these results, which is reduced by conversion of the adjoint representation to
\begin{equation}\label{Specificheskij_vid_tenzora}
e_{{{g}_{1}}{{g}_{2}}}^{s\left( 0 \right)}=\frac{1}{\sqrt{2}}\left( {{\delta }_{{{g}_{1}}1}}{{\delta }_{{{g}_{2}}1}}-{{\delta }_{{{g}_{1}}2}}{{\delta }_{{{g}_{2}}2}} \right)
\end{equation}
In addition, the vector ${{e}_{{{g}_{1}}}},$ which considered in the previous section, as seen from \eqref{eq:Kalibrovka_Higgs}, must also be reduced to a specific form
\begin{equation}\label{Specificheskij_vid_vectora}
{{e}_{{{g}_{1}}}}=\left( \begin{matrix}
0 & 0 & 1  \\
\end{matrix} \right)
\end{equation}
At the same time, both representations \eqref{Specificheskij_vid_tenzora} and \eqref{Specificheskij_vid_vectora} should be implemented in the same basis of the linear space on which the group representation is implemented $SU\left( 2 \right).$ This can be achieved by considering the process of constructing a two-part gauge field \eqref{eq:Lagrang_Eiler}-\eqref{eq:Videlenie_skalara }, which included components ${{A}_{{{a}_{1}},{{g}_{1}}=1}}\left( {{x}_{b}} \right),{{A}_{{{a}_{1}},{{g}_{1}}=2}}\left( {{x}_{c}} \right),b,c=1,2.$ These components are easy to express through the fields $W_{{{a}_{1}}}^{\pm }\left( {{x}_{b}} \right).$  It means that the Higgs boson will be considered not merely as a bound state of two gauge bosons, namely, as a bound state of ${{W}^{+}}$ and ${{W}^{-}-}$ boson, but not $Z-$boson. In this case, all components of the tensor \eqref{eq:Videlenie_skalara }, and hence its irreducible parts \eqref{eq:Rozklad_phi_na_nezvidni_tenzori} in which at least one of the indices is equal to three, are zero. Under these conditions, the vector ${{\phi }_{{{g}_{3}}}}\left( {{x}_{1}},{{x}_{2}} \right),$ which is determined by the ratio
\eqref{antisimetricnij_cherez_LeviChivitta} has only one component different from zero and this component corresponds to the value ${{g}_{3}}=3$. Now the vector ${{e}_{{{g}_{1}}}}$, which defined by \eqref{cherez_ro}, has the desired form \eqref{Specificheskij_vid_vectora}.

As non-zero components of the tensor $e_{{{g}_{1}}{{g}_{2}}}^{s\left( 0 \right)}$ have indices that are equal to 1 and 2, then its diagonalization can be achieved only by rotations in the plane of the space of internal indices, which is orthogonal to the vector ${{e}_{{{g}_{1}}}}.$ Therefore, these turns do not change the desired form of this vector. In addition, components  $e_{{{g}_{1}}{{g}_{2}}}^{s\left( 0 \right)}$ in which at least one index is 3 at such rotation will remain equal to zero. Then the diagonal form of this tensor with the necessity coincides with \eqref{Specificheskij_vid_tenzora}.

\section{Discussion of results and conclusions}
In the proposed model, the Higgs boson has a weak isospin, in contrast to the Standard Model, what refers to the experimental data \cite{Olive:2016xmw} on the channels of its decay. It is considered as the bound state of ${{W}^{+}}$ i ${{W}^{-}}-$bosons, which are particles with weak isospin 1. Therefore, such a bound state may have a weak isospin or 0, or 1, or 2, respectively, with three terms in the expansion on irreducible tensors \eqref{eq:Rozklad_phi_na_nezvidni_tenzori}. The decomposition channel for two photons \cite{Olive:2016xmw} suggests that the Higgs boson can be in a state with weak isospin 0. Channels of decomposition into two particles with weak isospin ${1}/{2}\;$, for example, an electron and a positron, add an opportunity to observe the Higgs boson in a state with weak isospin 1. And the four lepton channels give a value of 2. Consequently, known decay channels indicate that the Higgs boson is an inappropriate state for a weak isospin, but only its integer values of 0,1 or 2 can be obtained when it measured. This property is reproduced by the formula \eqref{eq:Rozklad_phi_na_nezvidni_tenzori}.

An essential feature of the proposed model is that the Higgs field self-interaction, which provides a non-zero vacuum value, is considered not as an independent non-gauge interaction, but as a manifestation of the self-interaction of a non-abelian gauge field. Namely, since the Higgs boson is regarded as a bound state of ${{W}^{+}}$ and ${{W}^{-}}-$bosons, the interaction between Higgs bosons is a consequence of a non-Abelian weak $SU\left( 2 \right)-$ interaction of gauge bosons. This is formally seen from the procedure for obtaining equations for a two-part gauge field in Section 3. The authors admit that this procedure is somewhat artificial, but it allows us to describe some important things in the experiment. In particular, with the help of a similar procedure in works \cite{Chudak:2016, Volkotrub:2015laa} it was described the confinement of quarks and gluons. As can be seen from the foregoing considerations, it allows the contribution to the Lagrangian with {\flqq unnatural \frqq} sign in front of the mass squared, and it is not simply to enter it, as in the Standard Model.

Unfortunately, the proposed model in this paper does not solve the problem of non-gauge input in the Standard Model of the Yukawa interaction of fermion fields with the Higgs field. In the physical point of view, this fermions interact with $ W- $ bosons, and the Higgs boson is formed, within the framework of the considered model, from $ W- $ bosons, such an interaction should exist. How to introduce it on the basis of the gauge principle is still unclear. Even if we solve this problem, since all fundamental fermion fields interact with  $ W- $ bosons, then by mediating the Higgs field they can interact with each other. On the one hand, this may allow to explain the neutrino mixing, which leads to neutrino oscillations \cite{SuperCamiokaNDEPhysRevLett.81.1562,SNOcollaborationPhysRevLett.87.071301,SNO_1_PhysRevLett.89.011301}. On the other hand, it is unclear why such oscillations are peculiar to neutrinos, that is why the oscillations between the electron and the muon do not stem from both of them through  $ W- $ bosons, which can interact with the Higgs field. This field can {\flqq mix\frqq} like corresponding neutrinos.

Another problem of the Standard Model which was not solved in this paper is due to the fact that it is impossible to make a gauge-invariant partition of gauge fields on charged fields of $W-$bosons and the insecure field of the $Z-$boson. When it moves to a new gauge, we need to go to the real fields ${{A}_{{{a}_{1}},{{g}_{1}}}}\left( x \right)$, make a gauge transform and then select fields $W-$ and $Z-$ bosons. Exactly in this sense we understood ${{W}^{+}}$ , ${{W}^{-}}$ and $Z-$ fields in this paper.

%
%\begin{flushright}
%{\footnotesize Received 07.05.09}
%\end{flushright}
%\end{thebibliography}

\bibliographystyle{aipnum4-1}
\bibliography{references-UTF8}

\end{document}